\documentclass[12pt,preprint,aps,nofootinbib,shownopacs]{revtex4}
\usepackage{graphicx}
\newcommand*{\be}{\begin{equation}}
\newcommand*{\ee}{\end{equation}}
\newcommand*{\bea}{\begin{eqnarray}}
\newcommand*{\eea}{\end{eqnarray}}

\newcommand{\nn}{\nonumber}
\newcommand{\Frac}[2]{\frac{\displaystyle{#1}}{\displaystyle{#2}}}
\newcommand{\lsim}{\raise0.3ex\hbox{$\;<$\kern-0.75em\raise-1.1ex\hbox{$\sim\;$}}}
\newcommand{\gsim}{\raise0.3ex\hbox{$\;>$\kern-0.75em\raise-1.1ex\hbox{$\sim\;$}}}
\newcommand{\eq}[1]{Eq.~(\ref{#1})}
\newcommand{\Ri}{{\rm R}}
\newcommand{\Le}{{\rm L}}

\begin{document}
\preprint{FTUV/06-0512}
\preprint{IFIC/06-05}
\preprint{FERMILAB-PUB-06-067-T}
\title{Analysis of enhanced $\tan \beta$ effects in MFV GUT scenarios}
\author{Enrico~Lunghi}
\email{lunghi@fnal.gov}
\affiliation{
Fermi National Accelerator Laboratory \\
P.O. Box 500, Batavia, IL, 60510-0500, USA
}
\author{Werner~Porod}
\email{porod@ific.uv.es}
\affiliation{IFIC - Instituto de F\'{\i}sica Corpuscular, CSIC, E-46071,
 Val\`encia, Spain.}

\author{Oscar~Vives}
\email{oscar.vives@uv.es}
\affiliation{Departament de F\'{\i}sica Te\`orica and IFIC, Universitat de Val\`encia-CSIC, E-46100, Burjassot, Spain.}

\begin{abstract}
We analyse a minimal supersymmetric standard model (MSSM) taking a
minimal flavour violation (MFV) structure at the GUT scale. We
evaluate the parameters at the electroweak scale taking into account
the full flavour structure in the evolution of the renormalization group
equations. We concentrate mainly on the
decay $B_s \to \mu^+ \mu^-$ and its correlations with other observables
like BR($b \to s \gamma$), BR($b \to s l^+ l^-$), $\Delta M_{B_s}$ and
the anomalous magnetic moment of the muon. We restrict our analysis to the
regions in
parameter space consistent with the dark matter constraints.
We find that the BR($B_s  \to \mu^+ \mu^-$) can exceed the current
experimental limit in the regions of parameter space which are allowed by all
other constraints thus providing an additional bound on supersymmetric
parameters. This holds even in the constrained MSSM. Assuming an hypothetical
measurement of BR$(B_s  \to \mu^+ \mu^-) \simeq 10^{-7}$ we analyse the
predicted MSSM spectrum and flavour violating decay modes of supersymmetric
particles which are found to be small.
\end{abstract}

\maketitle

\section{Introduction}

The Minimal Supersymmetric Standard Model (MSSM) is the simplest
supersymmetric (SUSY) structure that includes the SM gauge group and
matter content.  However, this definition does not unambiguously fix a
single supersymmetric model. As is well-known, SUSY cannot be an
exact symmetry of nature and must be broken. Therefore, to
specify completely the model, it is necessary to fix the soft
breaking terms: this amounts to 124 parameters at the electroweak
scale. Although these parameters would be fixed in the presence of a
truly fundamental theory, from the point of view of the effective
theory they are only constrained by low energy experimental data.  In
fact, our freedom in this enormous parameter space is already severely
limited by phenomenological constraints. The accumulating concordance
between the SM expectations and the vast range of experimental results
in FCNC and $CP$ violation point toward a low energy new physics which
is, at least approximately, flavour blind \cite{fcncreview}.  In this spirit,
the
so-called Minimal Flavour Violating (MFV) MSSM is an MSSM where the
only non-trivial flavour structures present are the usual Yukawa
couplings while all sfermion masses and trilinear couplings are
completely family universal \cite{MFV}. Notice that realistic flavour models have been
constructed where the deviations from this minimal flavour structure are
rather small \cite{flavour}.

Several interesting
flavour changing processes can take place for large values of $\tan
\beta$ at a rate much larger than the corresponding expectations from
the SM. A particularly interesting process in this scenario is the
decay $B_s \to \mu^+ \mu^-$ which has been shown to receive
enhancements of even three orders of magnitude with respect to the
standard model expectations for large $\tan \beta$ and small
pseudoscalar Higgs masses \cite{Choudhury:1998ze,Babu:1999hn}.  This
process has been carefully analysed in previous studies. In fact, two
different points of view have been taken in these analyses. On the one
hand, several groups examined this decay in the framework of an
effective MFV MSSM model defined at the electroweak scale where masses
and mixing angles are
uncorrelated~\cite{Chankowski:2000ng,Bobeth:2002ch,
Isidori:2001fv,Buras:2002wq,Buras:2002vd,Isidori:2006pk,Carena:2006ai}. In these works large
enhancement factors are relatively easy to find because several
correlations with other observables are lost.  On the other hand,
various other works repeated this analysis in a MFV MSSM defined at
the GUT scale in terms of a reduced number of parameters
\cite{Choudhury:1998ze,Dedes:2001fv,Arnowitt:2002cq,Dedes:2002zx,
Mizukoshi:2002gs,Baek:2002wm,Baer:2002gm,Ibrahim:2002fx,Blazek:2003hv,
Kane:2003wg,Ellis:2005sc}.  Most of these analysis take place in the
framework of a constrained MSSM (CMSSM) defined at $M_{\rm GUT}$ in
terms of $m_0, m_{1/2}, A_0, \tan \beta$ and ${\rm sgn}(\mu)$ and
neglecting flavour mixing in the RGE evolution.  In principle it is
possible to consider different family-universal masses for different
representations under the gauge group. However, to our knowledge, in
the literature so far only the masses of the Higgs multiplets are
allowed to be different from a common sfermion mass
\cite{Dermisek:2003vn,Baek:2004et,Ellis2006} with the only exception of
Reference~\cite{Bartl:2001}. In any of these GUT scenarios,
several relations between the sfermion masses, gaugino masses, Higgs
masses and mixing angles are obtained and although large enhancements
of the $B_s \to \mu^+ \mu^-$ branching ratio are still allowed, they
are clearly more difficult to obtain.

In this work we analyse this decay in the framework of a
completely generic MFV MSSM defined at the GUT scale. In this general
MFV scenario, the soft masses and trilinear terms for
different representations under the SM gauge group are different,
although family-universal. This scenario can be realised in different
ways, as RGE-induced splittings between different multiplets in the
running from $M_{\rm Planck}$ to $M_{\rm GUT}$ in $SU(5)$ or flipped
$SU(5)$ models \cite{Polonsky}, or in models of direct unification at the
Planck-scale \cite{Ibanez}. Another difference with respect to previous
analysis is the fact that, for the first time, we use the full two
loop renormalization group equations (RGE) \cite{Martin:1993zk} with the
complete flavour
structure of the different flavour matrices to determine the Yukawa
couplings and soft SUSY breaking parameters at the electro-weak scale.
Furthermore, we take into account all the relevant constraints in our analysis
including the dark matter relic abundance \cite{wmap3rdyear} together
with the updated
results on processes like $B_s$--$\overline B_s$ mixing
\cite{Abazov:2006dm,CDFBs:2006},
muon anomalous magnetic moment \cite{Bennett:2004pv},
$b\to s l^+ l^-$ \cite{Abe:2005.qqqq,Aubert:2004it} and the LEP and Tevatron
constraints from searches of
SUSY particles \cite{Eidelman:2004wy}. All the analysis is done with
updated values of the input
parameters and in particular we use the last value of the top quark mass
\cite{CDF-d0}.
Finally, we take into account the
latest constraints on the BR($B_s \to \mu^+ \mu^-$) from D\O~and CDF
\cite{last}.
In this framework, we study the neutral Higgs effects in the decay $B_s
\to \mu^+ \mu^-$ and the $B_s$ mass difference, comparing the results of
the CMSSM with our generic MFV scenario. We take special care
in checking the compatibility of these processes with other $\tan
\beta$ enhanced observables like the $b \to s \gamma$ decay or the muon
anomalous magnetic moment.

The structure of the paper is as follows. We start with a discussion of
the origin of Higgs-mediated flavour changing neutral currents (FCNC)
and study their phenomenological effects in the $B_s \to \mu^+ \mu^-$
decay and the $B_s$ mass difference which are enhanced by additional
$\tan \beta$ factors. In section \ref{sec:correl} we analyse the
correlations of these Higgs-mediated processes with other
$\tan \beta$ enhanced observables like the $b \to s \gamma$ decay, the
muon anomalous magnetic moment and the $b \to s l^+ l^-$ decay. In
section \ref{sec:bsmumu}  we describe the procedure used in our
numerical analysis and present our results for the $B_s \to \mu^+ \mu^-$ decay
in the various scenarios. Furthermore we discuss the various
correlations with other observables. Section \ref{sec:collider} is
devoted to the discussion of
the expected collider phenomenology if the decay $B_s \to \mu^+ \mu^-$
is found with a branching ratio close to the present bounds.
Finally in section  \ref{sec:conclu} we present our conclusions.

\section{Higgs mediated FCNC's}
\label{sec:HiggsFCNCs}

The MSSM is classically a type-II two-Higgs doublet model. This means
that one of the Higgs fields, $H_u$, couples only to the up quarks
while the other, $H_d$, couples only to the down quarks. In this way,
dangerous tree--level Higgs-mediated FCNC's are avoided. However, the
presence of the $\mu$-term in the superpotential breaks the symmetry
protecting this structure. Therefore quantum corrections modify this
type-II structure and generate couplings of $H_d^\dagger$ to up quarks
and of $H_u^\dagger$ to down quarks, hence reintroducing
Higgs-mediated FCNC's. We use in our numerical analysis the full
expressions for the one-loop vertex and self-energy corrections as
given in Reference \cite{Buras:2002vd} (see also
\cite{Dedes:2002er}). All our numerical results are valid for any
value of $\tan \beta$.  However, in the following we summarize the
main features of Higgs-mediated FCNCs, relevant for our analysis, in
the limit of large $\tan \beta$.  A more detailed discussion can be
found in Reference \cite{Buras:2002vd}.
\begin{figure}
\includegraphics[width=8cm]{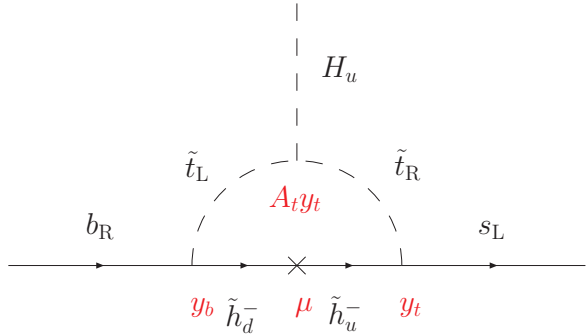}
\caption{Flavour changing coupling of the up Higgs-boson, $H_u$, to the
down-type quarks.}
\label{fig:Hudown}
\end{figure}
In Fig. \ref{fig:Hudown} we present an example of FCNC coupling of the
up-type Higgs, $H_u$, to the down quarks through a stop--chargino
loop. There are also similar couplings mediated by gluino loops and
the corresponding $H_d$ couplings to up quarks. The effective Yukawa
Lagrangian at 1 loop is,
\bea \label{L1loop}
{\mathcal L}_{\rm Yuk} = - \bar d_{\Le_i} Y^d_{ij} d_{\Ri_j} H_d +
\bar d_{\Le_i} \left(\Delta Y^d\right)_{ij} d_{\Ri_j} H_u^* + \bar
u_{\Le_i} Y^u_{ij} u_{\Ri_j} H_u + \bar u_{\Le_i} \left(\Delta
Y^u\right)_{ij} u_{\Ri_j} H_d^*,
\eea
where $Y^d_{ij}$ and $Y^u_{ij}$ include also the 1 loop corrections
to the $H_d$--down-quark and $H_u$--up-quark couplings \cite{Buras:2002vd}.
After spontaneous breaking of the electroweak symmetry, this
Lagrangian gives rise to new contributions to the quark mass
matrices. Working in the basis where the tree-level down-quark
Yukawa couplings are diagonal and neglecting the subdominant loop
corrections in the large $\tan \beta$ limit, the contributions to the down
quark mass matrix are:
\bea {\mathcal L}^d_{\rm
mass} =- \bar d_{\Le_i} \left(Y^d_{ii} \delta_{ij} \frac{v_d}{\sqrt{2}} +
\left(\Delta Y^d\right)_{ij} \frac{v_u}{\sqrt{2}}\right) d_{\Ri_j}+
{\rm H.c.}\nn \\
 = \frac{v_d}{\sqrt{2}}~ \bar d_{\Le_i} \left(Y^d_{ii}\delta_{ij}+
\tan \beta \left(\Delta Y^d\right)_{ij} \right) d_{\Ri_j}+{\rm H.c.},
\eea
where we can see the appearance of a $\tan \beta$ enhanced correction to the
down quark mass due to the presence of the wrong-type Higgs vacuum expectation
value (vev).
In the same basis, we obtain for the neutral Higgs
couplings to the down quarks \cite{Buras:2002vd}:
\bea \label{Lneutral}
{\mathcal L}_{S^0} = - \bar d_{\Le_i} \left(\Frac{\overline m_{d_j}}
{v_d~(1+\tilde\epsilon_j \tan \beta)}~ x^S_d~\delta_{ij} +
\left(\Delta Y^d\right)_{ij} ~x^S_u \right) d_{\Ri_j}~ S^0 +{\rm
H.c.},
\eea
with $S^0 = (H^0,h^0,A^0,G^0)$,
$x^S_d = (\cos \alpha,-\sin \alpha,i \sin \beta,-i \cos \beta)$ and
$x^S_u = (\sin \alpha,\cos \alpha,-i \cos \beta,-i \sin \beta)$. Here,
$\overline m_j$ denotes the running quark mass that can be extracted
by experiments, the parameter $\tilde \epsilon_j$ accounts for the
$\tan \beta$ enhanced corrections to the down quark eigenvalues
\cite{Buras:2002vd,Dedes:2002er} and it is given by
\bea \label{epsilons}
\tilde \epsilon_j &=&
\epsilon_0 + |K_{3j}|^2 y_t^2 ~\epsilon_Y ,\nn\\
\epsilon_0 &=& - 2\alpha_s ~\Frac{\mu}{3 \pi ~m_{\tilde g}}~
H_2\left(\frac{m_{d_{\Le}}^2}{m_{\tilde g}^2},\frac{m_{d_{\Ri}}^2}{m_{\tilde g}^2}
\right),\nn\\
\epsilon_Y &=& \Frac{A_t}{16 \pi^2 ~\mu}~
H_2\left(\frac{m_{t_\Le}^2}{\mu^2},\frac{m_{t_\Ri}^2}{\mu^2}\right),
\eea
where the loop function is defined in Appendix D. Using the fact
that\footnote{We have that $H_2(x,x)$ goes from $-0.18$
for $x=4$ to $-1.7$ for $x=0.1$}
$H_2(1,1)=-1/2$ and taking  the limit
$m_{\tilde g} \simeq \mu \simeq m_{\tilde t_a} \simeq - A_t$, we obtain $\epsilon_0\simeq {\rm sign}(\mu)\cdot 0.012$ and $\tilde
\epsilon_3 \simeq {\rm sign}(\mu)\cdot 0.015$.
Finally $\left(\Delta Y^d\right)_{ij}$ is given by:
\bea
\label{deltaY}
\left(\Delta Y^d\right)_{ij} =  y_{d_j} \left(\epsilon_0
\delta_{ij} +  y_t^2 K_{3 i}^* K_{3 j} ~\epsilon_Y \right),
\eea
with $y_{d_j}$ being the eigenvalues of the tree-level Yukawa matrix.

After diagonalizing the full one-loop down-quark mass matrix, the
couplings to the neutral Higgs-bosons are given by
\bea \label{Lneutral2}
{\mathcal L}_{S^0} &=& - \bar d_{\Le_i}~  \bigg(\Frac{\overline m_{d_j}}
{v_d~(1+\tilde\epsilon_j \tan \beta)}~ \left(x^S_d+ \tilde \epsilon_j x^S_u
\right)~\delta_{ij} +\Frac{\overline m_{d_j}}
{v_d~(1+\tilde\epsilon_j \tan \beta)^2}~ y_t^2 \lambda^{ij}_0
\nn \\&&
~~~~~~~~ \epsilon_Y~
\left(~x^S_u - x^S_d \tan \beta \right)\bigg) d_{\Ri_j}~ S^0 +{\rm
H.c.},
\eea
with $\lambda^{ii}_0= 0$ and $\lambda^{ij}_0= K_{3 i}^* K_{3 j}$.
Hence, from this equation we can see that we do have
Higgs-mediated FCNC's after the diagonalization of the one-loop mass matrix and
in particular $\tan \beta$ enhanced off-diagonal couplings proportional
to  $x^S_d$ which come from the rotation diagonalizing the one loop Yukawa
couplings.
In the following we study the phenomenological consequences of these
FCNC Higgs couplings in the down-quark sector\footnote{Other papers study
the large $\tan \beta$
effects in $\tau$ decays \cite{Brignole:2003iv,Paradisi:2005tk,
Masiero:2005wr}, flavour changing decays of the Higgs bosons
\cite{Curiel:2003uk,Kanemura:2004cn} and collider processes
\cite{Ibrahim:2004cf,Ibrahim:2004gb,Hollik:2005as}.}.

\subsection{The decay $B_s \to \mu^+ \mu^-$}

The present experimental bound on this decay from
the CDF and D\O~\cite{last,prev} collaborations is,
\bea
 {\rm BR}(B_s \to \mu^+ \mu^-) < 1.0 \times 10^{-7}~~~~ {\rm 95 \% C.L.}
\eea

This process is specially sensitive to the
presence of Higgs mediated FCNC's.
Using the Lagrangian given in
\eq{Lneutral2} we obtain the dominant $\tan \beta$-enhanced
neutral-Higgs terms to this decay which contribute mainly through two
effective operators:
\bea
{\cal H}_{\rm eff} = -\Frac{2 G_{\rm F}}{\sqrt{2}} \Frac{\alpha}{2 \pi \sin^2
  \theta_W} K_{tb}^* K_{ts} \left[ C_S~ \overline
m_b \left(\bar b_{\rm R} s_{\rm
      L}\right)\left(\bar l~ l\right)~ + ~C_P~\overline m_b \left(\bar b_{\rm R} s_{\rm
      L}\right)\left(\bar l~ \gamma_5~ l\right) \right].
\eea
The dominant contributions to the Wilson coefficients $C_S$ and $C_P$ in
the large $\tan \beta$ limit are
\bea
C_S&\simeq&\Frac{m_\mu \overline m_t^2}{4 M_W^2} \Frac{{A_t}~\tan^3
\beta~ H_2\left(m_{t_\Le}^2/\mu^2,m_{t_\Ri}^2/\mu^2\right) }{\mu ~
(1+\epsilon_0 \tan \beta)(1+\tilde \epsilon_3 \tan \beta)}
\left[\frac{\sin(\alpha-\beta)\cos\alpha}{M^2_{H^0}}-
\frac{\cos(\alpha-\beta)\sin\alpha}{M^2_{h^0}}\right],\nn\\
C_P&\simeq& -\Frac{m_\mu \overline m_t^2}{4 M_W^2} \Frac{{A_t}~\tan^3
\beta~ H_2\left(m_{t_\Le}^2/\mu^2,m_{t_\Ri}^2/\mu^2\right) }{\mu~
(1+\epsilon_0 \tan \beta)(1+\tilde \epsilon_3 \tan \beta)}
\left[\frac{1}{M^2_{A}}\right].
\eea
For large $\tan \beta$ we can neglect the $M_{h^0}^2$ contribution
and, taking $M_{H^0} \simeq M_{A}$, we find
that\footnote{However, a difference between these
masses is possible for light pseudoscalar Higgs and can give rise to
sizable effects
\cite{Paradisi:2005tk,Masiero:2005wr,Isidori:2006jh}. In our numerical
calculation, we use the full expression and we have found that for
$M_{A}\gsim 250$ GeV, $C_S\simeq C_P$ is a very good
approximation.} $C_S\simeq
C_P$.  Therefore, the leading contribution to the $b_\Ri
s_\Le \to \mu^+ \mu^-$ amplitude is given by:
\bea
\label{bsllampl1}
{\mathcal A} (b_\Ri s_\Le  \to \mu^+ \mu^-)
&\propto& \overline m_t^2 \Frac{\overline m_b m_\mu ~K_{tb}^* K_{ts}}
{(1+\epsilon_0 \tan \beta)
(1+\tilde \epsilon_3 \tan \beta)}~\Frac{A_t}{\mu} \Frac{\tan^3\beta}{M_A^2}
~H_2\left(m_{t_\Le}^2/\mu^2,m_{t_\Ri}^2/\mu^2\right).
\eea
Using this formula, we obtain an estimate of the branching ratio in
the large $\tan \beta$ region as,
\bea
\label{BRbsll}
{\rm BR}\left(B_s\to \mu^+ \mu^-\right) &=& 2.2 \times 10^{-6}
\left[\Frac{\tau_{B_s}}{1.5~{\rm ps}}\right]\left[\Frac{F_{B_s}}{230~{\rm MeV}}
\right]^2\left[\Frac{\overline m_t}{175~{\rm GeV}}\right]^4
\left[\Frac{\tan \beta}{50}\right]^6 \left[\Frac{350~{\rm GeV}}{M_A}\right]^4
\nn\\&&
\Frac{A_t^2}{\mu^2}~~
\Frac{\left(H_2\left(m_{t_\Le}^2/\mu^2,m_{t_\Ri}^2/\mu^2\right)\right)^2}
{(1+\epsilon_0 \tan \beta)^2(1+\tilde \epsilon_3 \tan \beta)^2}
\eea
and we see that the branching ratio for this decay scales like $
\tan^6 \beta/m_A^4$. As we will show below, given the new experimental
bounds from CDF and D\O,~this process is already able to probe a
sizable part of the large $\tan \beta$ parameter space.

\subsection{$B^0_{d,s}$ mass differences}

The CDF and D\O~collaborations \cite{CDFBs:2006} have recently presented the
first  experimental results on $B_s$ mass-difference:
\bea
\Delta M_{B_s} =
\left(17.33^{+0.42}_{-0.21}({\rm stat}) \pm 0.07 ({\rm syst})\right)~
{\rm ps}^{-1},
\eea
while the $B_d$ mass difference is
\bea
\Delta M_{B_d} =
\left(0.500^{+0.42}_{-0.21}({\rm stat}) \pm 0.07 ({\rm syst})\right)~
{\rm ps}^{-1}.
\eea

The basic formula for the $B_q$ mass difference is \cite{Buras:2002vd}
\bea
\label{basicDMBq}
\Delta M_{B_q} = \Frac{G_F^2 M_W^2}{6 \pi} M_{B_q} \eta_{B} F_{B_q}^2
\hat B_{B_q} |K_{tq}|^2 |F_{tt}^{q}|,
\eea
with
\bea
F_{tt}^{q} = \left[  S_0(m_t^2/M_W^2) + \frac{1}{4 r} C_{\rm new}^{\rm VLL}
  (\mu) \right] + \frac{1}{4 r} C_{1}^{\rm VRR} (\mu) + \bar P_1^{\rm LR}
C_{1}^{\rm LR} (\mu) +  \bar P_2^{\rm LR} C_{2}^{\rm LR} (\mu) + \nn \\
\bar P_1^{\rm SLL} \left[ C_1^{\rm SLL} (\mu) + C_1^{\rm SRR} (\mu) \right]
+\bar P_2^{\rm SLL} \left[ C_2^{\rm SLL} (\mu) + C_2^{\rm SRR} (\mu) \right].
\eea
The different Wilson coefficients are defined in ~\cite{Buras:2002vd} and
$r=0.985$ describes the $O(\alpha_s)$ radiative corrections to
$S_0(m_t^2/M_W^2)$ in the SM.

As shown in Refs.~\cite{Buras:2002vd,others}, the same
flavour-changing Higgs couplings generating the leading
contribution to the decay $B_s \to \mu^+ \mu^-$ at large $\tan \beta$
induce a Higgs--mediated double penguin contribution to
$B_{s,d}$--$\overline B_{s,d}$ mixing. In the large $\tan  \beta$ regime, the dominant contributions to
the $B_{s,d}$ mass differences are
given by the usual SM box and by the double penguin
contribution
\bea
\label{DeltaBq}
\Delta M_{B_q} &\propto & |K_{tq}|^2 \left[ S_0(m_t^2/M_W^2)~ -~
\Frac{\sqrt{2} G_F}{
\pi^2 M_W^2}
\Frac{\overline m_b\overline m_q m_t^4}{(1+\epsilon_0 \tan \beta)^2
(1+\tilde \epsilon_3 \tan \beta)^2}~
 \Frac{A_t^2 \tan^4\beta}{\mu^2 m_A^2}~\right. \times
\nn \\
&&\left.\qquad~~~
\left(H_2\left(m_{t_\Le}^2/\mu^2,m_{t_\Ri}^2/\mu^2\right)\right)^2\right],
\eea
with $q= s,d$ and $S_0(x)$ is defined in the appendices. From this expression
we can see that the
double penguin contribution is suppressed by an additional factor of
$G_F$ with respect to the SM contribution due to the fact that it is a
two loop contribution. However, it has a strong dependence on $\tan
\beta$ which can lead to a huge enhancement compensating this loop
suppression. Moreover, these double penguin contributions have always
opposite sign with respect to the Standard Model box contribution and,
thus reduce the mass difference for large $\tan
\beta$. This effect is specially important in the case of $\Delta
M_{B_s}$ yielding an additional constraint in the
large $\tan \beta$ regime. Notice that in the case of $\Delta
M_{B_d}$, the double penguin contribution is the same as in $\Delta
M_{B_s}$ with an additional suppression of $m_d/m_s$.  Given that this
contribution competes with the same SM function $S_0(x_t)$ it is clear
that this process will not be relevant to constrain the SUSY
contributions.

\section{Correlations with other processes}
\label{sec:correl}
Given the strong dependence of $B_s \to \mu^+ \mu^-$ on $m_A$ and
$\tan \beta$, we can expect that other processes, specially
$\tan \beta$ dependent processes, receive large SUSY contributions if BR($B_s
\to \mu^+ \mu^-$) is at the $10^{-7}$ level.  A typical example is the
decay $B \to X_s \gamma$ whose measured BR agrees very well with the
SM prediction. In turn this provides a stringent constraint to SUSY
contributions and given that the chargino amplitude is proportional to
$\tan \beta$ this holds in particular for the large $\tan \beta$
regime. The decay $B \to X_s \ell^+ \ell^-$ is closely related and the
combination of both processes has been recently shown to eliminate the
possibility of changing the sign of the dipole Wilson coefficient
$C_7$ in MFV scenarios \cite{misiak,HLMW}.
Another important constraint is provided by the muon
anomalous magnetic moment $a_\mu$ which is also $\tan \beta$ dependent
and constitutes a very important constraint specially for $\mu<0$.
\subsection{$B \to X_s \gamma$}
The Standard Model contributions to the decay $B\to X_s \gamma$ are
known at NLO \cite{Ciuchini:1997xe,Buras:2002tp}
and the excellent agreement with the
experimental results constrain strongly any extension of the SM.
Charged-Higgs contributions in two-Higgs doublet models are also known
at NLO \cite{Ciuchini:1997xe,Borzumati:1998tg,Borzumati:1998nx}. In the MSSM a
complete NLO calculation is still missing but the most important
contributions have been calculated in
Refs.~\cite{Bobeth:1999ww,Degrassi:2000qf,Carena:2000uj,Degrassi:2006eh}.
The MSSM contributes to
this process mostly through the dipole operators
\begin{eqnarray}
H_{\rm Dipole}^{b\rightarrow s \gamma} =
- {4 G_F\over \sqrt{2}} {K_{tb}^{\rm eff}} {K_{ts}^{\rm eff}}^*
\left[ C_7(\mu)\cdot
\frac{e \overline m_b}{16 \pi^2}\bar{s}_{\rm L} \sigma_{\mu \nu} b_{\rm R}
F^{\mu \nu} ~+~C_8(\mu)\cdot \frac{g_s \overline m_b}{16 \pi^2}\bar{s}_{{\rm L}\alpha}
T_{\alpha \beta}^a \sigma_{\mu \nu} b_{{\rm R} \beta} G^{a \mu \nu} \right] \;, \nonumber \\
 & &
\label{bsgOB}
\end{eqnarray}
The experimental world average from the CLEO~\cite{cleobsg}, Belle~\cite{bellebsg1,bellebsg2}
and BaBar~\cite{babarbsg1,babarbsg2} collaborations is given by~\cite{hfag}:
\begin{eqnarray}
{\rm BR} (B\to X_s \gamma)_{E_\gamma > 1.6 {\rm GeV}}
=
( 3.55 \pm 0.24_{-0.10}^{0.09}\pm 0.03) \times 10^{-4}\; .
\end{eqnarray}
This has to be compared with the standard model prediction~\cite{Hurth:2003dk}
\begin{eqnarray}
{\rm BR} [\bar B \to X_s \gamma]
& = & ( 3.61 \,\, \left. {}^{+0.24}_{-0.40} \right|_{m_c / m_b}
                     \pm 0.02_{\rm CKM} \pm 0.24_{\rm param.} \pm 0.14_{\rm scale} ) \times 10^{-4}  \,.
\end{eqnarray}
In the numerical analysis we utilize the formula presented in Ref.~\cite{Hurth:2003dk} in which
the branching ratio is explicit given in terms of arbitrary complex $C_7$ and $C_8$.

The two Wilson coefficients, $C_7(\mu)$ and $C_8(\mu)$, receive
contributions from W-boson, charged Higgs and chargino diagrams. It is
very interesting to discuss the relative signs of these three
contributions. Charged Higgs and W-boson diagrams have the same sign
and interfere always constructively. In contrast, the chargino
contribution can have either sign depending on the sign of the $\mu$
parameter. In the large $\tan \beta$ region the relative sign of the
chargino mediated diagram is given by $-\hbox{sign}(A_t \mu)$. In the
scenarios studied in this paper sign$(A_t)=- {\rm sign}(M_{1/2})$ as can be
seen in table IV of \cite{Bartl:2001}. For this
reason, $\mu>0$ implies destructive interference
while $\mu<0$ implies constructive interference of the chargino and
W--boson contributions.

Given that the SM contribution essentially saturates the experimental
bound, $\mu<0$ scenarios are tightly constrained and the sum of charged
Higgs and chargino contributions is restricted to be a small
correction. However, for positive $\mu$ the chargino
contribution has opposite sign with respect to the W-boson and charged
Higgs ones. In this case, a cancelation between sizable
charged Higgs and chargino contributions allows to keep the total
Wilson coefficient within a small margin from the SM expectation. In
principle it is possible for the chargino to cancel both the charged
Higgs and W-boson contributions and generate a total Wilson
coefficient of equal modulus and opposite sign to the usual SM
contribution. This is consistent with experimental data because the
BR($B\to X_s \gamma$)
depends essentially on $|C_7^{\rm eff}(\mu_b)|^2$ and therefore is not
sensitive to the sign of these coefficients. However, as we will see
later, combining this process with the decay $B \to X_s \ell^+ \ell^-$
it is possible to constrain the sign of the $C_7$ Wilson
coefficient~\cite{misiak,HLMW} and exclude this possibility in the generic MFV
scenarios studied.

The complete expressions for the LO contributions to the $C_7$ and
$C_8$ contributions that are used in our numerical evaluation are
well-known and can be found in Ref.~\cite{Bertolini:1990if,chomisiak}.
For the
following discussion, it is enough to examine the $C_7$ coefficient as
$C_8$ behaves in a similar way,
\begin{eqnarray}
C_7^W (m_W) &=& {m^2_t\over 4 m_W^2}  f_1\left(m_t^2/m_W^2\right),
\\ C_7^{H^+} (m_W) &=& {1\over 6}
\left\{{1 \over 2}~{m^2_t\over m^2_{H^\pm}}~{1\over \tan^2 \beta}~
f_1\left(m_t^2/m_{H^\pm}^2\right)~ +~ f_2\left(m_t^2/m_{H^\pm}^2\right)\cdot\right.
\nn\\
&&\left. \left(1 - \Frac{\left(\epsilon_0^\prime+\tilde \epsilon_3\right) \tan
\beta}{1+\tilde \epsilon_3 \tan \beta} + y_b^2 y_t^2\Frac{\epsilon_Y
\epsilon_Y^\prime~ \tan^2 \beta}{\left(1+\tilde \epsilon_3 \tan \beta\right)
\left(1+\epsilon_0 \tan \beta\right)}\right)\right\}
,
\\ C_7^{\tilde \chi} (m_W) &=&
{1\over3}
~\sum_{\alpha,\alpha'}\sum_{i=I}^2 \sum_{a=1}^6~ {K_{\alpha b}^{\rm eff}
{K^{\rm eff}_{\alpha' s}}^* \over K_{tb}^{\rm eff} {K_{ts}^{\rm eff}}^*}~
{m_W^2\over m_{{\tilde \chi}^\pm_I}^2} \left\{
-{1\over 2} G^{(\alpha',a)I*} G^{(a,\alpha)I}
f_1\left(m_{\tilde u_a}^2/m_{{\tilde \chi}^\pm_I}^2\right) ~+ \right. \nonumber \\ &&
\left.
{m_{{\tilde \chi}_I}\over m_W}~
{1\over \sqrt{2} \cos \beta (1+\tilde \epsilon_3 \tan \beta)}~
G^{(\alpha',a)I*}  H^{(a,\alpha)I}~
f_2\left(m_{\tilde u_a}^2/m_{{\tilde \chi}^\pm_I}^2\right)
\right\},
\label{SUSYWC}
\end{eqnarray}
where, for simplicity we have only included in this equation the
leading contributions in the limit of unbroken $SU(2)\times U(1)$
symmetry \cite{Buras:2002vd}. In the numerical analysis we use the
complete expressions with full flavour dependence. The parameters
$\epsilon_0^\prime$ and $\epsilon_Y^\prime$ are given by
\bea
\label{epsilonspr}
\epsilon_Y^\prime = \Frac{A_b}{16 \pi^2 ~\mu}
H_2\left(\frac{m_{\tilde b_\Le}^2}{\mu^2},\frac{m_{\tilde b_\Ri}^2}{\mu^2}\right),~
\epsilon_0^\prime = \Frac{- 2\alpha_s\mu}{3 \pi ~m_{\tilde g}}
H_2\left(\frac{m_{\tilde s_\Le}^2}{m_{\tilde g}^2},\frac{m_{\tilde t_\Ri}^2}{m_{\tilde g}^2}
\right).
\eea
The main $\tan \beta$ effects are then present in the charged Higgs
couplings and in the chargino couplings to right-handed down squarks.
$K_{\alpha q} G^{(\alpha,k)I}$ represents the coupling of the chargino
$I$ and the squark $k$ to the left--handed down quark $q$; and
$m_q/(\sqrt{2} m_W \cos \beta) K_{\alpha q} H^{(\alpha,k)i}$ the
coupling of the chargino $i$ and of the squark $k$ to the
right--handed down quark $q$.  These couplings, in terms of the
standard mixing matrices defined in the appendices are~\cite{haber}
\begin{eqnarray}
G^{(\alpha, k) I} &=& \left( \Gamma_{U L}^{k \alpha} V_{I 1}^{*} -
\frac{m_{\alpha}}{\sqrt{2} m_W \sin \beta} \Gamma_{U R}^{k \alpha}
V_{I 2}^{*}\right)\nonumber \\
H^{(\alpha,k)I} &=& - U_{I 2} \Gamma_{U L}^{k \alpha}.
\end{eqnarray}
The explicit expressions for the loop functions can be found in the
appendices.

In the large $\tan \beta$ regime, it is interesting to expand the
$C_7^{\tilde \chi}$ coefficient assuming the smallness of the off-diagonal
entries with respect to the corresponding diagonal elements both in
the stop and chargino (more precisely in $M_{\tilde \chi} M_{\tilde \chi}^\dagger$) mass
matrices. In this approximation we obtain,
\bea C_7^{\tilde \chi} (m_W)&
\simeq&\Frac{- \mu A_t m_t^2}{6 m_W^2 \sin^2 \beta}~ \Frac{\tan \beta}
{1+\tilde \epsilon_3 \tan \beta}~
{m_W^2\over m_{{\tilde \chi}^\pm_2}^2} {f_2\left(m_{\tilde
t_L}^2/m_{{\tilde \chi}^\pm_2}\right) - f_2\left(m_{\tilde
t_R}^2/m_{{\tilde \chi}^\pm_2}\right) \over m_{\tilde t_L}^2 - m_{\tilde
t_R}^2}
\nn\\
&\simeq& \Frac{\mu A_t m_t^2}{6 m_W^2 \sin^2 \beta}~\Frac{\tan \beta}
{1+\tilde \epsilon_3 \tan \beta}~
{5~ m_W^2 \over 12~ m_{\tilde t}^4}
\eea
where in the second line we replaced the combination of loop functions
by its value in the limit in which stop left, stop right and chargino
have the same ``average'' SUSY mass, $m_{\tilde t}^2$. In this limit
the SUSY contribution to the amplitude is
\bea \label{bsllampl} {\mathcal
A} (b_\Ri \to s_\Le \gamma) &\propto &{4 G_F\over \sqrt{2}} K_{tb}^{\rm eff}
{K_{ts}^{\rm eff}}^* \overline m_b ~
\Frac{5~\mu A_t m_t^2 \tan\beta}{72~m_{\tilde t}^4~(1+\tilde \epsilon_3
\tan \beta)} \; .
\eea
If we compare now this amplitude with the corresponding amplitude of
the $B_s \to \mu^+ \mu^-$ decay, \eq{bsllampl}, we see that, although
both them are enhanced by powers of $\tan \beta$, these two amplitudes
behave differently: ${\mathcal A} (b_\Ri s_\Le \to l^+ l^-)$ scales as
$\tan^3\beta/m_A^2$ while the $b \to s \gamma$ amplitude scales as
$\tan\beta/m_{\tilde t}^2$ (assuming $\mu$, $A_t$ and $m_{\tilde t}$
of the same order). Taking into account this difference we can analyze
if it is possible to obtain a large contribution for $B_s \to \mu^+
\mu^-$ while at the same time satisfy the stringent bounds on the $b
\to s \gamma$ decay. Obviously, the answer to this question depends on the sign
of $\mu$.

In the $B\to X_s \gamma$ decay with $\mu < 0$ chargino contributions
to $C_7$ and $C_8$ have the same sign as charged Higgs and W-boson
contributions.  If we require a sizable neutral Higgs contribution to
$B_s \to \mu^+ \mu^-$ we need a light pseudoscalar Higgs, which
implies also a light charged Higgs, and large $\tan \beta$.  A
light charged Higgs enhances $C_7^{H^+}$
while large $\tan \beta$ enhances $C_7^{\tilde \chi}$ . As
both contributions compete to fill the narrow space left over by the
SM contribution in the $B\to X_s \gamma$ amplitude these two
requirements are clearly incompatible.  The only possible solution is
to increase the squark masses (in particular the stop mass) to suppress the
chargino contribution to $C_7$ while keeping $m_A$ as low and $\tan
\beta$ as large as possible. In this case we are essentially in a two-Higgs
doublet model without SUSY and from the existing constraints on the charged
Higgs mass \cite{Ciuchini:1997xe,Borzumati:1998tg,Borzumati:1998nx}
we infer that the pseudoscalar must be equally heavy.
Moreover, the constraints
from the muon anomalous magnetic moment, that are analyzed below, make
the $\mu<0$ situation even more difficult. Therefore it is not possible to
reach very large values of BR$(B_s \to \mu^+ \mu^-)$, as we will see in the
numerical analysis of the next section.

The case $\mu > 0$ is much more interesting phenomenologically. A large
BR$(B_s \to \mu^+ \mu^-)$ still
requires a light charged Higgs and large $\tan \beta$, however the
chargino contributions interfere now destructively with the charged Higgs
and W-boson ones and thus
these two conditions are fully compatible because $|C_7^{\tilde \chi} +
C_7^{H^+}| \lsim 0.1 C_7^{\rm SM}$.
As we will see in the next
section, this cancelation must be quite precise and this constraint
remains very strong. In the flavour-blind models defined at the GUT
scale, in particular in the CMSSM, is not always easy to obtain the
required values of $m_A$,
$m_{\tilde t}$ and $\tan \beta$ to achieve this cancelation. The
extent to which this cancelation can be realized will be addressed in
the next section where we perform a full analysis of the parameter
space.

\subsection{$B\to X_s \ell^+ \ell^-$}
\label{sec:bsll}
The effective Hamiltonian that describes $b\to s \ell\ell$ transitions
in the Standard Model is
\bea
H_{\rm eff} & = & -\frac{4 G_F}{\sqrt{2}} K_{tb}^{*} K_{ts}^{*}
                  \left(
                  \sum_{i=1}^{10} C_i (\mu) \; O_i(\mu) +
                  \sum_{i=3}^{6}  C_{iQ} (\mu) \; O_{iQ}(\mu) +
                                  C_b \; O_b (\mu)
                  \right)\; ,
\eea
where we adopt the same definitions as in Ref.~\cite{HLMW}. The
operators $O_{iQ}$ and $O_b$ are required only for the inclusion of
electro--weak corrections. In particular the most relevant Wilson
coefficients are $C_2 (\mu_b)$ (that does not receive sizable new
physics contributions), $C_7 (\mu_b)$, $C_8(\mu_b)$, $C_9(\mu_b)$ and
$C_{10}(\mu_b)$. The explicit definitions of $O_7$ and $O_8$ are given
in Eq.~(\ref{bsgOB}) and the semileptonic operators read
\bea
O_9 =  \frac{\alpha_{em}}{4 \pi} \;
       ({\bar s}_L \gamma_\mu b_L) \; (\bar \ell \gamma^\mu \ell )
\; , \;\;\;\;
O_{10} =  \frac{\alpha_{em}}{4 \pi} \;
       ({\bar s}_L \gamma_\mu b_L) \; (\bar \ell \gamma^\mu \gamma_5 \ell)\; .
\eea
The branching ratio ${\rm BR} (B\to X_s \ell\ell)$ is known at NNLO in
QCD and at NLO in QED (only terms enhanced by large logarithms --
i.e. $\log m_W/m_b$ and $\log m_b/m_\ell$ -- are included).

Intermediate $c\bar c$ resonances produce peaks in the dilepton
invariant mass ($4 m_\ell^2 < s < m_B^2$) spectrum and disrupt
quark--hadron duality. For this reason one has to consider the low-
and high-$s$ regions separately. The very low-$s$ region ($s<1
\;{\rm GeV}^2$) is dominated by the quasi-real photon emission and
contains the same amount of information we already extract from $b\to
s \gamma$. In the high-$s$ region ($s>14.3\; {\rm GeV}^2$) the
integrated branching ratio is quite small; moreover, the $1/m_b$
expansion breaks down and reliable predictions for the spectrum suffer
large uncertainties. The low-$s$ region ($1\;{\rm GeV}^2 < s< 6 \;{\rm
GeV}^2$) is dominated by the Wilson coefficients $C_9$ and $C_{10}$,
has sizable integrated branching ratio and is very sensitive to the
$C_7 - C_9$ interference. In the following, we utilize the latter to
put constraints on the Wilson coefficients $C_{7,8,9,10}$.

We calculate the integrated branching ratio in the low-$s$ region
following Ref.~\cite{HLMW}:
\bea\label{numform}
\hskip -3cm
{\rm BR}_{\ell\ell} & = & \Big[\;
2.2306-0.0005 \; {\cal I} (R_{10})+0.0005 \; { \cal I}( R_{10}
R_{8}^*)+0.0906 \; {\cal I} (R_{7})+0.0223 \; { \cal I} (R_{7}
R_{8}^*)
\nonumber\\ & &
+0.0050 \; { \cal I}( R_{7} R_{9}^*)+0.0086 \; {\cal I}
(R_{8})+0.0258 \; {\cal I} (R_{8} R_{9}^*)-0.0615 \; {\cal I} (R_{9})
\nonumber\\ & &
-0.5389
\; {\cal R} (R_{10})+0.1140 \; {\cal R} (R_{7})+0.0154 \; { \cal R}( R_{7}
R_{10}^*)+0.0687 \; { \cal R} (R_{7} R_{8}^*)
\nonumber\\ & &
-0.8414 \; { \cal R} (R_{7}
R_{9}^*)-0.0036 \; {\cal R} (R_{8})+0.0019 \; { \cal R}( R_{8}
R_{10}^*)-0.0980 \; { \cal R} (R_{8} R_{9}^*)
\nonumber\\ & &
+2.6260 \; {\cal R}
(R_{9})-0.1074 \; {\cal R}( R_{9} R_{10}^*)+10.6042 \; |R_{10}|^2+0.2837 \;
|R_7|^2
\nonumber\\ & &
+0.0039 \; |R_8|^2+1.4645 \; |R_9|^2\; \Big] \times 10^{-7} \; ,
\eea
where $R_i = C_i^{\rm tot}(\mu_0)/C_i^{\rm SM}(\mu_0)$ are the
next-to-leading order Wilson coefficients (i.e. they do not include
$O(\alpha_s)$ corrections) at the high scale normalized to their SM
values. When dealing with the dipole operators we utilize the standard
definition of the scheme-independent effective
coefficients~\cite{HLMW}. In the numerical analysis we set to zero
supersymmetric contributions to the NLO matching conditions.

The standard model and supersymmetric contributions to the leading
order matching conditions of the semileptonic operators
are~\cite{HLMW} (we do not give explicit formulae for the gluino,
neutralino and neutral Higgses contributions because their effects are
negligible compared to the charged Higgs and chargino ones):
\bea
C_9^{W} & = & \frac{1}{s_W^2} Y(x_t) + W(x_t) +\frac{4}{9} -
              \frac{4}{9} \log \frac{\mu_0^2}{m_t^2} \;, \\
C_{10}^{W} & = & - \frac{1}{s_W^2} Y(x_t)  \;, \\
C_9^{H^\pm} & = &
\frac{4 s_W^2 - 1}{ 8 s_W^2} \frac{x_t}{\tan^2 \beta} \;
     f_5(m_t^2/m_{H^\pm}^2)
+\frac{1}{18}\frac{1}{ \tan^2 \beta} \; f_6(m_t^2/m_{H^\pm}^2) \; , \\
C_{10}^{H^\pm} & = &
\frac{1}{8 s_W^2}\frac{x_t}{ \tan^2 \beta} \; f_5(m_t^2/m_{H^\pm}^2) \; , \\
C_{9}^{{\tilde \chi}^\pm} & = &
    \sum_{\alpha,\alpha'} \sum_{A,B=1}^6 \sum_{I,J=1}^2 \;
    {K^{\rm eff}_{\alpha b} {K^{\rm eff}_{\alpha' s}}^* \over
     K_{tb}^{\rm eff} {K_{ts}^{\rm eff}}^*} \;
    G^{(\alpha',A)I*} G^{(B,\alpha)J} \;
    \Bigg[ {1- 4s_W^2 \over s_W^2} \; \Bigg(
\nonumber \\
& &
c_2(m^2_{{\tilde \chi}^\pm_I},m^2_{\tilde u_A},m^2_{\tilde u_A}) \left( \Gamma_{UL}
\Gamma_{UL}^\dagger \right)_{AB} \delta_{IJ}
+ {m_{{\tilde \chi}_I} m_{{\tilde \chi}_J} \over 2}
c_0(m^2_{\tilde u_A},m^2_{{\tilde \chi}^\pm_I},m^2_{{\tilde \chi}^\pm_J}) \delta_{AB}
U_{I1}^* U_{J1} \nonumber \\
& &
- c_2(m^2_{\tilde u_A},m^2_{{\tilde \chi}^\pm_I},m^2_{{\tilde \chi}^\pm_J})
\delta_{AB} V_{I1}^* V_{J1}
\Bigg) \;
- {2\over 9} {m_W^2\over m_{\tilde u_A}^2} f_7(m^2_{{\tilde \chi}_I^\pm}/m^2_{\tilde u_A})
 \delta_{AB} \delta_{IJ} \nonumber \\
& &
+ {2\over s_W^2} m_W^2 d_2(m^2_{{\tilde \chi}^\pm_I},m^2_{{\tilde \chi}^\pm_J},m^2_{\tilde u_A},
m^2_{\tilde \nu_1})
\delta_{AB} V_{I1}^* V_{J1}
    \Bigg]
\;, \\
C_{10}^{{\tilde \chi}^\pm} & = &
    -\frac{1}{2 s_W^2} \;
    \sum_{\alpha,\alpha'} \sum_{A,B=1}^6 \sum_{I,J=1}^2 \;
    {K^{\rm eff}_{\alpha b} {K^{\rm eff}_{\alpha' s}}^* \over
     K_{tb}^{\rm eff} {K_{ts}^{\rm eff}}^*} \;
    G^{(\alpha',A)I*} G^{(B,\alpha)J} \;
    \Bigg[ \nonumber \\
& &
c_2(m^2_{{\tilde \chi}^\pm_I},m^2_{\tilde u_A},m^2_{\tilde u_A}) \left( \Gamma_{UL}
\Gamma_{UL}^\dagger \right)_{AB} \delta_{IJ}
+ {m_{{\tilde \chi}_I} m_{{\tilde \chi}_J} \over 2}
c_0(m^2_{\tilde u_A},m^2_{{\tilde \chi}^\pm_I},m^2_{{\tilde \chi}^\pm_J}) \delta_{AB}
U_{I1}^* U_{J1} \nonumber \\
& &
+ \left[2 m_W^2 d_2(m^2_{{\tilde \chi}^\pm_I},m^2_{{\tilde \chi}^\pm_J},m^2_{\tilde u_A},
m^2_{\tilde \nu_1}) - c_2(m^2_{\tilde u_A},m^2_{{\tilde \chi}^\pm_I},m^2_{{\tilde \chi}^\pm_J})
\right] \delta_{AB} V_{I1}^* V_{J1} \;
    \Bigg] \;.
\eea
A scan of the CMSSM and MFV parameter space results in tiny
deviations of the Wilson coefficients $C_9$ and $C_{10}$ from their SM
values: $|C_9^{\rm SUSY}/C_9^{\rm SM}| < 0.3/1.65$ and $|C_{10}^{\rm
SUSY}/C_{10}^{\rm SM}| < 0.8/4.45$.

\begin{figure}
\begin{tabular}{lr}
\includegraphics[width=7cm]{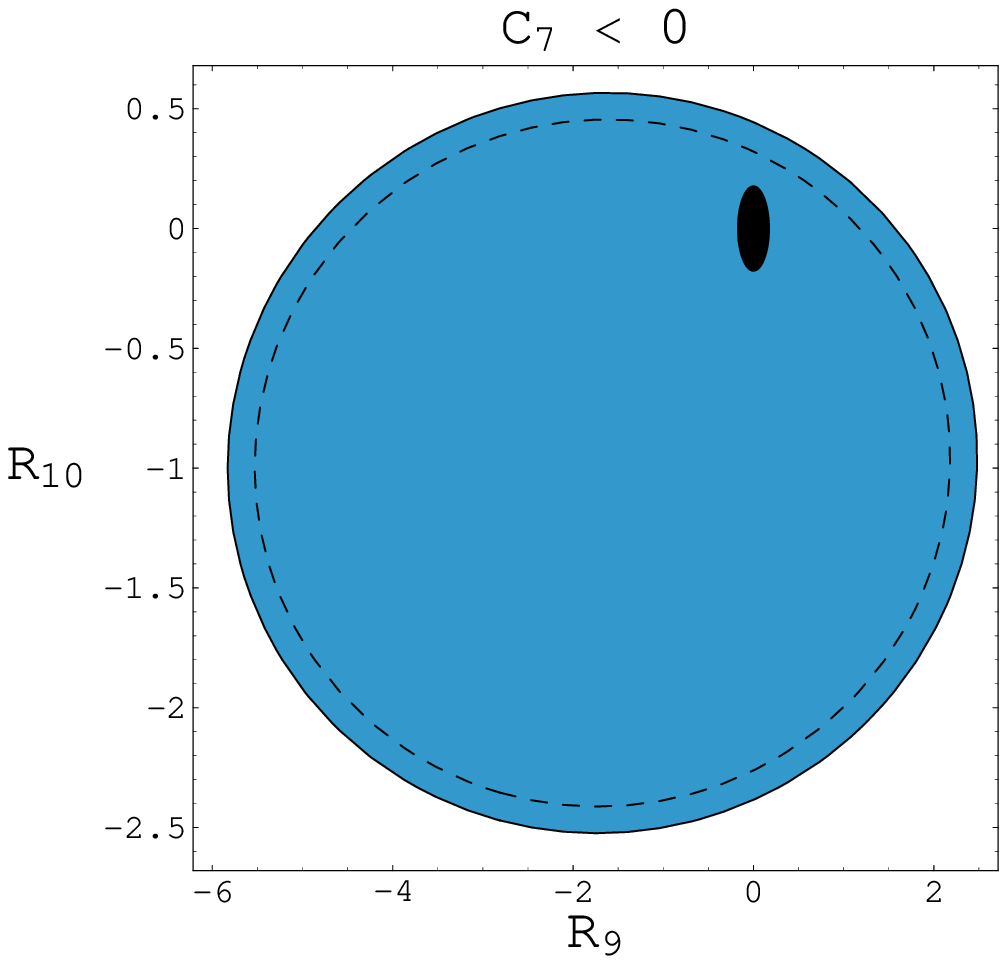}
&
\includegraphics[width=7cm]{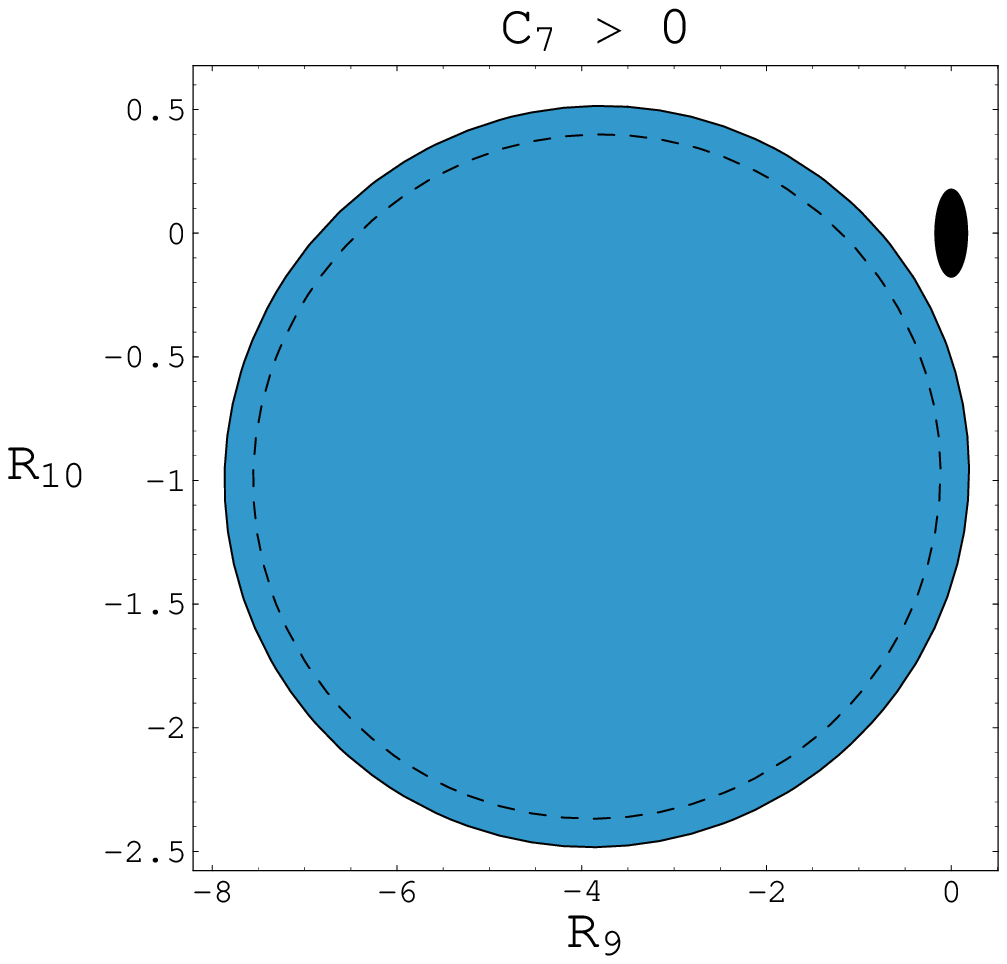}
\cr
\end{tabular} \caption{Impact of the ${\rm BR} (B\to X_s \ell\ell)$
measurement on the $R_9$-$R_{10}$ plane. The black disk is the
region accessible in an MFV model, i.e. it includes the possible
SUSY contributions both in the CMSSM as in the generic MFV scenario that
we analyse here.} \label{fig:r9r10}
\end{figure}

This decay has been observed by Belle~\cite{Abe:2005.qqqq} and
BaBar~\cite{Aubert:2004it}. In the low--$s$ region the experimental
results read as
\bea
{\rm BR} (B\to X_s \ell^+\ell^-) &=& (1.493 \pm 0.504^{+0.411}_{-0.321})
\times 10^{-6} \;\;\; ({\rm Belle}) \; ,\\
{\rm BR} (B\to X_s \ell^+\ell^-) &=& (1.8 \pm 0.7\pm0.5)
\times 10^{-6} \;\;\; ({\rm BaBar}) \; .
\eea
This leads to a world average
\bea
{\rm BR} (B\to X_s \ell^+\ell^-)_{\rm WA} &=& (1.60 \pm 0.51)\times 10^{-6} \; .
\eea
That is in perfect agreement with the SM prediction, that reads:
\bea
{\rm BR} (B\to X_s \ell^+\ell^-)_{\rm SM}  & = &
\Big[
1.58
\pm 0.07_{\rm scale}
\pm 0.06_{m_t}
\pm 0.025_{m_c}
\pm 0.015_{m_b} \nonumber \\
&&  \pm 0.023_{\alpha_s(M_Z)}
\pm 0.015_{\rm CKM}
\pm 0.025_{{\rm BR}_{sl}}
\Big] \times 10^{-6} \nonumber \\
&=& (1.58 \pm 0.10) \times 10^{-6} \;.
\eea
In order to assess the impact of this measurement on the
supersymmetric parameter space, we perform a model independent
analysis. As a first step we extract the allowed ranges for the Wilson
coefficients $C_7$ and $C_8$ from the measured $B\to X_s \gamma$
branching fraction. This gives two thin stripes in the $C_7 - C_8$
plane that correspond to the $C_7 <0$ and $C_7>0$ scenarios. We
further restrict $C_8$ to be within a factor of 10 around the SM
value. As a second step, we consider the two strips separately, vary
$C_7$ and $C_8$ in the allowed range, and plot the resulting
constraint in the $C_9 - C_{10}$ plane. The whole analysis is
performed at the $90 \% \; C.L.$ and theory errors are included. Our
results are summarized in fig.~\ref{fig:r9r10}. The most striking
feature of these plots is that, in the $C_7>0$ scenario, the SM point
is excluded and contributions in MFV supersymmetric scenarios are not
large enough to bring theory and experiment into agreement; thus we
conclude that in such models the sign of $C_7$ has to be SM--like (
i.e. negative). For a similar discussion see also \cite{misiak}.

\subsection{Muon anomalous magnetic moment}
In SUSY theories, $a_{\mu^+}$ receives contributions via vertex diagrams with
${\tilde \chi}^0$--$\tilde \mu$ and ${\tilde \chi}^\pm$--$\tilde \nu$
loops~\cite{moroi}.
The chargino diagram strongly dominates in almost all the parameter
space. For simplicity, we will present here only the dominant part of
the chargino contribution (the complete expressions that we use in the
numerical simulation can be found in Ref.~\cite{moroi}):
\begin{equation}
\delta a_{\mu^+}^{{\tilde \chi} \tilde\nu} \simeq -{g_2^2\over 8 \pi^2}
{m_\mu^2\over m_{\tilde\nu}^2}
 \sum_{i=1}^2
 \frac{m_{{\tilde \chi}_i} \hbox{Re} (U_{i2} V_{i1})}
      {\sqrt{2} M_W \cos \beta}
 f_3 \left({m_{{\tilde \chi}_i}^2\over m_{\tilde\nu}^2}\right),
\label{gm2char}
\end{equation}
where the loop function $f_3$ is given in the appendices. In fact, if
we expand this function in the small off-diagonal
elements of the chargino matrix we obtain for the dominant $\tan
\beta$ enhanced contribution,
\bea
\delta a_{\mu^+}^{{\tilde \chi} \tilde\nu} \simeq {g_2^2\over 32 \pi^2}
{m_\mu^2 \over m_{\tilde\nu}^2} {\hbox{Re} (\mu) M_2 \tan \beta \over m_{\tilde\nu}^2}\; ,
\label{gm2app}
\eea
where we have chosen the convention $M_2 > 0$ as we do in the
rest of the paper.
The most relevant feature of Eqs.~(\ref{gm2char}) and (\ref{gm2app}) is
that the sign of $\delta a_{\mu^+}^{{\tilde \chi} \tilde\nu}$ is fixed by
$\hbox{sign}[\hbox{Re}(U_{12} V_{11}) ] = - \hbox{sign}[\hbox{Re}(\mu)]$.
The experimental result for this observable is at present~\cite{BNLexp},
\bea
a_{\mu}^{\rm exp} = 11\, 659\, 208\, (6) \times 10^{-10},
\eea
while the theoretical expectations within the SM are \cite{Jegerlehner:2003qp},
\bea
a_{\mu}^{\rm SM}({\rm e^+ e^-}) = 11\, 659\, 181\, (8) \times 10^{-10},\\
a_{\mu}^{\rm SM}({\rm \tau}) = 11\, 659\, 196\, (7) \times 10^{-10}.\nn
\eea
Comparison with the experimental results implies that these results
strongly favour the $\mu>0$ region in a MFV scenario. This is specially true
for the theoretical prediction based on $e^+ e^-$ annihilation that is
smaller than the experimental result by $2.7 \sigma$. In the case of the
prediction based on $\tau$ decay the difference is reduced to $1.4 \sigma$
\cite{BNLexp} but it still requires a positive correction and disfavours
strongly a sizable negative contribution. Therefore this
has important consequences in all our observables.
As we saw in the previous section, $\mu > 0$ is precisely the
phenomenologically interesting region for the combination of the
$B_s \to \mu^+ \mu^-$ and $B \to X_s \gamma$ decays. On the other hand
the $a_{\mu}$ measurement makes even more difficult to find interesting effects
in $B_s \to \mu^+ \mu^-$ for $\mu < 0$. In the numerical analysis, presented
in the next section, we use a conservative $a_{\mu}$ bound at $3 \sigma$ with
the  $e^+ e^-$ based estimate of the SM contribution and we still find
that with $\mu < 0$ we can only find BR$(B_s \to \mu^+ \mu^-)$ at the level
of $10^{-8}$ with very heavy SUSY spectrum. In the $\mu > 0$ region, the
conservative $3 \sigma$ bound allows for a vanishing supersymmetric contribution.

\section{Numerical results}
\label{sec:bsmumu}

As said in the introduction, in this work we concentrate on a
version of the MSSM with MFV soft-breaking terms at the GUT
scale.  This means that the only non-trivial flavour structures in the
model are the usual SM Yukawa couplings; all soft-breaking
masses are family universal and the trilinear couplings are proportional to
the corresponding Yukawa matrices. The RGE evolution between the GUT and
electroweak
scales introduces small non-universal entries in the soft-breaking terms.

In this framework we  analyse two different versions of the MSSM:
the CMSSM and the most general MFV MSSM. The full spectrum of the CMSSM is
determined (assuming vanishing SUSY phases) by five parameters:
$M_{1/2}$, $m_0$, $A_0$, $\tan \beta$ and sign($\mu$). Therefore, the SUSY
parameters at the electroweak scale, including couplings, masses and mixing
angles, show several interesting correlations. Furthermore, these five parameters
are strongly constrained by the
experimental bounds on masses and different FCNC processes. In this
restricted framework it is very interesting to investigate how large
BR($B_s \to \mu^+ \mu^-$) $\propto \tan^6\beta/m_A^4$ can be, while
remaining in agreement with all the other experimental constraints.

We can ask the same question in a more general flavour-blind MSSM,
where we allow for different, but generation-diagonal, parameters for the
fields with different gauge quantum numbers. Here the number of parameters
is thirteen:
$M_{1/2}$, $m_{\tilde Q,0}$, $m_{\tilde u_R,0}$, $m_{\tilde d_R,0}$,
$m_{\tilde L,0}$,
$m_{\tilde e_R,0}$, $m_{H_d,0}$, $m_{H_u,0}$, $A_{u,0}$, $A_{d,0}$,
$A_{e,0}$, $\tan \beta$ and
sign($\mu$), where the $0$ in the subscripts indicates that these parameters
are defined at $M_{\rm GUT}$. The precise definition of $M_{\rm GUT}$ and the
numerical evaluation of the parameters at the electroweak scale will be
described below.
Clearly in this model we have more freedom than in the CMSSM,
implying milder correlations between the masses of different particles; thus,
we can expect larger rates for the Higgs-mediated processes consistent with
other FCNC constraints.

In the numerical analysis presented below we define our model, CMSSM or
most general MFV, in terms of the GUT scale parameters listed above.
We scan the values of these parameters in the following ranges:
 $M_{1/2} \le 1$~TeV,
$m_{i,0} \le 2$~TeV ($i=Q,u_R,d_R,L,e_R,H_u,H_d$),
$A^2_{u,0} \le 3 (m^2_{\tilde Q,0} + m^2_{\tilde u_R,0} + m_{H_u,0}^2)$,
$A^2_{d,0} \le 3 (m^2_{\tilde Q,0} + m^2_{\tilde d_R,0} + m_{H_d,0}^2)$,
$A^2_{e,0} \le 3 (m^2_{\tilde L,0} + m^2_{\tilde e_R,0} + m_{H_d,0}^2)$;
$|\mu|$ is calculated from the requirement of correct electroweak
symmetry breaking. The bound on the $A_{i,0}$ parameters is set to
avoid charge and/or colour breaking minima.

At the low scale, we impose the following constraints on each point:
\begin{itemize}
\item Lower bound on the light and pseudo--scalar Higgs masses~\cite{unknown:2006cr};
\item The LEP constraints on the lightest chargino  and sfermion
      masses~\cite{Eidelman:2004wy};
\item The LEP and Tevatron constrains on squarks and gluino masses~\cite{Eidelman:2004wy};
\item Agreement with the experimental data on low energy processes like
${\rm BR}(B \to X_s \gamma)$, $\delta a_\mu$, ${\rm BR}(B_s\to \mu^+\mu^-)$,
${\rm BR}(B \to X_s \ell^+\ell^-)$, $\Delta M_{B_s}$~\cite{cleobsg,bellebsg1,bellebsg2,babarbsg1,babarbsg2,Bennett:2004pv,last,prev,Abe:2005.qqqq,Aubert:2004it,Abazov:2006dm,CDFBs:2006};
\item Correct dark matter relic abundance from stable neutralinos
  \cite{wmap3rdyear}\footnote{This constraint assumes a standard thermal
history of the universe and can be easily evaded in non standard cosmological
  models.}.
\end{itemize}
We consider MSSM with conserved R-parity and thus, the lightest neutralino (that
coincides with the lightest supersymmetric particle - LSP) provides an excellent
dark matter candidate. In fact, the
recent precise determination of the cold dark matter density of the
universe is a very strong constraint on the SUSY parameter space. In
most of the parameter space of the MSSM the neutralino abundance
generates a too large density of cold dark matter. Only certain
regions of the parameter space, in which neutralinos are efficiently
annihilated, are in agreement with the
experimental results.

The most precise results on the cold dark matter density come from the
WMAP experiment~\cite{wmap3rdyear}:
\bea
\label{wmap}
\Omega_{\rm CDM} h^2 = 0.1047 ^{+0.007}_{-0.013}~~ \mbox{at 68\% C.L.} \eea
This constraint selects very narrow strips in the MSSM parameter
space~\cite{refs}. On the other hand it has been shown~\cite{kraml}
that these allowed regions are extremely sensitive to small
differences in the computation of SUSY masses. This is specially
true in the large $\tan \beta$ region that we analyze in this paper,
where differences in the stability of radiative symmetry breaking or
in the Higgs masses are due to the different approximations used in
the RGE programs. In view of all this, we prefer to take a
conservative attitude with respect to the dark matter constraint
and, allowing the possibility of other cold dark matter candidates,
impose only the loose 99\% C.L. upper bound,
\bea
\Omega_{\rm CDM} h^2 \leq 0.13
\eea

In this section we analyze the different parameters (masses,
mixing matrices, \ldots) that enter the $B_s \to \mu^+ \mu^-$ amplitude,
and the impact of the various constraints that we have discussed above.
We start with the analysis of this process in the CMSSM and then
generalize it to the most general MFV model at the GUT
scale.

\subsection{Calculation of the parameters at the electroweak scale}

\begin{table}[t] \small
\begin{center}
\begin{tabular}{|c|c||c|c|}
\hline
$m_e$    & 5.110 $10^{-4}$ &  $m_t^{pole}$ & 172.9  \\
$m_\mu$  & 0.1057        & $m_b(m_b)$ & $4.25$ \\ \cline{3-4} \cline{3-4}
$m_\tau$ & 1.777          & $m_Z$ & 91.1876 \\             \cline{1-2}
$m_u(Q)$ & $ 3 \cdot 10^{-3}$  & $G_F$ & 1.1664 $\cdot 10^{-5}$ \\
$m_d(Q)$ & $7 \cdot 10^{-3}$ &  $ 1/\alpha$ & 137.036 \\
$m_s(Q)$ & 0.12 & $\Delta \alpha^5_{had}$ & 0.02769 \\
$m_c(m_c)$ & 1.2 & $\alpha_s^{\overline{MS}}(m_Z)$ & 0.1172 \\ \hline
\hline \multicolumn{4}{|c|}{
$s_{12} = 0.224  $,  $s_{23} = 4.13 \cdot 10^{-2}  $,
$s_{13} = 3.63 \cdot 10^{-3}  $,  $\delta = 1.13$} \\ \hline
\end{tabular}
\end{center}
\caption{
       Numerical values of the SM input. Masses are given
             in GeV, for the leptons and the $t$ quark the pole masses,
             for the lighter quarks the ${\overline{MS}}$ masses either at
          the mass scale itself, for $c$, $b$, or, for $u$, $d$, $s$,
          at the scale $Q=2$ GeV. $s_{ij}$ are the sines of the CKM
        mixing angle and $\delta$ the CKM phase. }
\label{tab:sminput}
\end{table}

In the numerical calculations, we use the following procedure.
The masses and mixing angles at the electroweak scale are calculated
in an iterative way: First the gauge and Yukawa couplings are calculated
from the input values given in Table~\ref{tab:sminput}
at the electroweak scale taking into account the shift from the
${\overline{MS}}$
to the  $\overline{DR}$ scheme as well as the effect of the SUSY thresholds.
In case of the SUSY thresholds the complete flavour structure of the sfermions
is taken into account in the calculation. Afterward these couplings
are evolved to the $GUT$ scale which is defined by the requirement
$g_1 = g_2$ where $g_1 = \sqrt{5/3} g'$ is the properly normalized $U(1)$
coupling and $g_2$ is the $SU_L(2)$ coupling. We do not require at
$M_{\rm GUT}$ that $g_3$ is equal to $g_1=g_2$ but assume that this
differences is accounted for by unknown thresholds due to very heavy
particles with masses of the order
 $M_{\rm GUT}$. At $M_{\rm GUT}$ the SUSY breaking boundary conditions are set.
At this scale the trilinear couplings are obtained multiplying the
$A$-parameters with the Yukawa matrices
taking into account the {\it complete} flavour structure. Also in the RGE
evolution, the complete flavour structure is taken into account and we
evolve the parameters at the two-loop order \cite{Martin:1993zk}.
This clearly induces off-diagonal elements in the sfermion mass
parameters.  The calculation
of the SUSY masses is carried out in the $\overline{DR}$ scheme and the
formulae for the one-loop masses are generalisations of the ones given in
Ref.~\cite{Pierce:1996zz} taking into account the complete flavour
structure and will be presented elsewhere \cite{PorodNew}.
In the case of the $\mu$ parameter and the masses of the neutral
Higgs bosons the
two-loop corrections as given in
Refs.~\cite{Degrassi:2001yf,brignole,Brignole:2002bz,Dedes:2003km,DedesSlavich,%
ADKPS} are
added. Note that in the one-loop contributions we use the full flavour
structure, while for the two-loop part we use the approximation that only
the sfermions of the
third generation contribute. The obtained spectrum is then used to recalculate
the thresholds to gauge and Yukawa couplings restarting the complete procedure.
This is repeated until the relative difference of all couplings and masses
between subsequent iterations is at most $10^{-5}$.

The calculation of the Yukawas, in particular $Y_d$, is important
for the calculation of low energy observables. In the calculation
of the Yukawa couplings (see Eq.~(\ref{L1loop}))
 we take into account the thresholds of
supersymmetric particles as given in \cite{Buras:2002vd}.
Due to the RGE running,
off-diagonal elements in the squark sector are induced which remain
non-zero when going to the super-CKM basis. For this reason, not only
the chargino contributions to the off-diagonal terms but also gluino and
neutralino contributions are taken into account although the later ones
are negligible. The gluino
contribution amounts typically to a 10\% correction of the chargino
contribution, which is precisely the order of magnitude one would expect
for the two-loop corrections.

\begin{table}
\begin{center}
\begin{tabular*}{0.9\textwidth}{@{\extracolsep{\fill}}|c|c|c|c|c|c|c|c|c|}
\hline
$M_{Bs}$ &  $\tau_{Bs}$ &
$\eta_B$~ &
$f_{Bs} \hat{B}_{Bs}^{1/2}$ &$\tau(B_s)$ & $\bar{P}_1^{LR}$ & $\bar{P}_2^{LR}$ &
$\bar{P}_1^{SLL}$ & $\bar{P}_2^{SLL}$ \\ \hline 5.3696 GeV & 1.461 ps~ & 0.55~ &
$(294\pm 33)$ MeV & 1.461 ps~ &
 -0.71   & 0.9 & -0.37 &  -0.72 \\
\hline
\end{tabular*}
\end{center}
\caption{Numerical values for the B-physics parameters using the notation
of ref.~\cite{Buras:2002vd}. The lattice parameter has been taken from
Ref.~\cite{Mackenzie:2006un}.}
\label{tab:Binput}
\end{table}

For the calculation of the low energy observables, the complete formulas
are used, including all contributions stemming
from $H^+$, $\tilde \chi^\pm_i$, $\tilde \chi^0_j$, $\tilde g$, $h^0$,
$H^0$,  $A^0$ in case of $B$ physics observables and the contributions due
to $\tilde \chi^\pm_i$, $\tilde \chi^0_j$ in case of $a_\mu$.
More precisely, in case of $B_s \to \mu^+ \mu^-$ we combine
the formulas of   \cite{Buras:2002vd}, \cite{Bobeth:2001jm} and
\cite{Goto:1996dh}.
However, we use only the lowest order contributions to $C_S$, $C_P$,
$C'_S$, $C'_P$, $C_{10}$ and $C'_{10}$ because the complete one-loop QCD
corrections to these Wilson coefficients
are not available in the literature. In the case of $\Delta M_{B_q}$ we
have used the formulas of Refs.~\cite{Baek:2001kh,Buras:2002vd}.
In the case of $b\to s \gamma$ we have used the formulas given in
\cite{Bertolini:1990if}
and we have added the leading SM QCD corrections as described in
\cite{Hurth:2003dk}. For the SUSY contributions to the anomalous magnetic
moment of the muon we have used the formulas given in
\cite{Ibrahim:1999aj}. The hadronic parameters we use in all these
calculations are collected in Table~\ref{tab:Binput}. In most of the cases we
neglect the uncertainty in these parameters with the exception of $f_{B_s} \hat
B_{B_s}^{1/2}$ that has a large impact in the $B_s$ mass difference. In the
calculation of the $B_s$ mass difference we compare below the results obtained
fixing this parameter to the value in Table~\ref{tab:Binput} and the
results varying $f_{B_s} \hat B_{B_s}^{1/2}$  within the one-sigma range
$f_{B_s} \hat B_{B_s}^{1/2} = \left(0.295 \pm 0.036 \right)$ GeV
\cite{Ball:2006xx}.

\subsection{$B_s \to \mu^+ \mu^-$ in the CMSSM}
As we have seen in previous sections, $\tan \beta$ enhanced Higgs FCNC
at low energies are inversely proportional to the pseudoscalar Higgs
mass and are essentially generated via a stop-chargino loop.
In the context of a CMSSM defined at $M_{\rm GUT}$ it is
natural to ask what are the possible values of these masses and couplings
($m_A$, $\tan \beta$, \ldots) that can be reached consistently with
all the phenomenological bounds.

We will analyze this problem discussing separately the case of $\mu>0$
and $\mu<0$. In fact sign($\mu$) is a key parameter in the analysis of
the phenomenology of the large $\tan \beta$ limit for two main
reasons.  First, as well-known and explained in detail section
\ref{sec:correl}, the sign of the SUSY contributions to several low
energy processes like $b\to s \gamma$ and the muon anomalous magnetic
moment are determined by the sign of the $\mu$ parameter; therefore
the stringent experimental constraints tend to favour the positive sign of
$\mu$
specially in the large $\tan \beta$ limit.  In addition, as we
have seen in section \ref{sec:HiggsFCNCs}, the sign of the chargino
one-loop corrections to the down-quark Yukawa matrices and in
particular of the $\epsilon_0$ and $\tilde \epsilon_j$ parameters are
also fixed by sign($\mu$). This implies that, in the basis of diagonal
tree-level down Yukawa matrices, these one-loop corrections tend to
reduce or increase these diagonal entries depending on the sign of
$\mu$ and hence they give rise to a different phenomenology.
In the $\mu>0$ case, we have that both $\epsilon_0$ and $\tilde \epsilon_j$
are positive; therefore, Eqs.~(\ref{L1loop})--(\ref{Lneutral2})
imply that the diagonal Higgs couplings to the down quarks
are smaller than the naive expectation, $\overline m_d/v_1$. As a consequence,
for instance, FCNC couplings are smaller and effects of down-quark Yukawa matrices
in the RGE's are reduced.

\begin{figure}
\includegraphics[width=8cm]{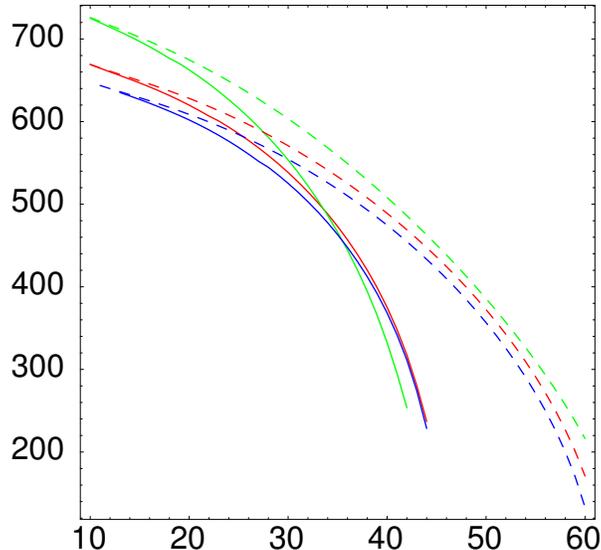}
\caption{Values of $M_A$ as a function of $\tan \beta$ in the CMSSM. The
SUSY parameters are fixed as $m_{1/2}=300$ GeV, $m_0=500$ GeV and
the blue lines correspond to $A_0=500$ GeV  blue, green lines to $A_0=0$ GeV
and red lines to $A_0=- 500$ GeV. Full lines correspond to $\mu<0$ and dashed
lines to $\mu>0$.}
\label{fig:mAtgb}
\end{figure}

The requirement of correct electroweak symmetry breaking sets an upper
bound on $\tan \beta$. This is mainly a consequence of
two difficulties: Firstly, at large $\tan \beta$, both $m^2_{H_u}$ and
$m^2_{H_d}$ become negative around the electroweak scale and
the pseudoscalar Higgs becomes tachyonic, as can be seen in
Fig.~\ref{fig:mAtgb}.  Secondly, the bottom Yukawa becomes non-perturbative
near the GUT scale. Furthermore, in the region with $M_0 \ll M_{1/2}$ there is the
additional problem of the lighter stau becoming tachyonic. Therefore,
the points of parameter space that succeed in generating a correct electroweak
symmetry breaking with large $\tan \beta$ are strongly reduced.
 These problems are clearly softened if the entries in the down-quark Yukawa
matrices are small as it is the case with $\mu>0$. In fact, for positive
$\mu$, values of $\tan \beta$ up to about
60 are still reachable in the CMSSM.

\begin{figure}
\begin{tabular}{lr}
\includegraphics[width=8cm]{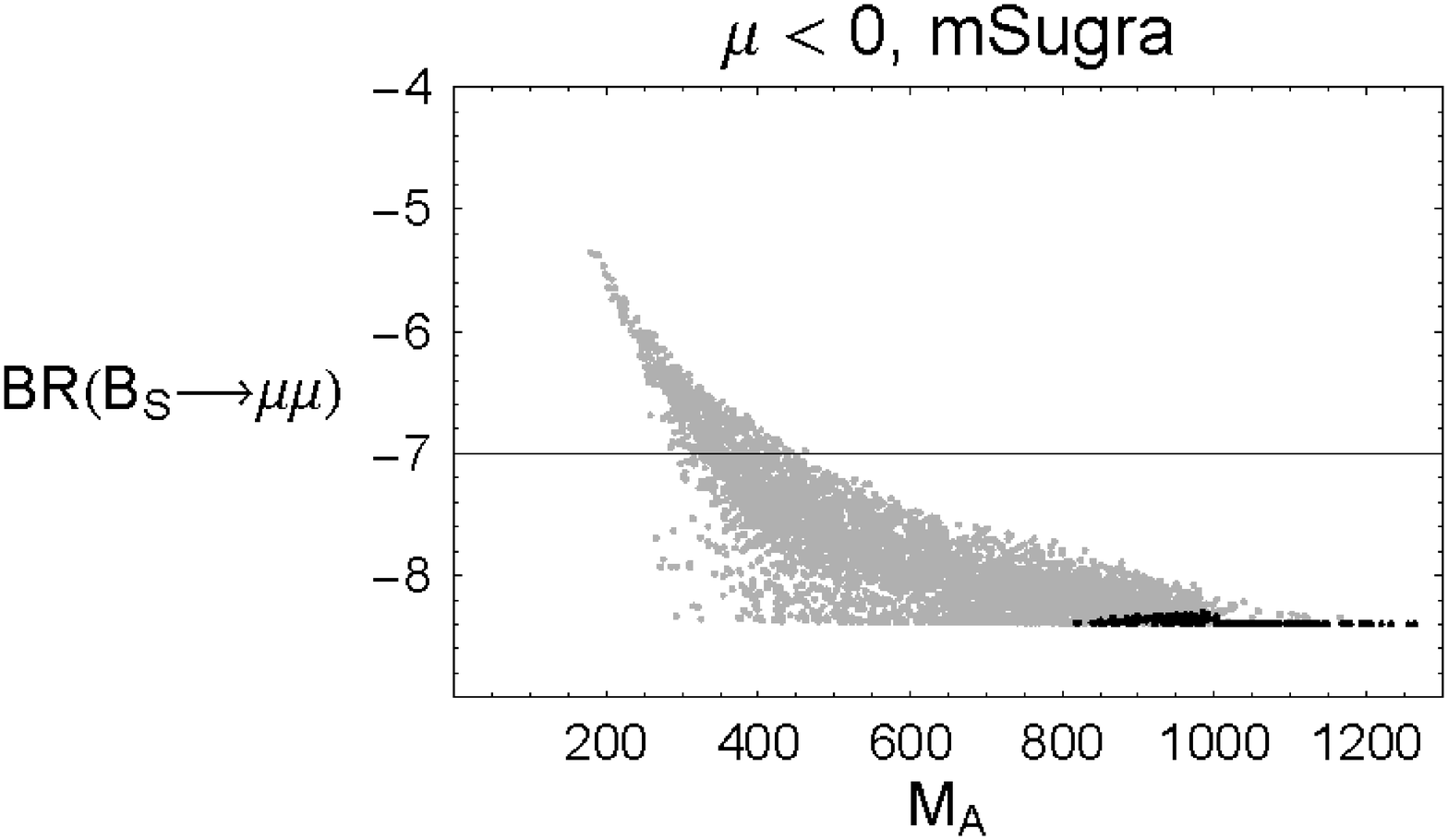}&
\includegraphics[width=8cm]{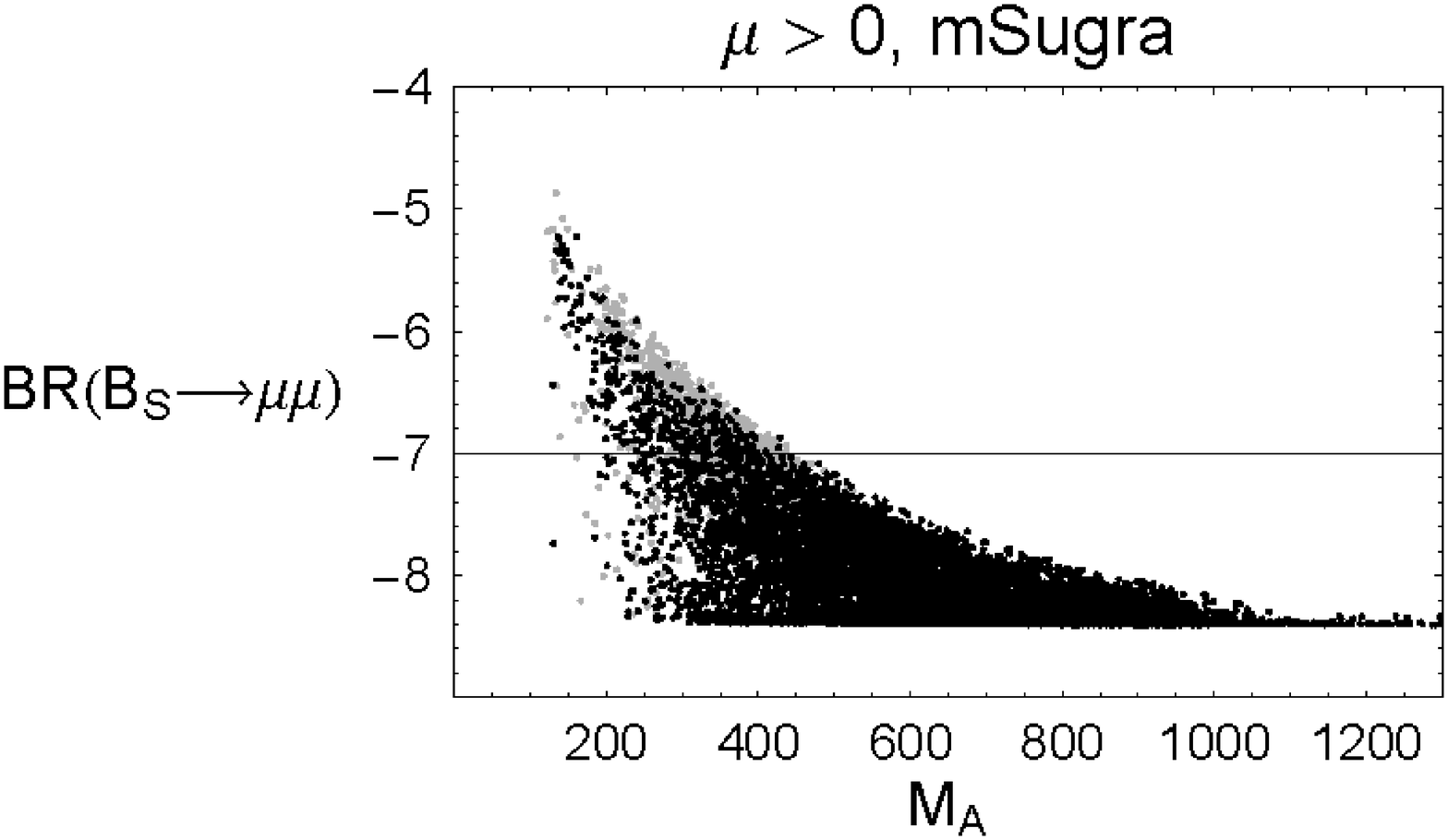}
\cr
\includegraphics[width=8cm]{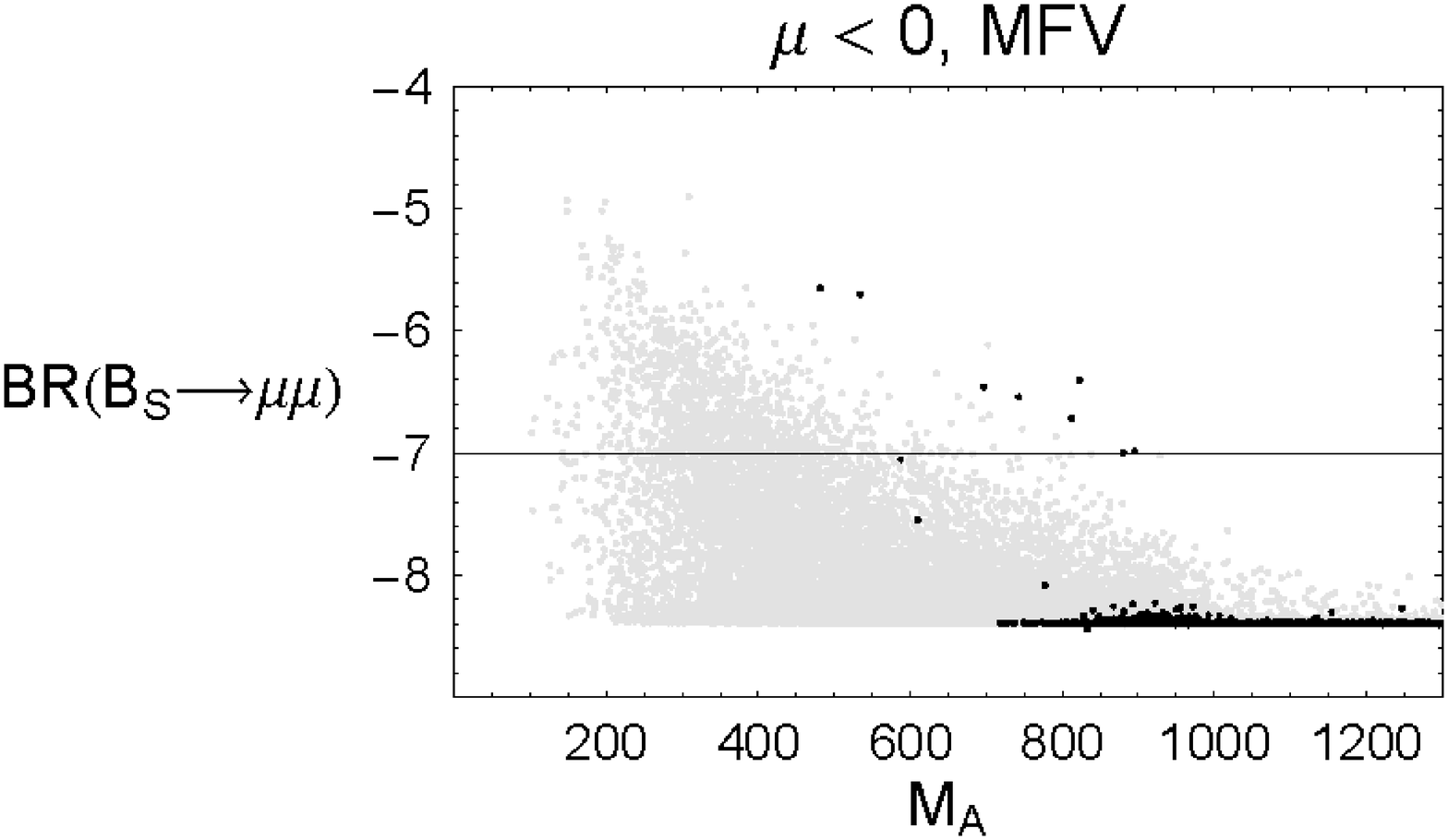}&
\includegraphics[width=8cm]{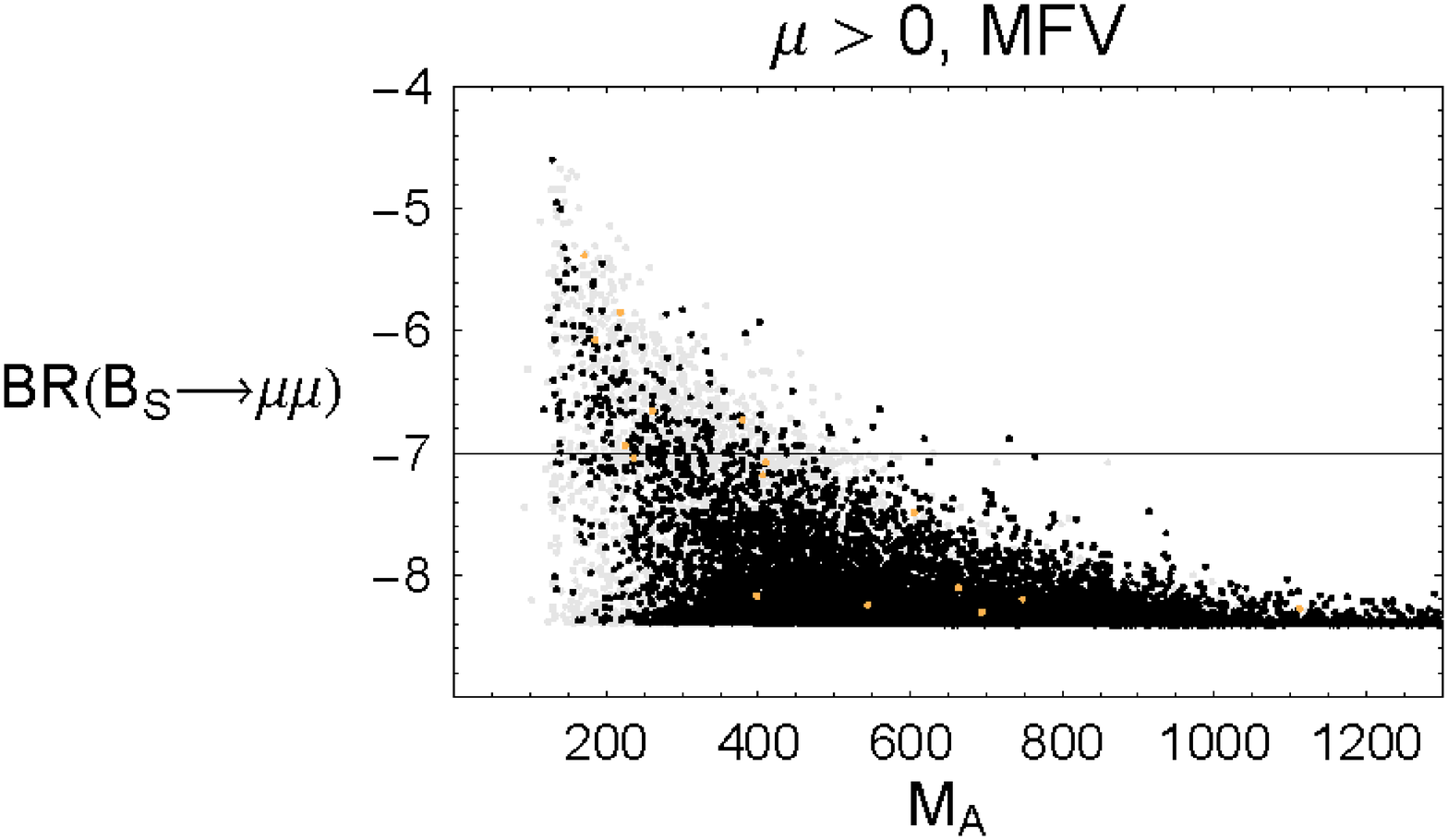}
\cr
\end{tabular}
\caption{Correlation between $M_A$ and ${\rm BR} \left(B_s\to \mu^+
\mu^-\right)$. All
points satisfy the dark matter constraint. The grey points do
not survive constraints from $B\to X_s\gamma$ and $a_\mu$.  Orange
dots correspond to scenarios in which $C_7^{\rm eff}(\mu_b)>0$
although they do not satisfy the BR($b \to s l^+ l^-$) constraint.}
\label{fig:mabsmm}
\end{figure}

In the same way we must analyze the possible values of $M_A$.
The mass of the pseudoscalar Higgs is approximately degenerate to the
mass of the heavy scalar Higgs and both of them are related to the value of
$|\mu|$. As already pointed out, at large values of $\tan \beta$ it is
possible to reach smaller values of $M_A$ due to due the fact that
$M^2_{H_d}$ gets negative which also is the reason for the smaller value
of $|\mu|$. This can be seen in Fig.~\ref{fig:mAtgb} where we show the
possible values of $M_A$ in the CMSSM as a function of $\tan \beta$ for fixed
values of $M_{1/2}=300$ GeV and $M_0=500$ GeV.
As expected, we see the different behaviour of $M_A$
for positive or negative $\mu$ at large $\tan \beta$.  For positive
$\mu$ the value of $M_A$ is larger than the corresponding value for
negative $\mu$ and the same value of $\tan \beta$ due to the smaller
effect of the down-quark Yukawas on the RGE running of $m_{H_d}$.  In
this plot, we do not include the experimental constraints on the SUSY
parameter space from direct and indirect processes. In fact, making a
general scanning of all the SUSY parameters, for $\mu>0$, we can still
find values of $M_A$ as low as 150 GeV corresponding to $\tan \beta
\geq 50$ consistently with all constraints, see for instance
Fig.~\ref{fig:mabsmm}.

\begin{figure}
\begin{tabular}{lr}
\includegraphics[width=8cm]{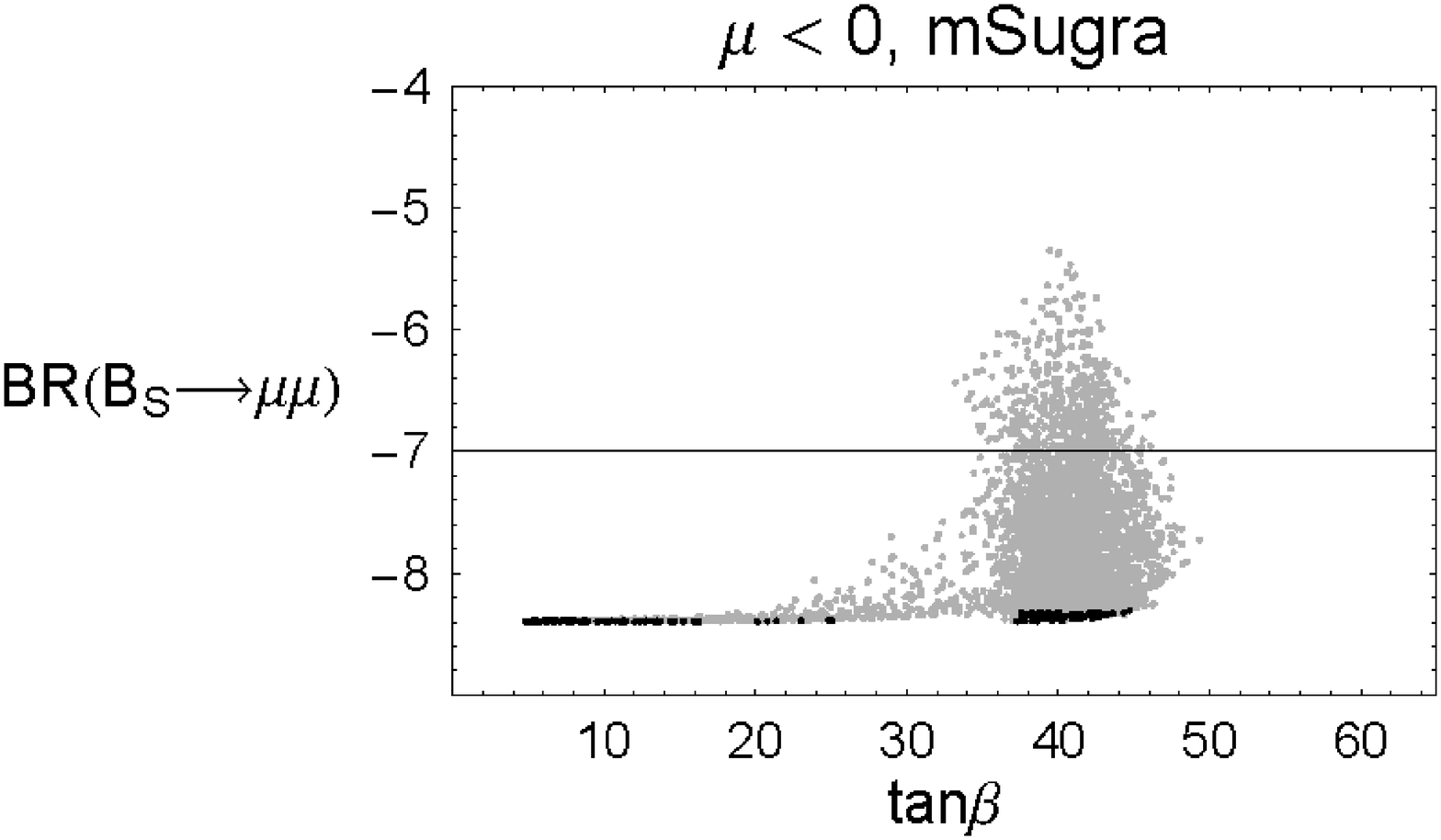}&\includegraphics[width=8cm]{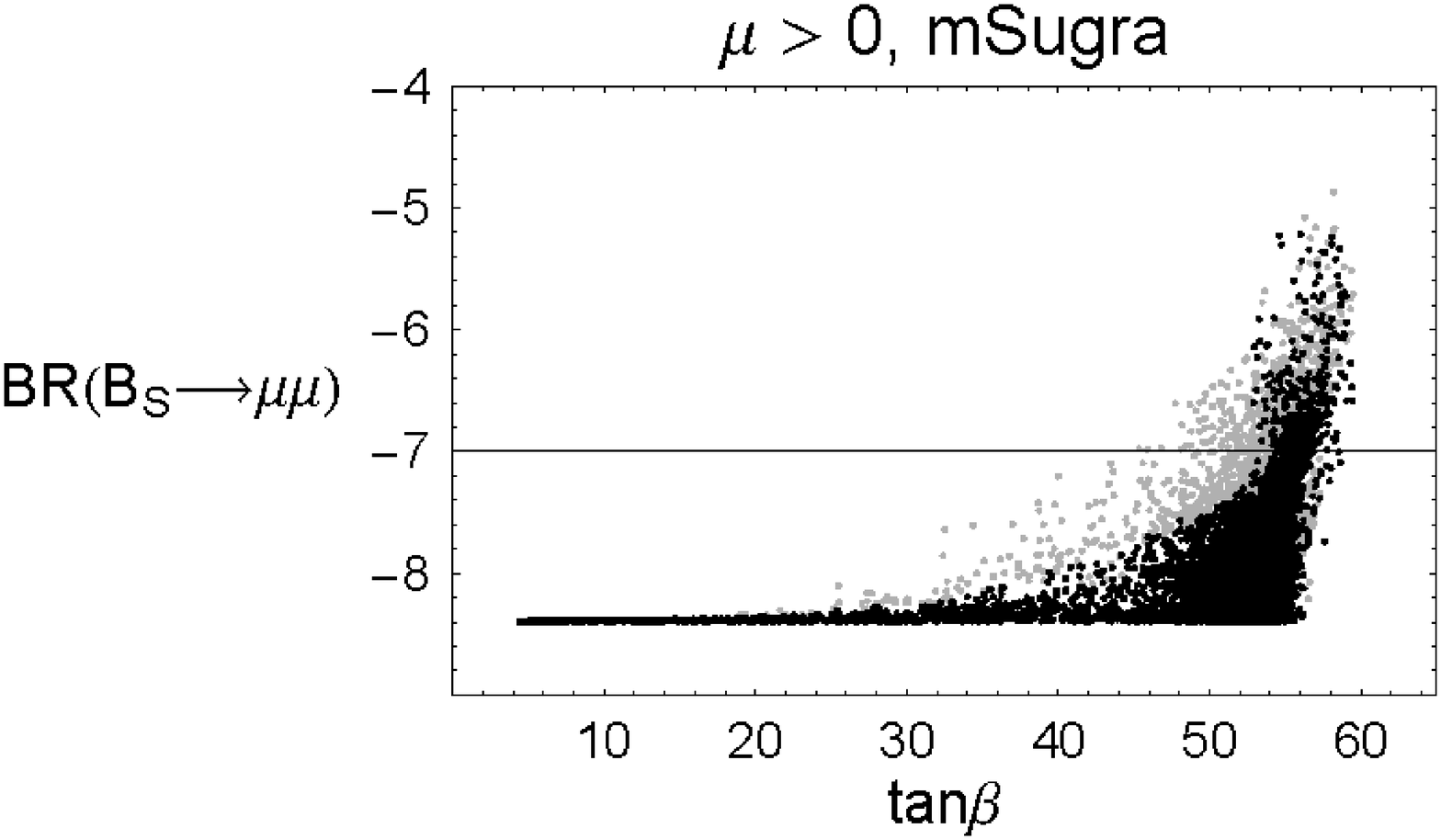}
\cr
\includegraphics[width=8cm]{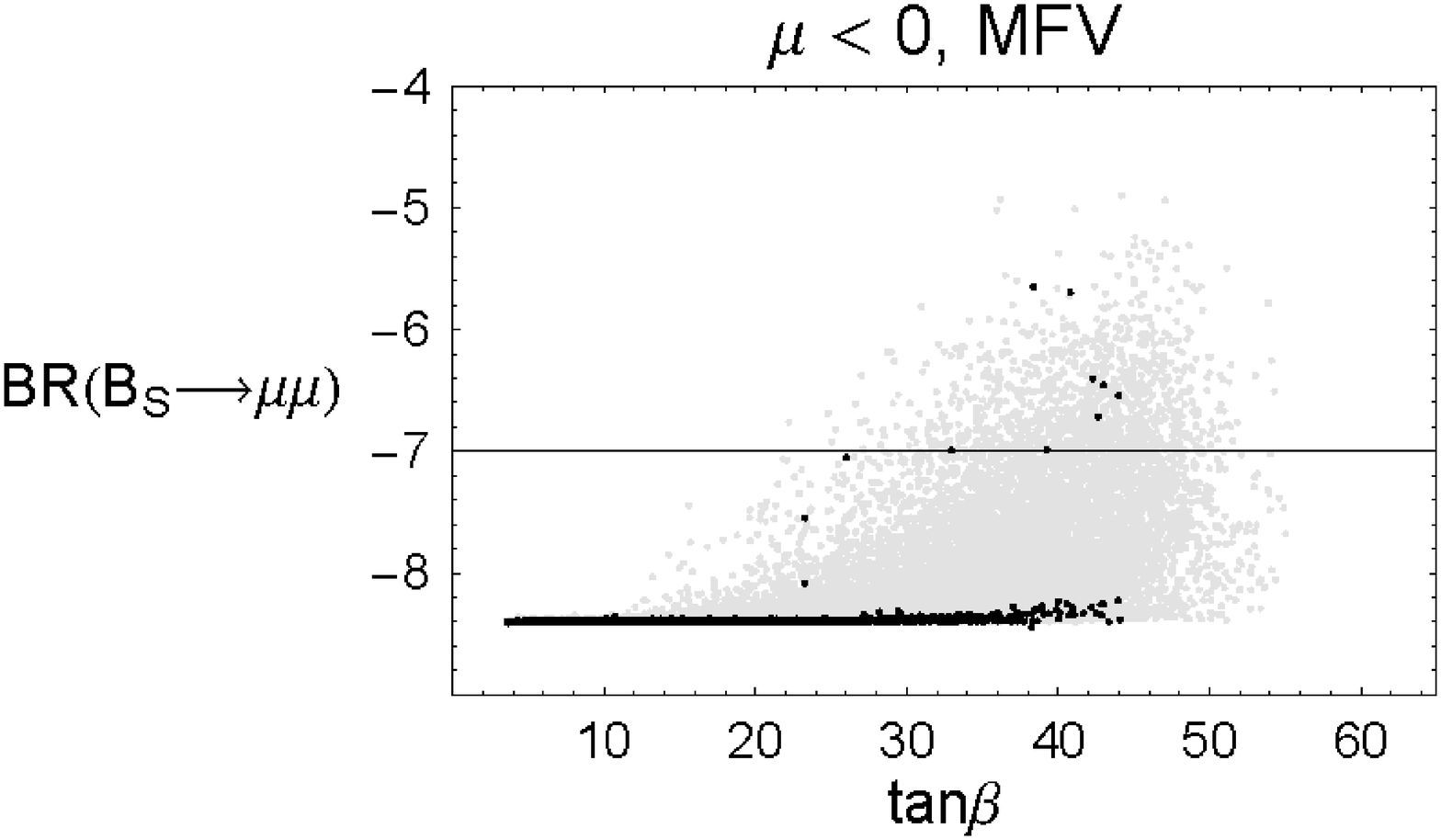}&\includegraphics[width=8cm]{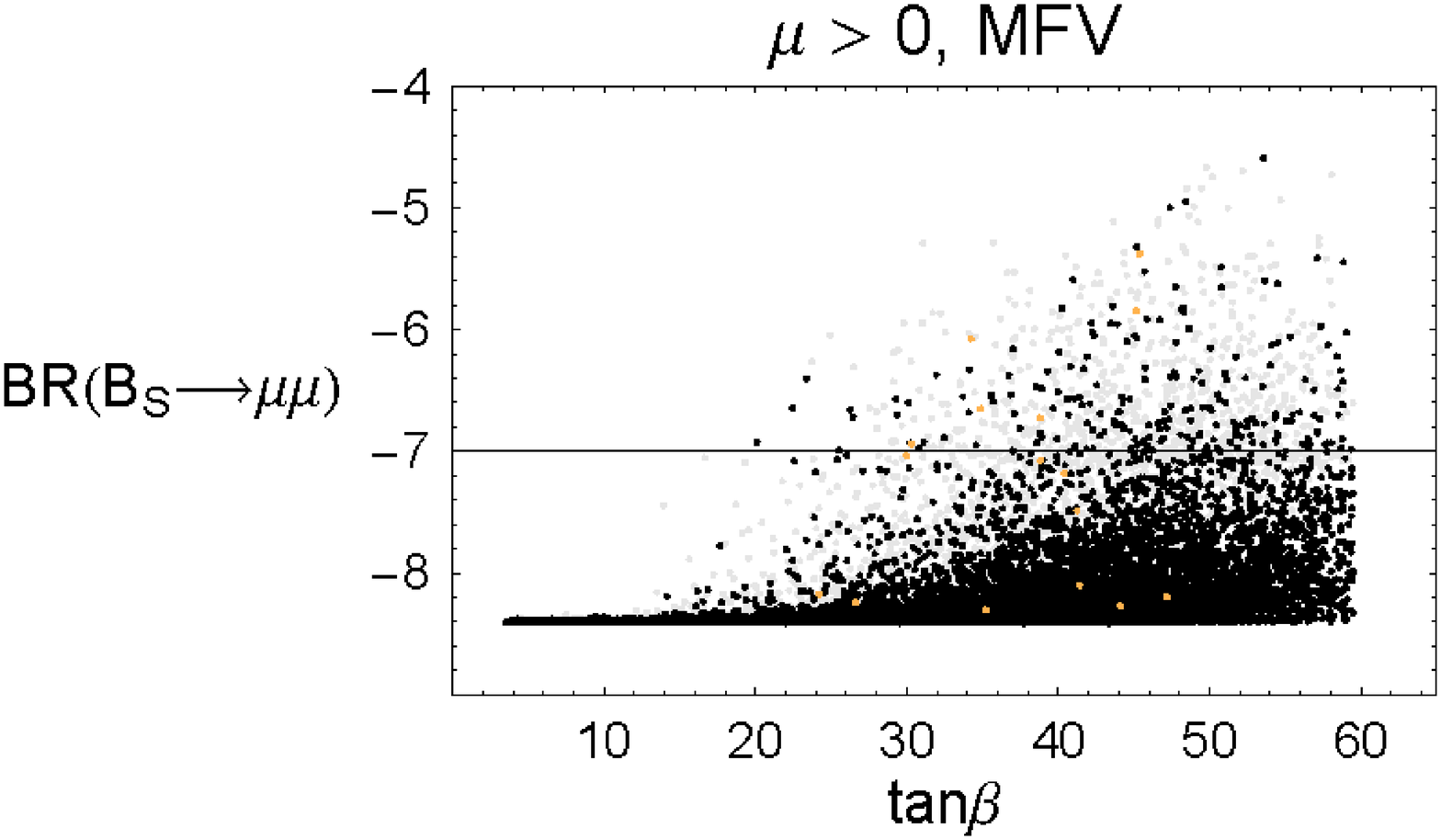}
\cr
\end{tabular}
\caption{BR($B_s \to \mu^+ \mu^-$) as a function of $\tan \beta$ in the CMSSM (upper plots) and MFV (lower plots). See the caption in Fig.~\ref{fig:mabsmm}. }
\label{fig:BRtgb}
\end{figure}

Inserting $M_A=150$ GeV, $\tan \beta = 60$  and $\mu>0$
in the approximate formula given in \eq{BRbsll} we obtain
\bea
\label{estim1}
{\rm BR}\left(B_s\to \mu^+ \mu^-\right) &=& 1.9 \times 10^{-4}
\Frac{A_t^2}{\mu^2}~~
\Frac{\left(H_2\left(m_{t_\Le}^2/\mu^2,m_{t_\Ri}^2/\mu^2\right)\right)^2}
{(1+\epsilon_0 \tan \beta)^2(1+\tilde \epsilon_3 \tan \beta)^2}.
\eea
Taking $m_{\tilde g} \simeq \mu \simeq m_{\tilde t}$ and using
$H_2(1,1)\simeq - 1/2$ we have $\epsilon_0 \simeq 0.012$ and
$\tilde \epsilon_3 \simeq 0.015$. Therefore, a
BR$\left(B_s\to \mu^+ \mu^-\right)$ between $10^{-7}$ and $5 \times 10^{-6}$
would be the maximal value we can get in this model.
In any case, we expect these regions of large $\tan \beta$ to be
strongly constrained by other FCNC processes like $b \to s \gamma$
and $\delta a_\mu$. In Figs.~\ref{fig:mabsmm} and \ref{fig:BRtgb} we
present the results for this branching ratio obtained using the complete
formulae and taking
into account all the constraints.
Although many points that give rise to a large  BR$(B_s \to \mu^+
\mu^-)$ are already excluded by the stringent bounds from FCNC processes,
cancelations between different contributions in $b\to s \gamma$ still allow a large
BR$(B_s \to \mu^+ \mu^-)$.

\begin{figure}
\begin{tabular}{c}
\includegraphics[width=8cm]{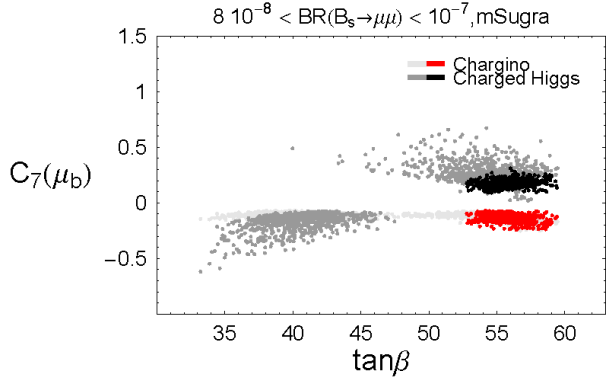} \cr
\end{tabular}
\caption{Charged Higgs and chargino contributions to the $C_7$
Wilson coefficient in CMSSM as function of $\tan \beta$
corresponding to large BR($B_s \to \mu \mu$) and $\mu>0$. Red points
refer to chargino and black points to charged Higgs contributions.
Orange points are points where $C_7^{\rm tot}$ changes sign although they
are not allowed by the  BR($b \to s l^+ l^-$) constraint.
Light grey points do not survive the FCNC constraints.} \label{fig:C7}
\end{figure}

In Figure~\ref{fig:C7} we compare the
chargino and charged Higgs contributions to the Wilson coefficient
$C_7(M_W)$ for the case of a BR($B_s \to \mu^+ \mu^-$) just below the
current experimental bound. With positive $\mu$, charged Higgs and
chargino contributions tend to compensate and we can find many points in
agreement with the $B\to X_s \gamma$ constraints. Moreover
$M_{H^+}^2 \simeq M_A^2 + M_W^2$, thus a small pseudoscalar mass implies
also a small charged Higgs mass. In fact, the
experimentally allowed values for $C_7^{\rm tot}$ range from $\sim -
0.23$ to $\sim -0.08$ while the contribution from the W-boson is about
$-0.17$. This agrees with Fig. \ref{fig:C7} where we can see that
the chargino and charged Higgs contributions must be strongly correlated:
in fact, their absolute value can be as large as 0.26 while their sum
must range between -0.05 and +0.10.
Therefore we see that, with positive $\mu$, it is possible
to have a light pseudoscalar and large $\tan \beta$ generating a large
BR($B_s \to \mu \mu$) while at the same time keep the sum of chargino and charged
Higgs contributions to
$b\to s \gamma$ under control.
Fig.~\ref{fig:C7} also shows a couple of points for which the chargino
contribution is sufficiently large to change the whole sign of the total Wilson coefficient
at the low scale and still remain in agreement with experiments. From the
discussion at the end of Sec.~\ref{sec:bsll},
it is clear that such points are excluded by the present measurement of ${\rm BR} (B\to X_s \ell^+\ell^-)$
because their
contributions to $C_9$ and $C_{10}$ (that lie within the black ellipsis drawn in Fig.~\ref{fig:C7})
are too small.

\begin{figure}
\begin{tabular}{lr}
\includegraphics[width=8cm]{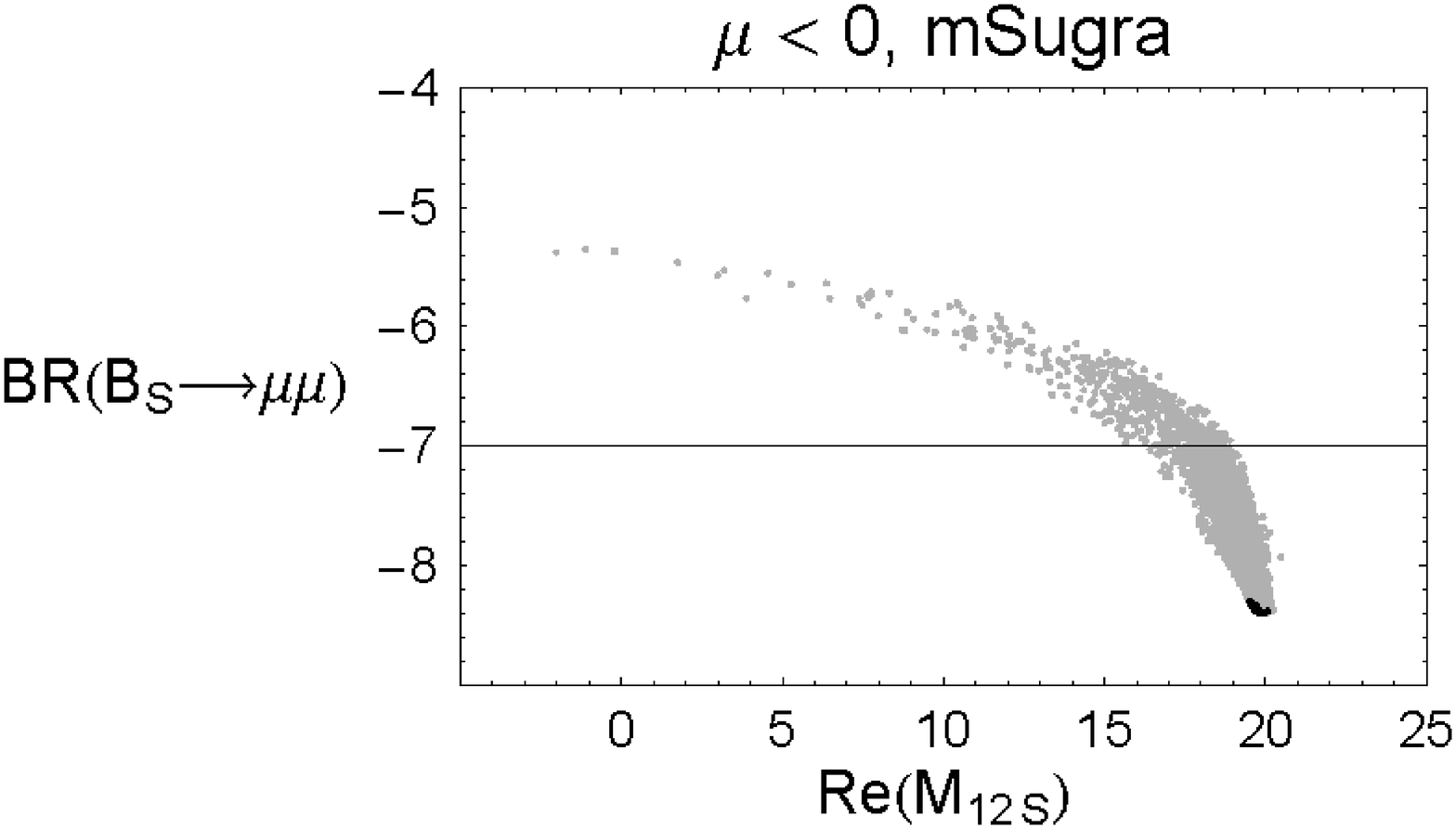}&\includegraphics[width=8cm]{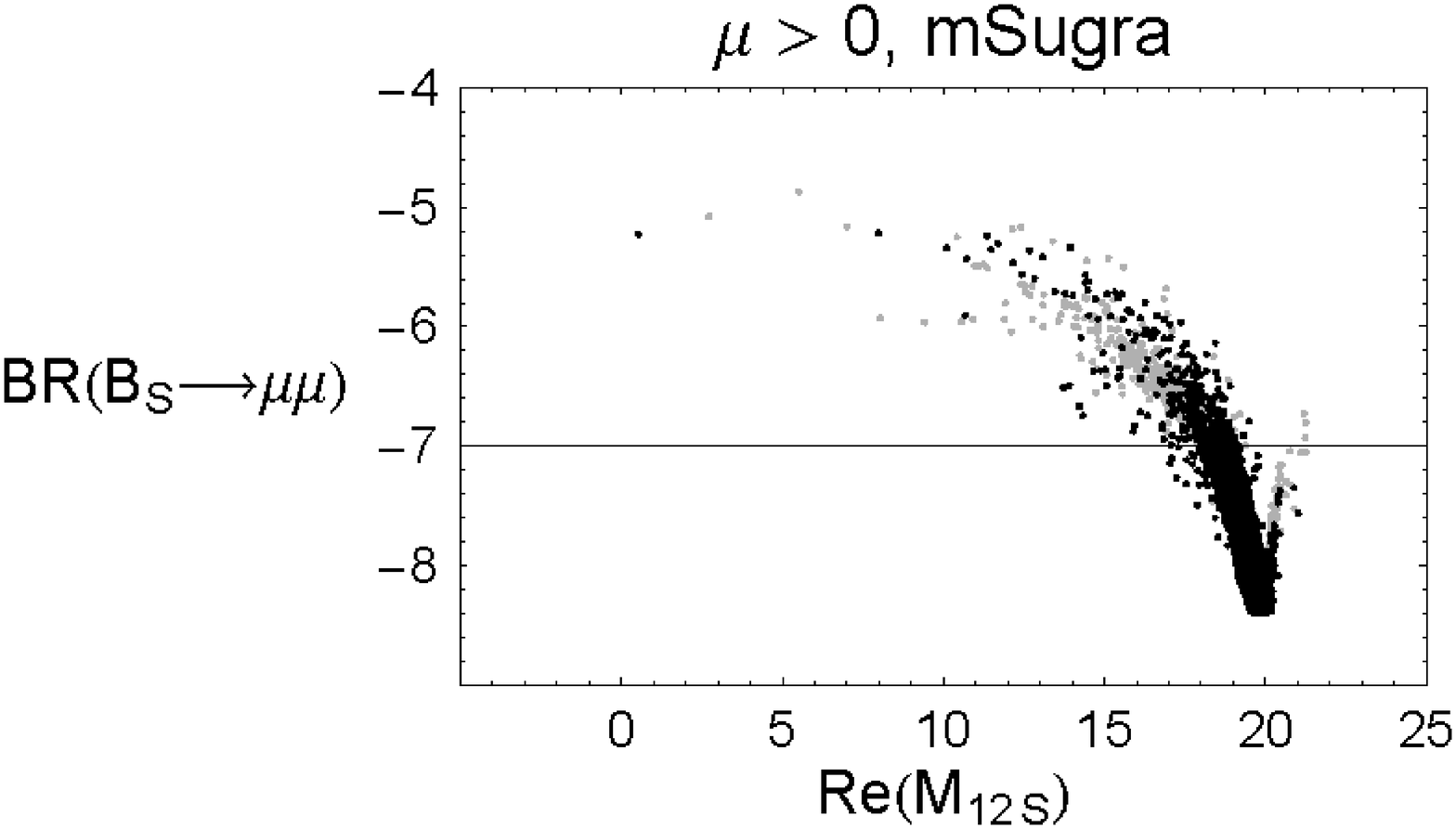}
\cr
\includegraphics[width=8cm]{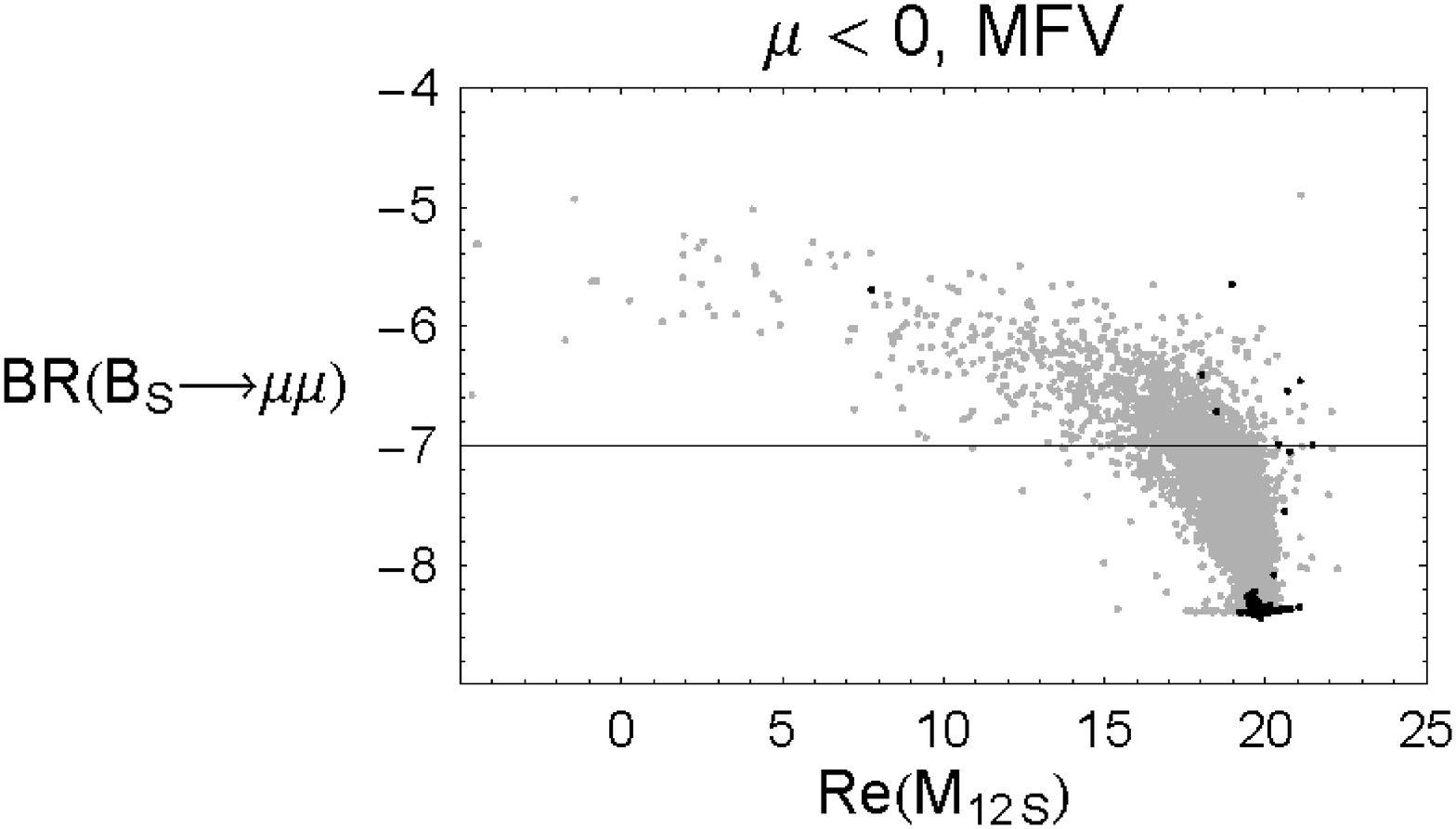}&\includegraphics[width=8cm]{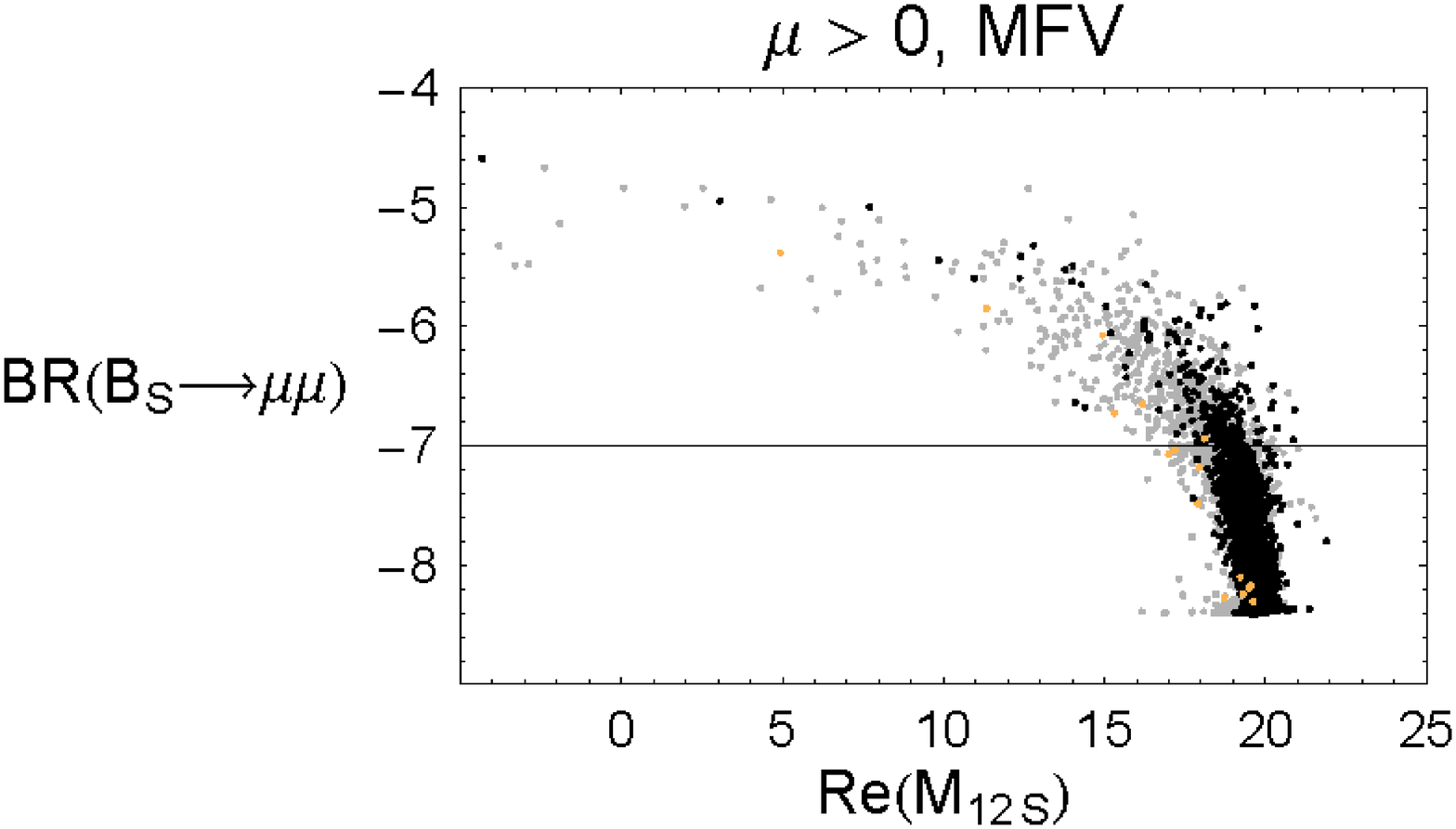}
\cr
\end{tabular} \caption{Correlation between ${\rm Re} M_{12}^s$ and ${\rm BR} \left(B_s\to \mu^+
\mu^-\right)$. See the caption in Fig.~\ref{fig:mabsmm}.}
\label{fig:DMBS}
\end{figure}

\begin{figure}
\begin{tabular}{lr}
\includegraphics[width=8cm]{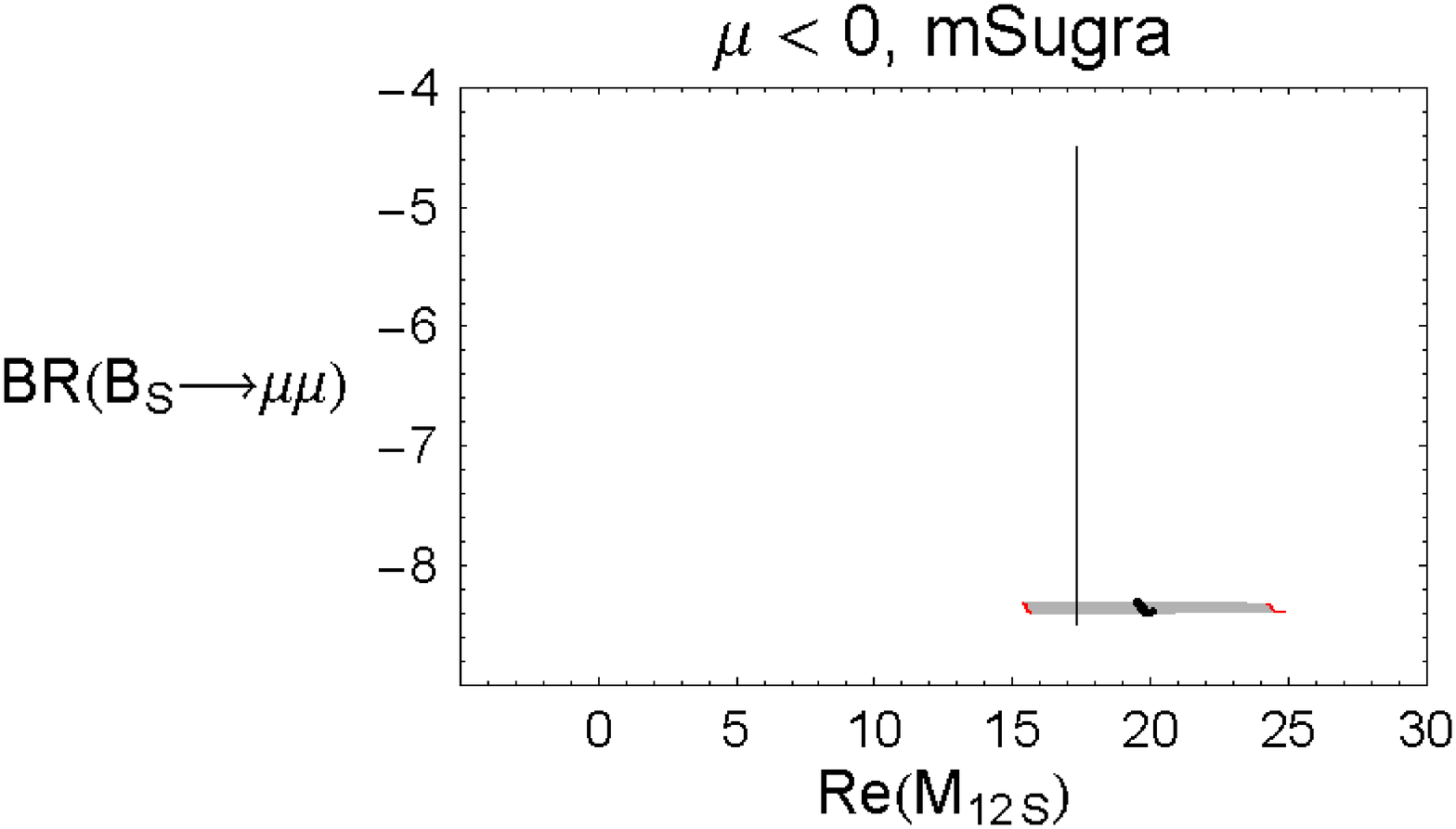}&\includegraphics[width=8cm]{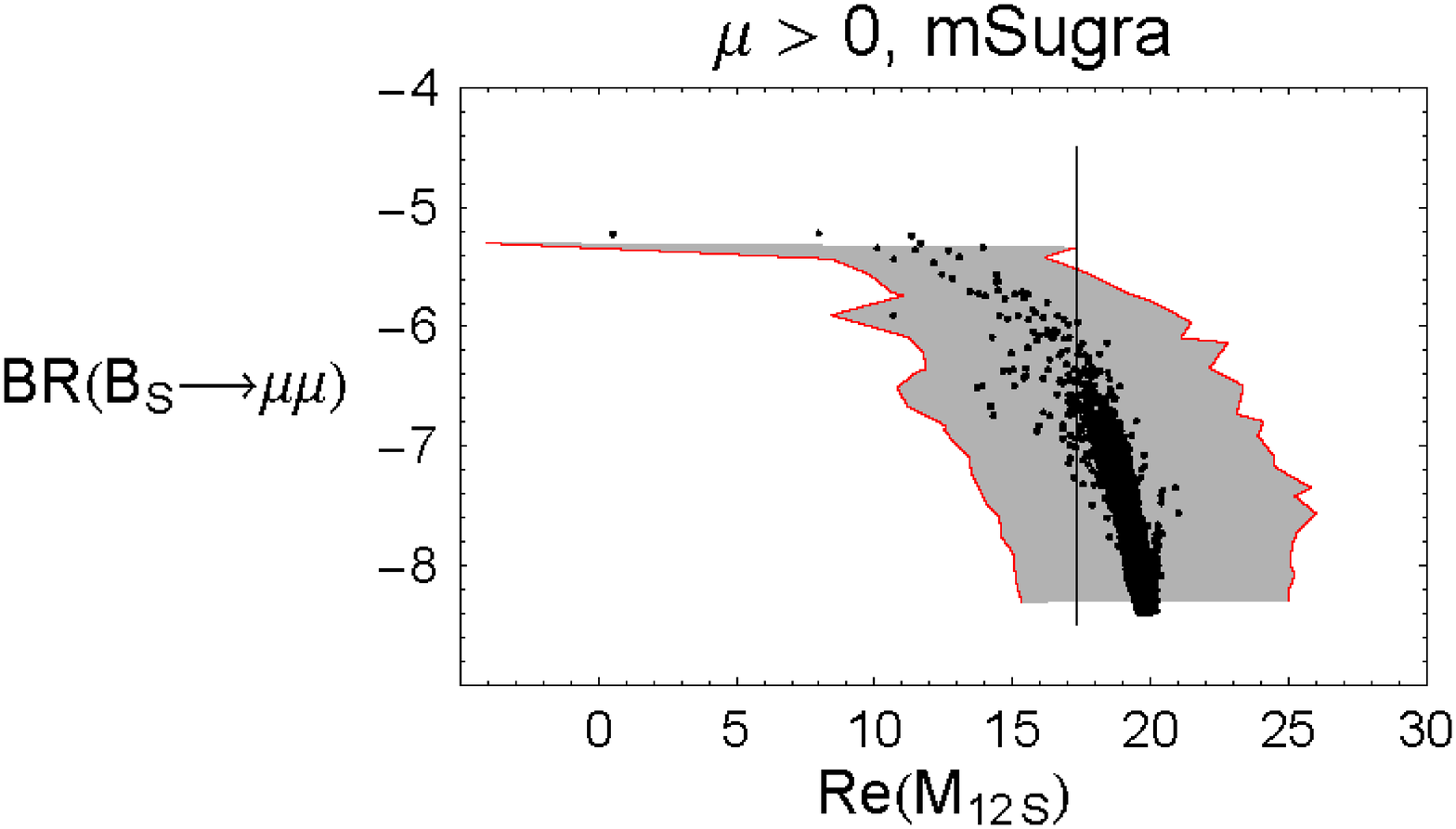}
\cr
\includegraphics[width=8cm]{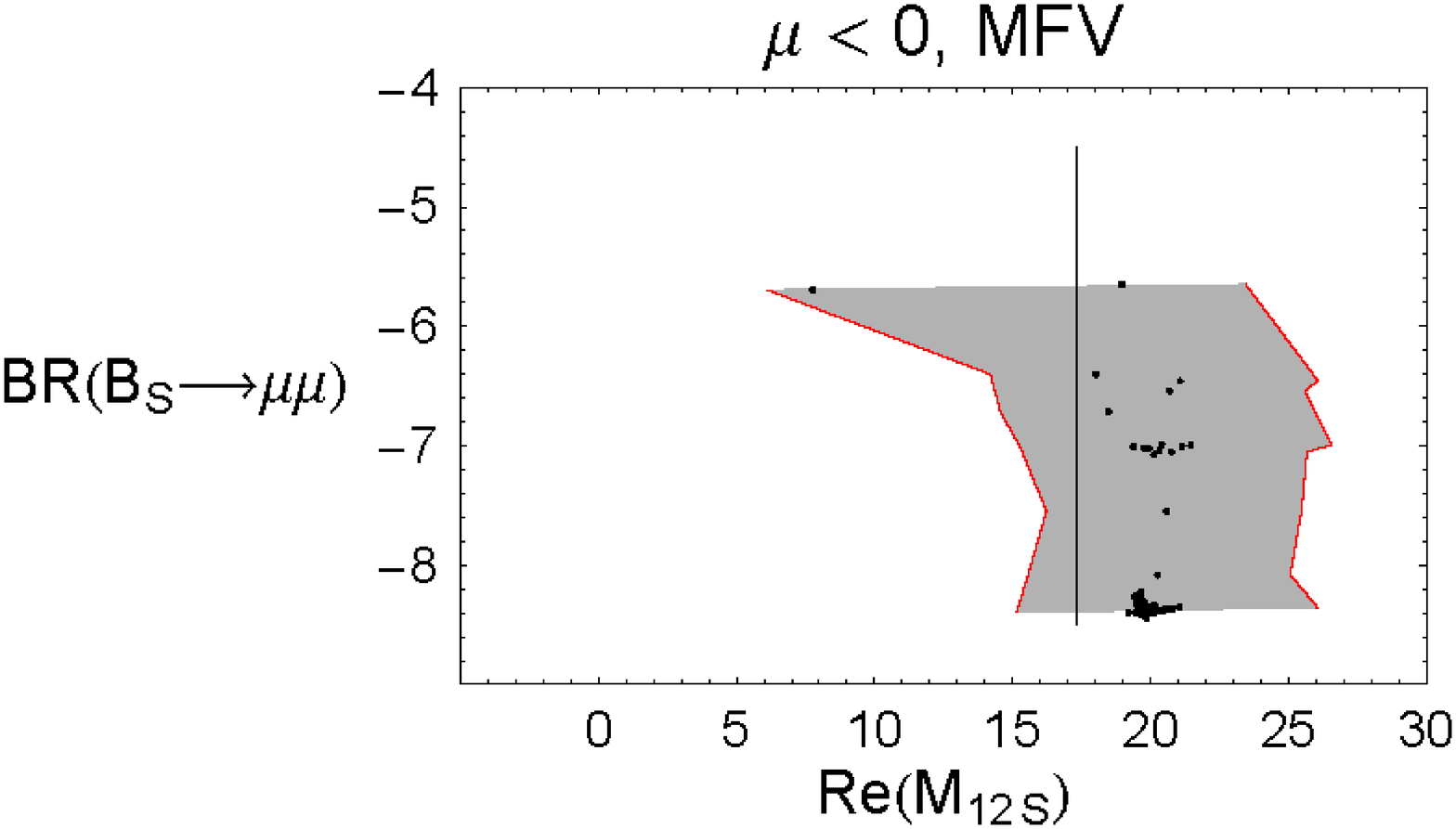}&\includegraphics[width=8cm]{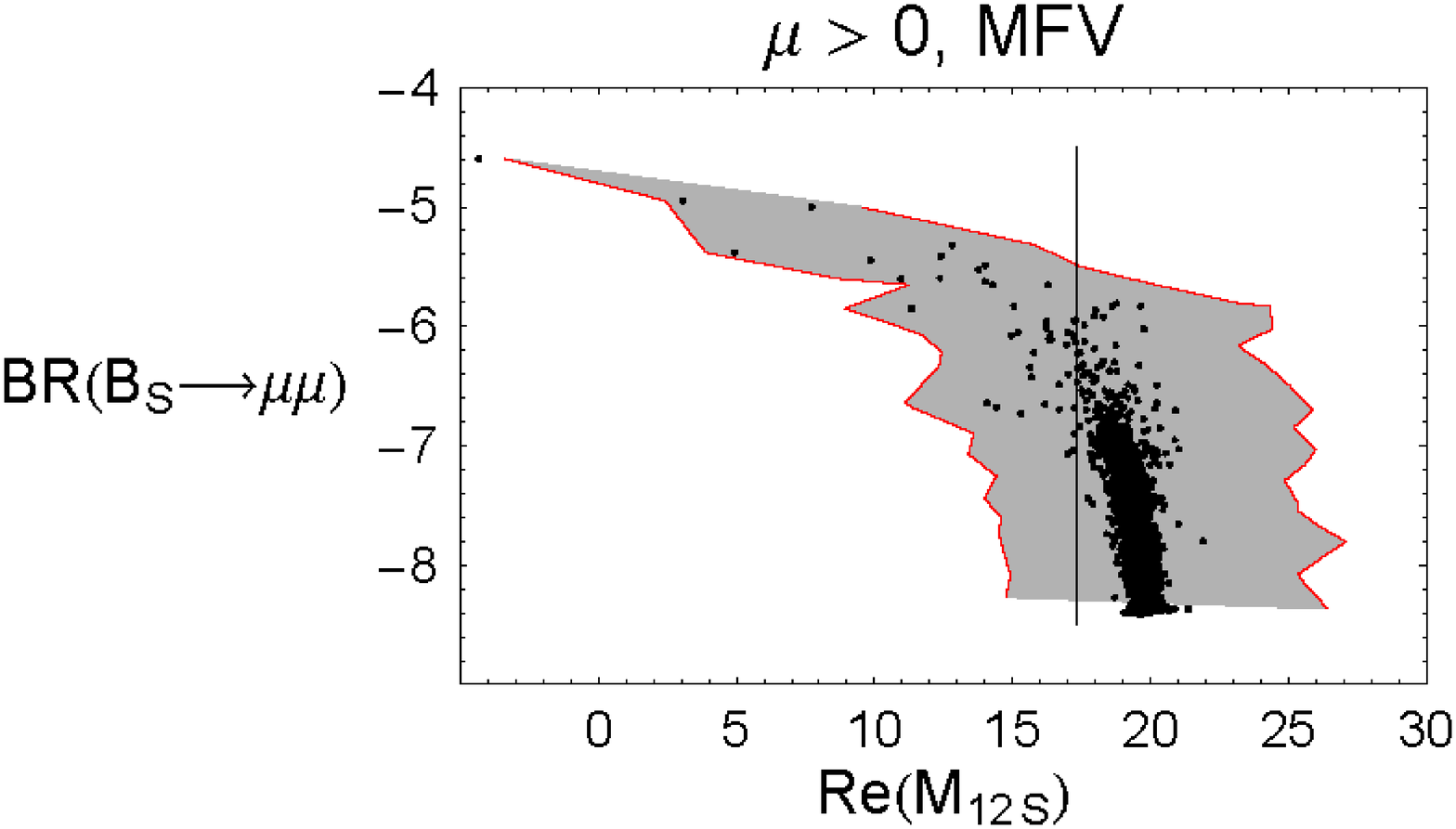}
\cr
\end{tabular} \caption{Correlation between ${\rm Re} M_{12}^s$ and ${\rm BR} \left(B_s\to \mu^+
\mu^-\right)$ including theory errors. The points satisfy all the constraints.
The gray shaded area corresponds to theory uncertainties. The line is the experimental value of
$\Delta MB_s$}.
\label{fig:DMBS2}
\end{figure}

A third process, closely related to the $B_s \to \mu \mu$ decay, is
the $B_s - \bar B_s$ mass difference. As discussed in section
\ref{sec:HiggsFCNCs},
both processes involve the same FCNC neutral Higgs coupling,
although they have a different $M_A$ and $\tan \beta$ dependence.
Note that in this case, the FCNC Higgs coupling enters twice as a
double-penguin contribution and this implies that the change of sign($\mu$)
does not affect the relative sign of the SUSY contribution to $\Delta M_{B_s}$
with respect to the SM one, that turns out to be always negative.
However,
the size of the FCNC Higgs couplings is smaller for $\mu>0$ than for $\mu<0$.
Recently the D\O~collaboration was able to set for the first time an
upper limit on $\Delta M_{B_s}$:
 $\Delta M_{B_s}= \left(19 \pm 2\right)~{\rm ps}^{-1}$ at 90 \%
C.L.~\cite{Abazov:2006dm}. It is necessary to stress that this error is
absolutely non-Gaussian and that there is a 5\% probability that
$\Delta M_{B_s}$ lies
anywhere between 21 $ps^{-1}$ and infinity. Shortly afterwards the
CDF collaboration
improved this bound \cite{CDFBs:2006}: $\Delta M_{B_s}=
\left(17.33^{+0.42}_{-0.21}({\rm stat}) \pm 0.07 ({\rm syst})\right)~
{\rm ps}^{-1}$. Unfortunately this very precise experimental measure is
not fully effective in constraining the SUSY parameter space due to the large
uncertainty in the theoretical input parameters, mainly $f_{B_s}^2 B_{B_s}$
\cite{Ball:2006xx}.
In Fig.~\ref{fig:DMBS} we compare Re~$M_{12}^s$ with BR($B_s \to \mu \mu$).
In fact, $\Delta M_{B_s}=|M_{12}^s|$, but in our model, as in the SM,
the phase of $M_{12}^s$ is at the per cent level and thus it is a
very good approximation to the total $\Delta M_{B_s}$. Moreover,
plotting Re~$M_{12}^s$ allows us to control the change of sign of
this amplitude. In this plot we fix the hadronic parameters to the values
shown in Table~\ref{tab:Binput}.
As we can see in the second plot of Fig. \ref{fig:DMBS}
corresponding to $\mu>0$, in the CMSSM and in the parameter space
allowed by FCNC constrains it is not possible to change
significantly the value of $\Delta M_{B_s}$ in the region with BR$(
B_s \to \mu^+ \mu^-) \leq 10^{-6}$. Consequently no change of sign
of Re~$M_{12}^s$ is possible in the CMSSM with $\mu>0$. Given
the present experimental bound on BR$( B_s \to \mu^+ \mu^-)
\leq 10^{-7}$, we conclude that double penguin contributions to
$\Delta M_{B_s}$ can not lower this mass difference below $17~{\rm
ps}^{-1}$. To compare this value with the experimental measure by the CDF
collaboration we must take into account the uncertainty in the hadronic
parameters.
In figure \ref{fig:DMBS2} we present the range of values we can reach varying
the value of $f_{B_s} B_{B_s}^{1/2}$ in the interval
$\left[0.259,0.331\right]$ GeV \cite{Ball:2006xx}. As we can see, after
taking into account the theoretical uncertainty, the $\Delta M_{B_s}$ constraint
can never compete with the bound on BR($B_s \to \mu^+ \mu^-$).

In the case of negative $\mu$, we have that both
$\epsilon_0$ and $\tilde \epsilon_j$ are negative.
Eqs.~(\ref{L1loop}--\ref{Lneutral2}) imply that the diagonal
Higgs couplings to the down quarks are larger than the naive expectation,
$\overline m_d/v_1$, and of the corresponding couplings for $\mu>0$. Hence,
FCNC couplings are larger and the effects of down-quark Yukawa matrices in
the RGE's is enhanced. This implies that the previously encountered problems
with electroweak symmetry breaking occur now already at $\tan \beta \simeq 45$.
Regarding the possible values of $M_A$, we see in Fig.~\ref{fig:mAtgb}
that, for negative $\mu$, $M_A$ tends to be much smaller than in the $\mu>0$
case due to the effects of the large down-quark Yukawas
on $m_{H_d}$. Taking into account the constraints from
BR$(B \to X_s \gamma)$ and $a_\mu$ we can only find allowed
points with $M_A\gsim 750$ GeV and $\tan \beta\simeq 40$.

Taking $\mu<0$, $x\simeq y \simeq 1$ and $m_{\tilde g} \simeq \mu \simeq m_{\tilde t}$
in \eq{estim1}, we get
$\epsilon_0 \simeq - 0.012$ and $\tilde \epsilon_3 \simeq -0.015$;
thus, in principle, we obtain an extra enhancement from the denominator
of \eq{estim1} instead of the suppression we get in the case of
$\mu >0$. However using the allowed masses (as obtained above) we find
that the maximum branching ratio we can get is about  $\simeq 10^{-7}$.
In the numerical analysis, the maximum branching ratio we obtain is always below
$10^{-8}$ as we can see in Fig.~\ref{fig:BRtgb}, because we have never
obtained simultaneously these extreme values for the masses and the quantities
$\epsilon_0$, $\tilde \epsilon_j$ while being consistent
with all the different constraints. In particular the  $b\to s \gamma$
constraint requires heavy squarks and Higgs-bosons and $\delta a_\mu$ requires
in additions heavy sleptons. Therefore, the SUSY contributions to
BR($B_s \to \mu \mu$) remain small for negative $\mu$ in the CMSSM.

Regarding the $B_s -\bar B_s$ mass difference, the double
penguin contribution has opposite sign to the SM contribution; hence,
reducing the predicted value of $\Delta M_{B_s}$. As we explained above,
Higgs FCNC couplings are now larger for $\mu<0$ and we expect larger
effects than for $\mu>0$. In fact we see in Fig.~\ref{fig:DMBS}
that with negative $\mu$ it is still possible to change the sign
of Re~$M_{12}^s$, although only very rarely. However, once we impose
the indirect constraints we see that no change in $\Delta M_{B_s}$
from the SM expectations is possible.
This result seems to differ from the analogous
plots (Fig. 23) of Ref. \cite{Buras:2002wq}. In fact, both figures would
perfectly agree if the $b \to s \gamma$ constraints are not imposed.
This is only due to
the difference in the MSSM models considered in both works.
Buras {\it et al.} analyze a general MSSM defined at the electroweak
scale with the different parameters unrelated and only constrained
by low energy experimental observables. In this framework they
have the freedom to assume a relative sign between $A_t$ and $\mu$
independently of the sign of $\mu$. In our model, all the parameters are
defined at  $M_{\rm GUT}$
and the sign of $A_i$ at the electroweak scale is always negative
due to the dominant (negative) gaugino contribution in the RGE evolution.
Therefore our results would agree without the $b \to s \gamma$ and
$\delta a_\mu$ constraints: taking into account these
constraints, large effects are not possible in this GUT inspired model.

Finally we analyze the effect of dark matter constraints on this process.
Although the lightest neutralino of the CMSSM is a good dark matter
candidate, it is well-known that in most of the parameter space of this model
a too large dark matter density is generated. Only under certain specific
conditions a correct value of $\Omega_{{\tilde \chi}^0} h^2$ is generated after
annihilation of the excess neutralinos among themselves or
coannihilation with other light sparticles. The regions where these
conditions are possible are:
\begin{enumerate}
\item The bulk annihilation region, at low values of $m_{1/2}$ and $m_0$,
where neutralinos annihilate in pairs at a sufficient rate via t-channel
slepton exchange.
\item The stau coannihilation region, where neutralinos coannihilate
with staus, given that $m_{{\tilde \chi}^0}\simeq m_{\tilde \tau}$.
\item The focus point region, at large $m_0$ where the value of $|\mu|$
is small and the neutralinos have a significant higgsino component enhancing
the annihilation into $WW$ and $ZZ$ pairs.
\item The funnel region, which occurs at very large $\tan \beta\sim 45$--$60$
and the pseudoscalar mass $m_A \sim 2 m_{{\tilde \chi}^0}$ so that neutralinos
annihilate into a fermion pair through a pseudoscalar-Higgs boson
in the s channel.
\end{enumerate}
\begin{figure}
\begin{tabular}{lr}
\includegraphics[width=8cm]{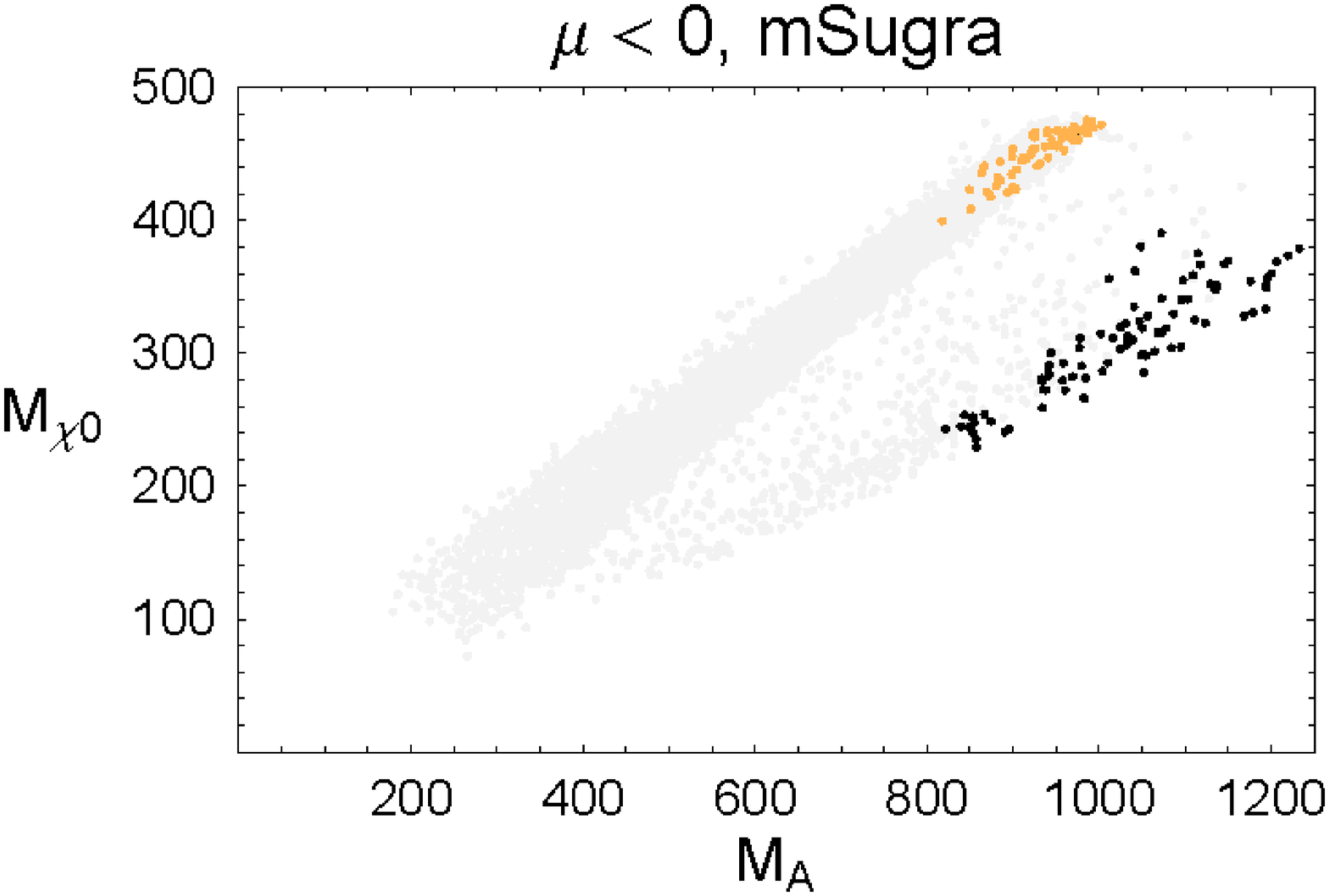}&\includegraphics[width=8cm]{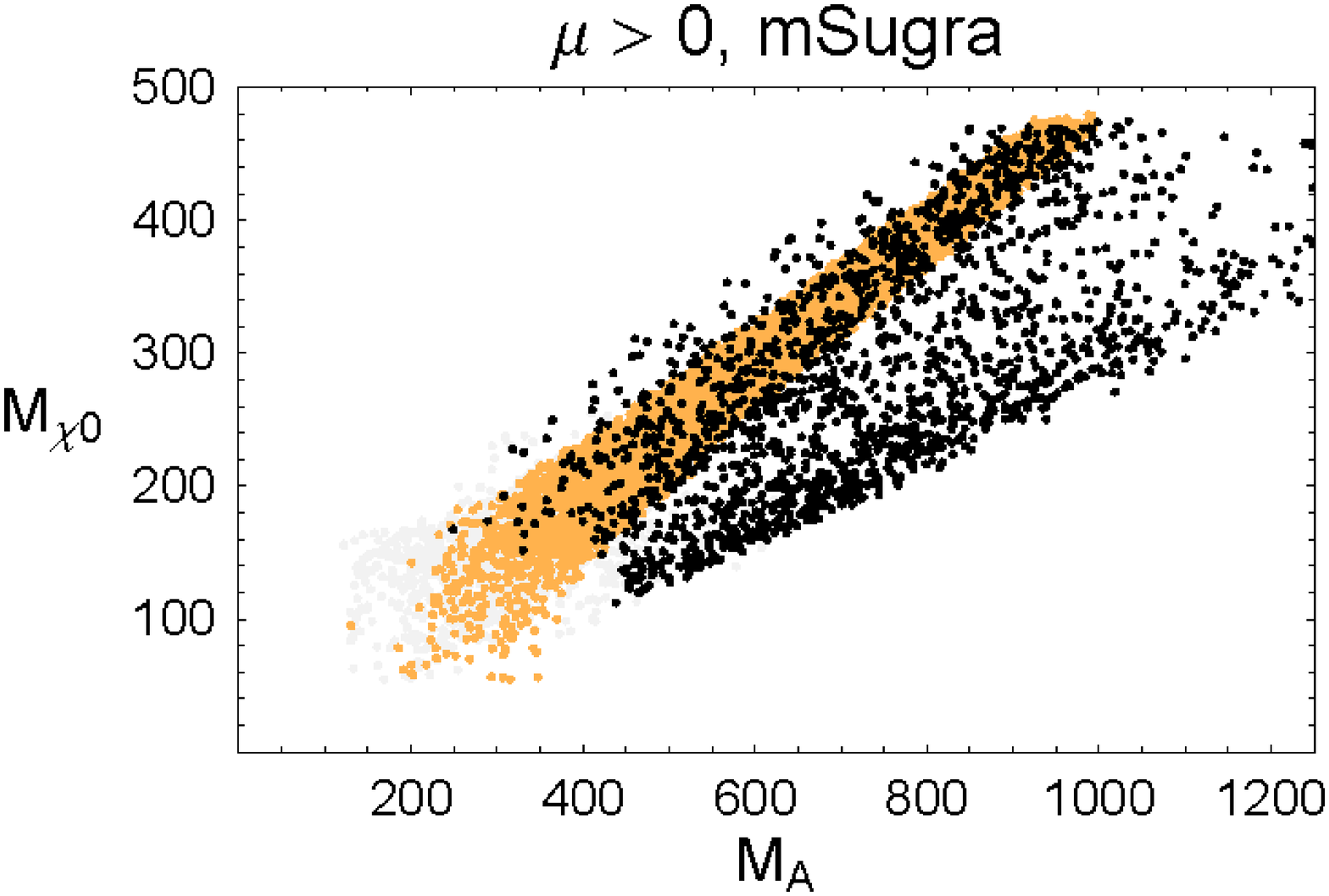}
\cr
\includegraphics[width=8cm]{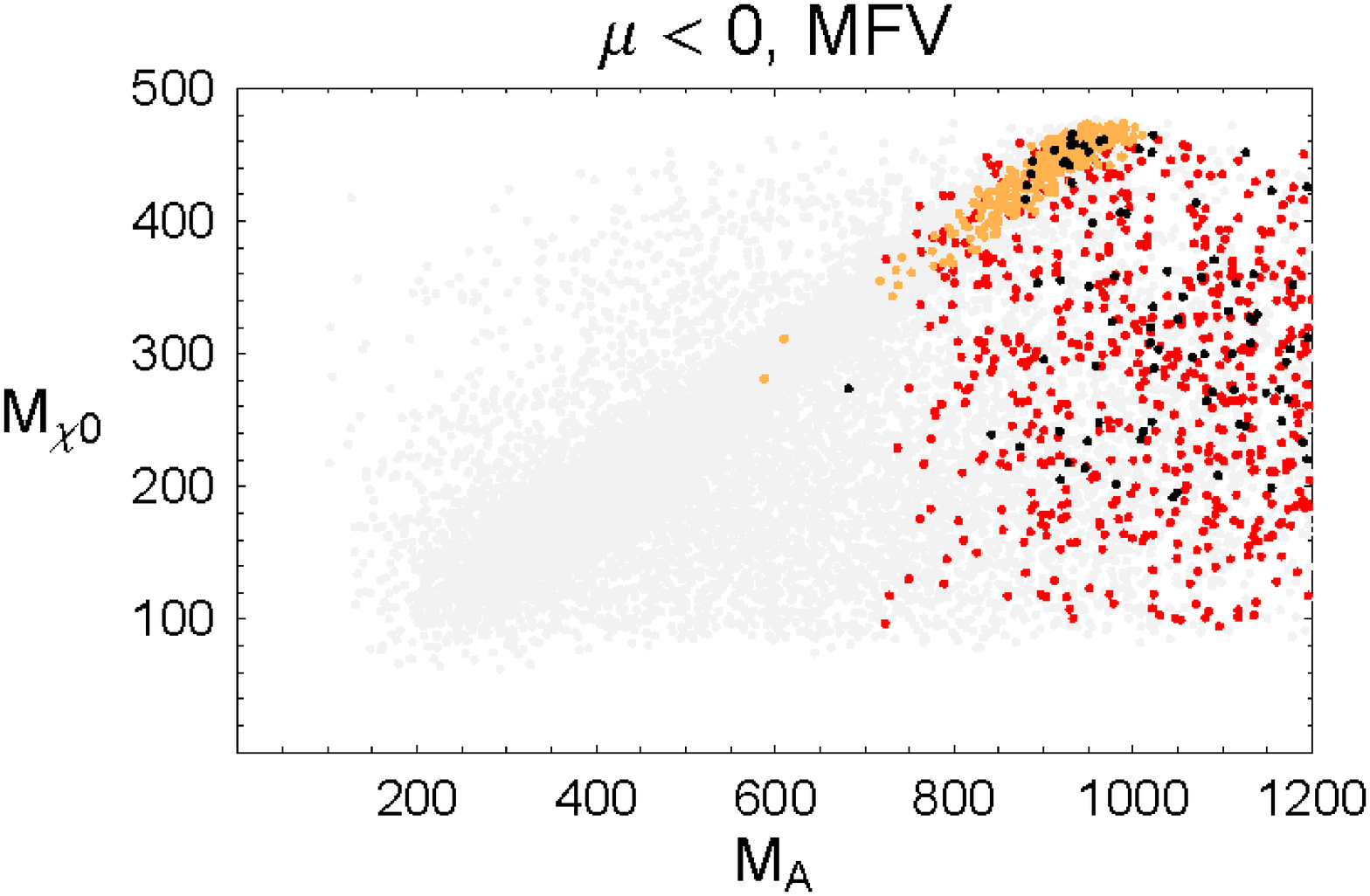}&\includegraphics[width=8cm]{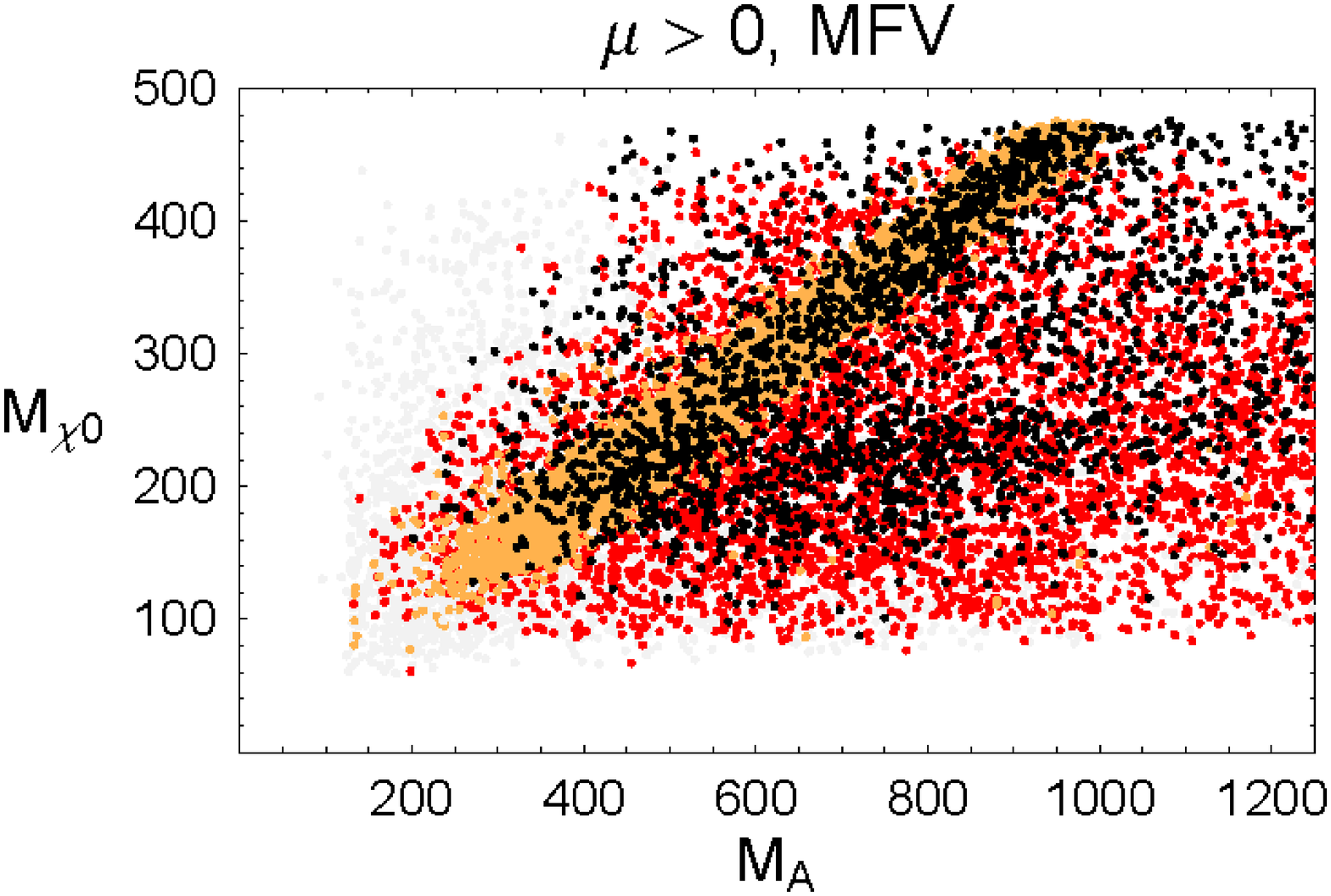}
\cr
\end{tabular} \caption{Correlation between $M_A$ and $M_{{\tilde \chi}^0}$. Blue
and orange points corresponds to stau co-annihilation and funnel,
respectively. In the MFV case, the dark matter constraint can be also
fulfilled by a neutralino with a sizeable higgsino component (red points). }
\label{fig:mamn}
\end{figure}
Given that in this paper we are mainly interested in the large $\tan
\beta$ region, it is clear that we can expect the funnel region to
play a special role in our analysis. In Figs.~\ref{fig:mamn} and
\ref{fig:mstaumn} we plot the allowed points in the
$m_A$--$m_{{\tilde \chi}^0}$ and $m_{\tilde \tau}$--$m_{{\tilde \chi}^0}$
planes. Here we select points in the stau coannihilation region (black
points) for which the difference between $m_{\tilde \tau}$ and
$m_{{\tilde \chi}^0}$ is less than $10\%$. Similarly we select the funnel
points when $\left(M_A - 2 m_{{\tilde \chi}^0}\right)/\Gamma_A \leq 6$. Taking
into account that $\Gamma_A \propto M_A \tan ^2\beta \left(3 m_b^2 +
m_\tau^2 \right)$, we have that at large $\tan \beta$ the difference
between neutralino and pseudoscalar masses can be larger. In
Fig.~\ref{fig:mamn}, we see that the funnel region is wider at low
pseudoscalar masses because low pseudoscalar
masses correspond to large $\tan \beta$. In the same way,
Fig.~\ref{fig:mstaumn} shows clearly all the coannihilation points
concentrated around the $M_A \simeq 2~ m_{{\tilde \chi}^0}$ line. It is
interesting to notice that we have points where both the funnel and
coannihilation mechanisms are active (although we plot them as
coannihilation points, i.e. black points). In fact, in these points
the cross sections for ${\tilde \chi}^0 \tilde \tau_1^+ \to \gamma \tau^+$
and/or $\tilde \tau_1^+ \tilde \tau_1^- \to f \bar f$ constitute more
than 10\% of the total annihilation of SUSY particles. This means that
these points correspond to the
coannihilation region. Moreover the ratio of ${\tilde \chi}^0 {\tilde \chi}^0
\to b \bar b$ and ${\tilde \chi}^0 {\tilde \chi}^0 \to \tau^+ \tau^-$
channels is proportional to $\left(m_b/m_\tau\right)^2$ which means they
are mediated by a Higgs particle and hence they belong also to the funnel
region.

In the CMSSM we scan values of $m_0$ and $M_{1/2}$ up to 1 TeV and thus we
do not enter the focus point region. In any case, points
belonging to the focus point region have very heavy sfermion and pseudoscalar
Higgs masses and therefore they have no effect on the processes we are
analysing here.
The bulk region is already very constrained in the CMSSM
scenario due to the bounds on the masses of the lighter chargino and the
light scalar Higgs boson and pure bulk annihilation is nearly excluded.
 \begin{figure}
 \begin{tabular}{lr}
 \includegraphics[width=8cm]{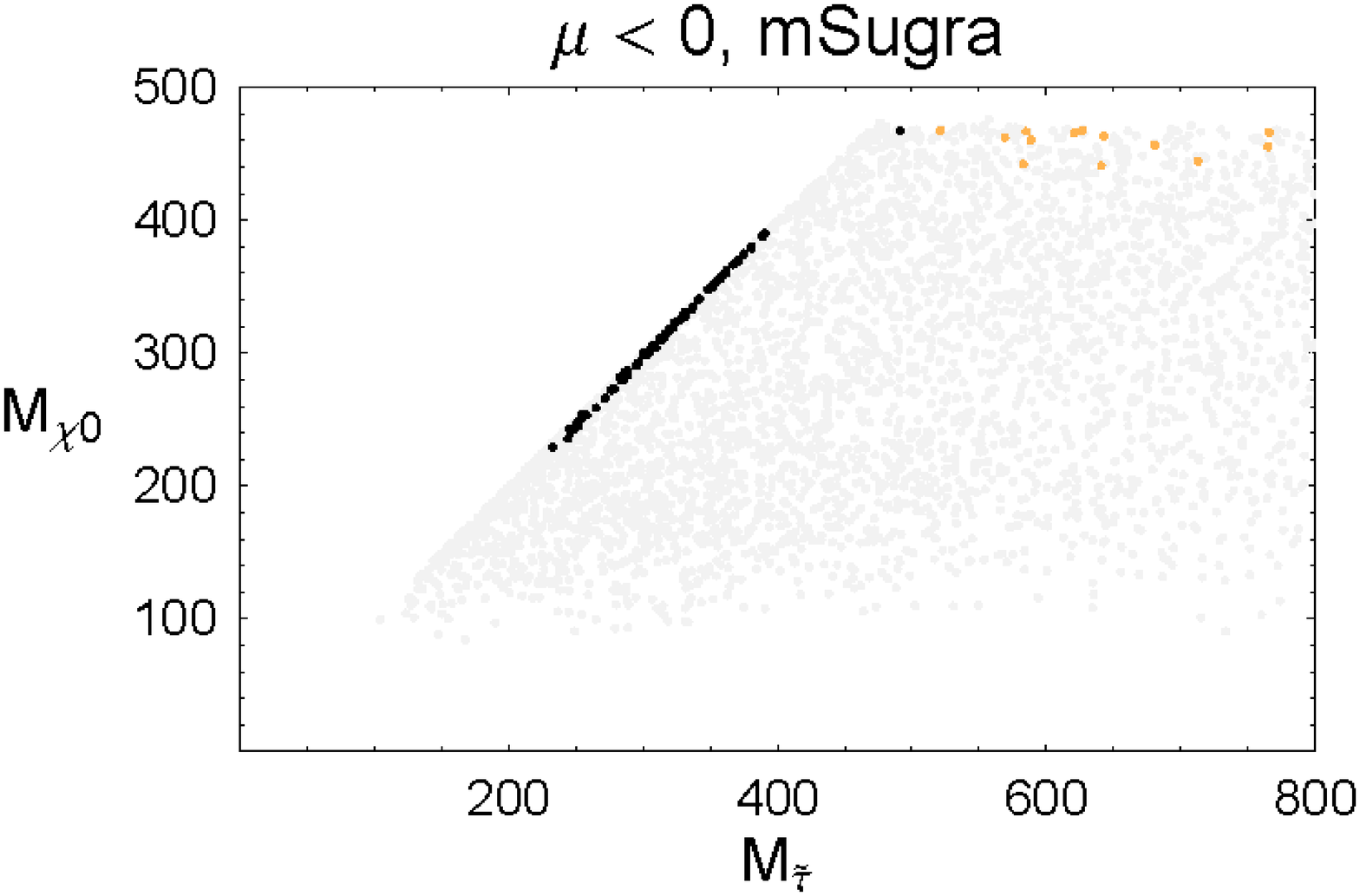}&\includegraphics[width=8cm]{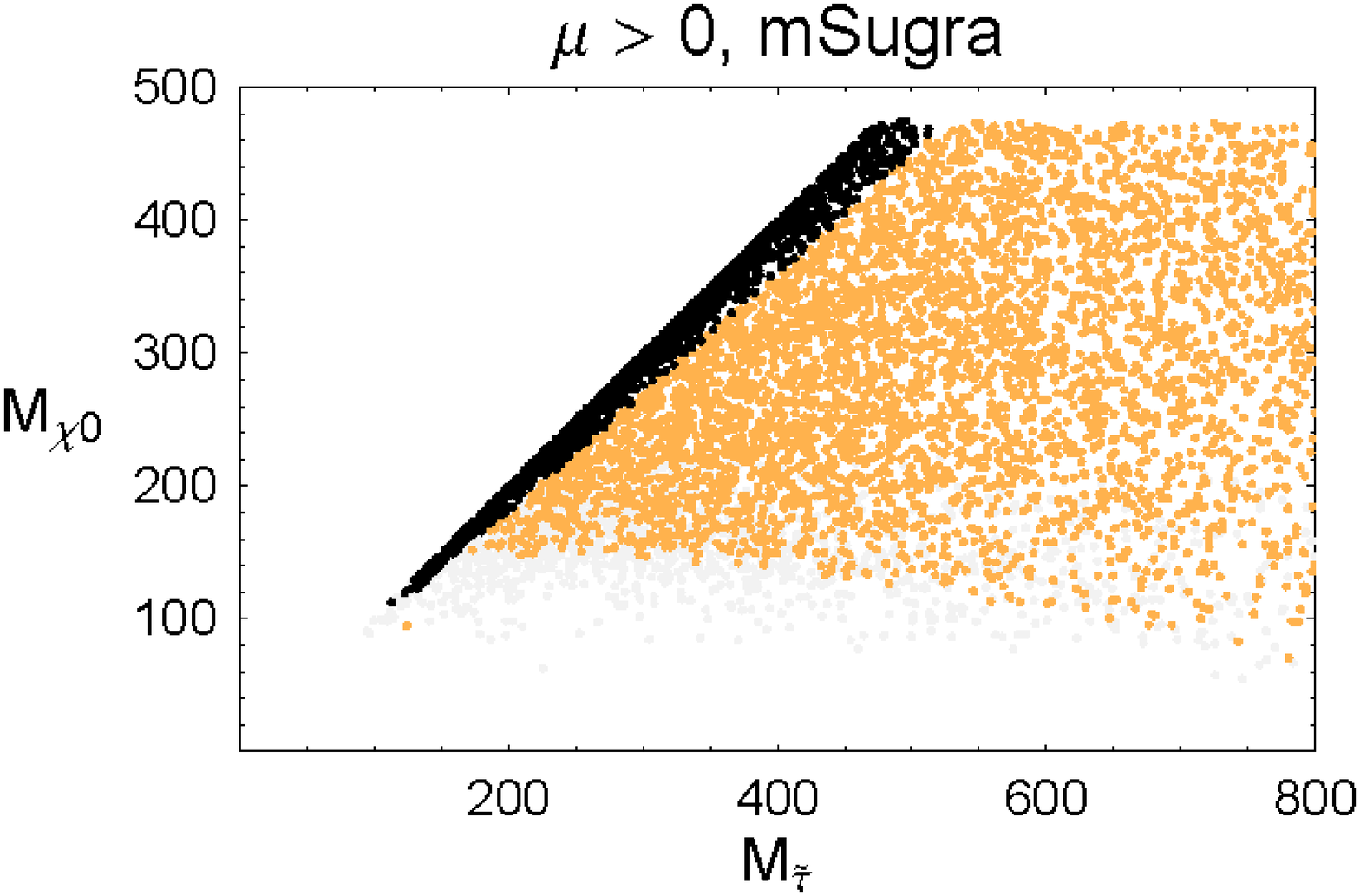}
 \cr
 \includegraphics[width=8cm]{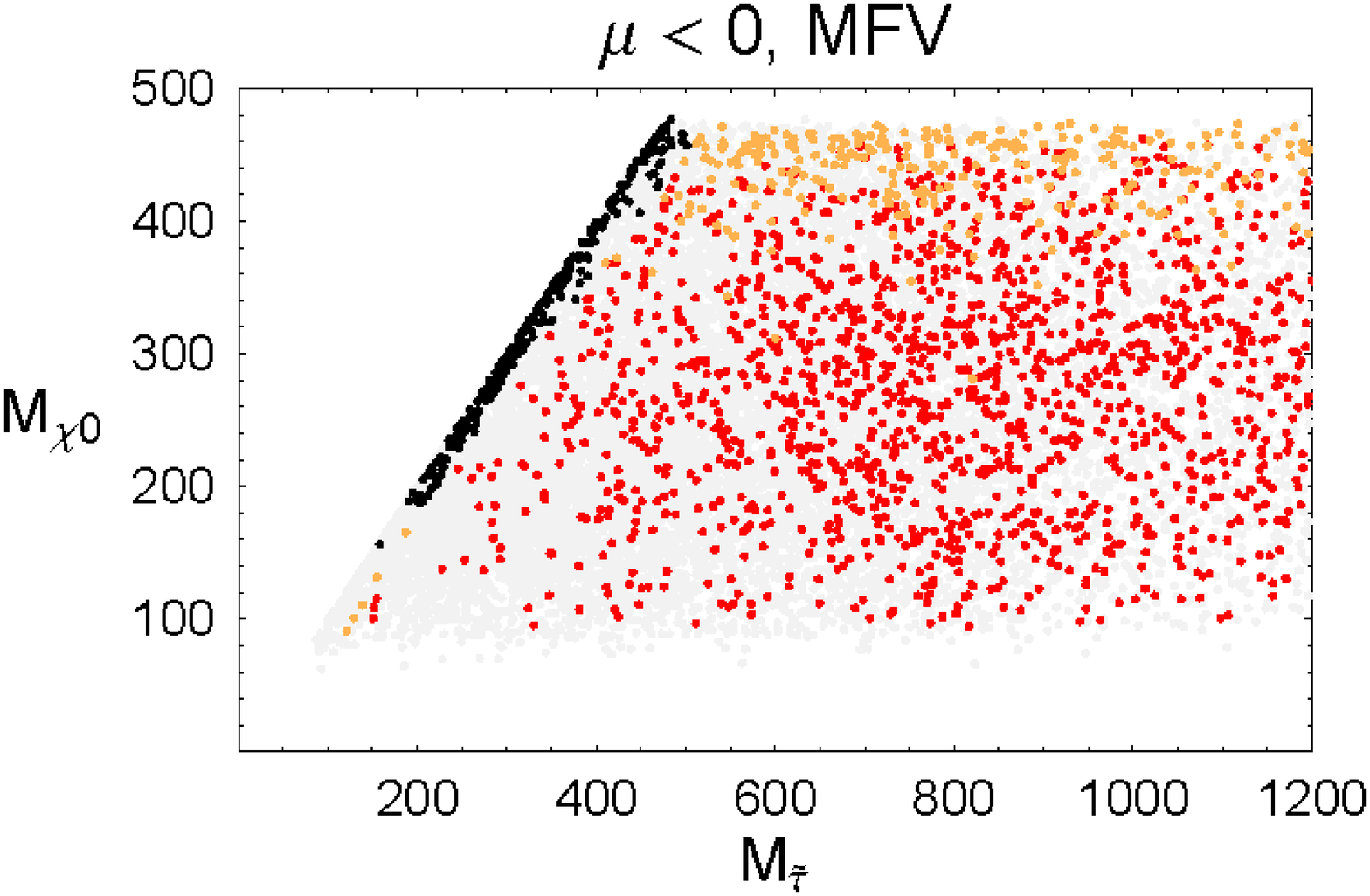}&\includegraphics[width=8cm]{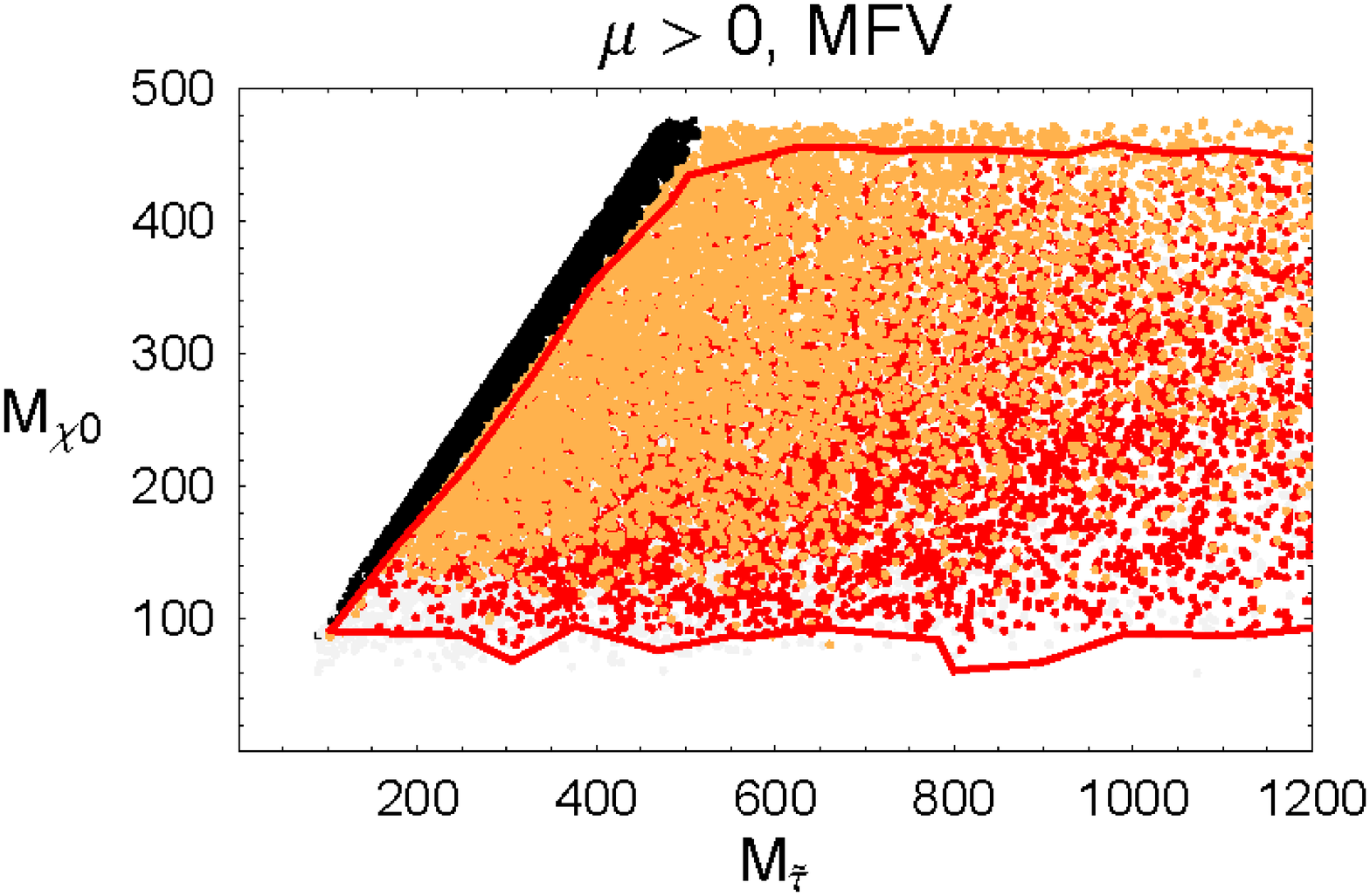}
 \cr
 \end{tabular} \caption{Correlation between $M_{\tilde \tau}$ and $M_{\tilde \chi^0}$. See the caption in Fig.~\ref{fig:mamn}.}
 \label{fig:mstaumn}
 \end{figure}

\begin{figure}
\begin{tabular}{lr}
\includegraphics[width=8cm]{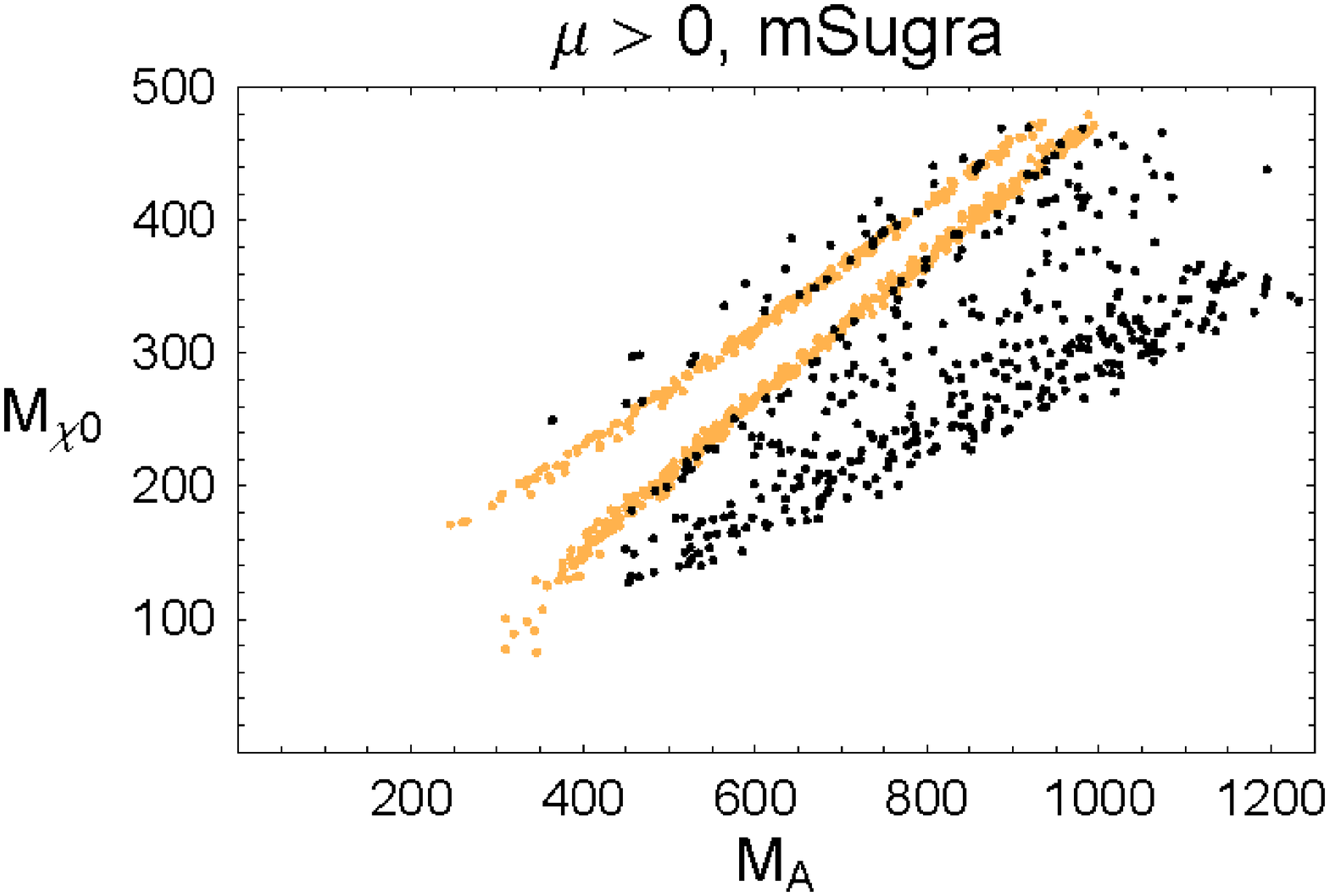}
&
\includegraphics[width=8cm]{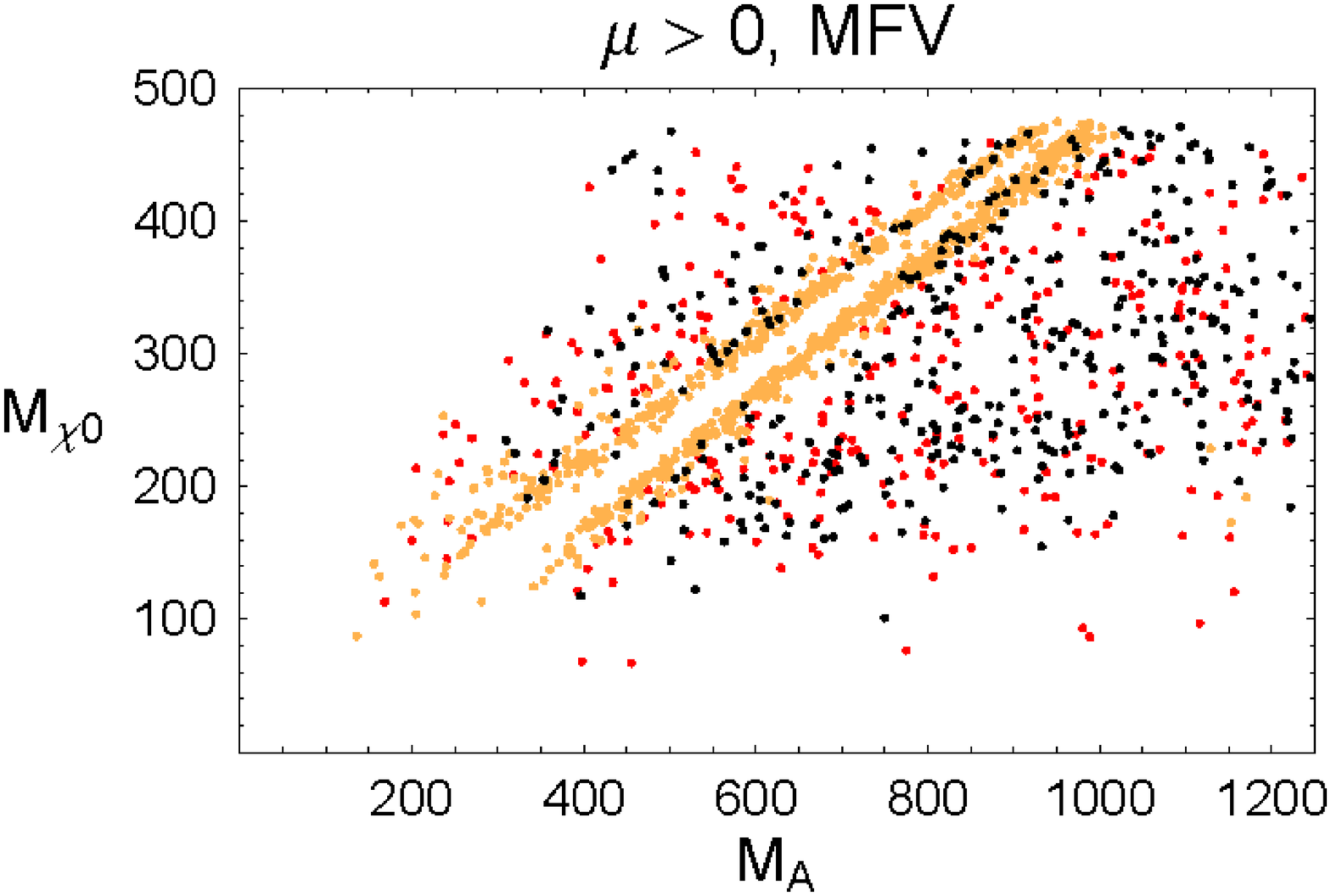}
\cr
\end{tabular} \caption{Impact of the one sigma dark matter bound on
the correlation between $M_A$ and $M_{\tilde \chi^0}$.}
\label{fig:mamnstrictdm}
\end{figure}

In Figure \ref{fig:mamnstrictdm} we present for completeness the correlation
between $M_A$ and $M_{\tilde \chi^0}$ imposing the full one-sigma
bound on the dark matter abundance from \eq{wmap}. In this case the number of
allowed points is reduced but the main features of our analysis remain valid.

 \subsection{$B_s \to \mu^+ \mu^-$ in the general MFV MSSM}
 In this section we analyze the effects of flavour changing
 Higgs couplings in a general MFV MSSM defined at the GUT scale.
 In this model we assign different soft masses and trilinear
 couplings to the different representations under the SM gauge group
 and thus we have 13 independent
 parameters. The main difference with the CMSSM analyzed in the
 previous section concerns precisely the Higgs soft masses that are not
 related with the squark or slepton masses. In this way we can obtain
 lighter Higgs masses while at the same time the sfermion masses are
 heavy enough to satisfy the stringent FCNC constraints.

 \begin{figure}
 \includegraphics[width=8cm]{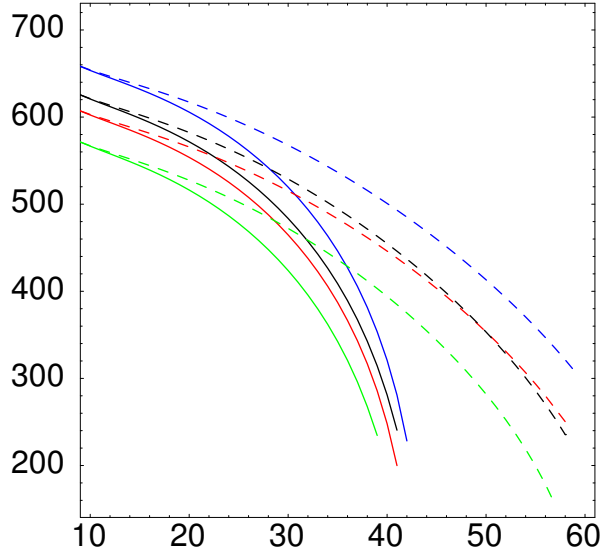}
 \caption{Values of $m_A$ as a function of $\tan \beta$ in the CMSSM.
 In this plot all sfermion masses are equal to 500 GeV and $m_{1/2}=300$ GeV.
 The red lines correspond to $m_{H_1}=m_{H_2}=150$ GeV, green lines
 $m_{H_1}=150$ and $m_{H_2}=300$ GeV, blue lines $m_{H_1}=300$ GeV and
 $m_{H_2}=150$ GeV and black lines to $m_{H_1}=m_{H_2}=300$ GeV.
 $A_0=0$ GeV for all the lines. Full lines
 correspond to $\mu<0$ and dashed lines to $\mu>0$.}
 \label{fig:mAtgbMFV}
 \end{figure}
 Again we distinguish the cases of $\mu>0$ and $\mu<0$. Analogously to the
 CMSSM case, $\mu>0$ implies that both $\epsilon_0$ and $\bar \epsilon_j$ are
 negative. Therefore Yukawa couplings are reduced and their effects in FCNC and
 the RGE evolution is smaller.

 The values that can be obtained obtain for $m_A$ and $\tan \beta$ in this
 general MFV model are similar to the values obtained in the
 CMSSM case, as can be seen in Fig.~\ref{fig:mAtgbMFV}.
 Therefore, in principle, similar vales for the BR($B_s\to \mu^+ \mu^-$) are
 possible in both models. However,
 as we will show here, after taking into account the FCNC constraints,
 combinations of ($m_A$, $\tan \beta$) that would be forbidden in the CMSSM are
 now allowed in the MFV model. Thus it is easier to obtain larger values for
 BR($B_s\to \mu^+ \mu^-$) consistently with the FCNC constraints.
 This can be seen in Figs.~\ref{fig:mabsmm} and~\ref{fig:BRtgb} where for
 positive $\mu$ we obtain similar values for  BR$(B_s\to \mu^+\mu^-)$ as
 in the CMSSM case although the dependence of  BR$(B_s\to \mu^+\mu^-)$ on
 $M_A$ and $\tan \beta$ is less defined than in the CMSSM case. In these
 figures we find larger values for the branching ratios with respect to the
 CMSSM case with smaller values of $M_A$ or $\tan \beta$.

 \begin{figure}
 \begin{tabular}{lr}
 \includegraphics[width=8cm]{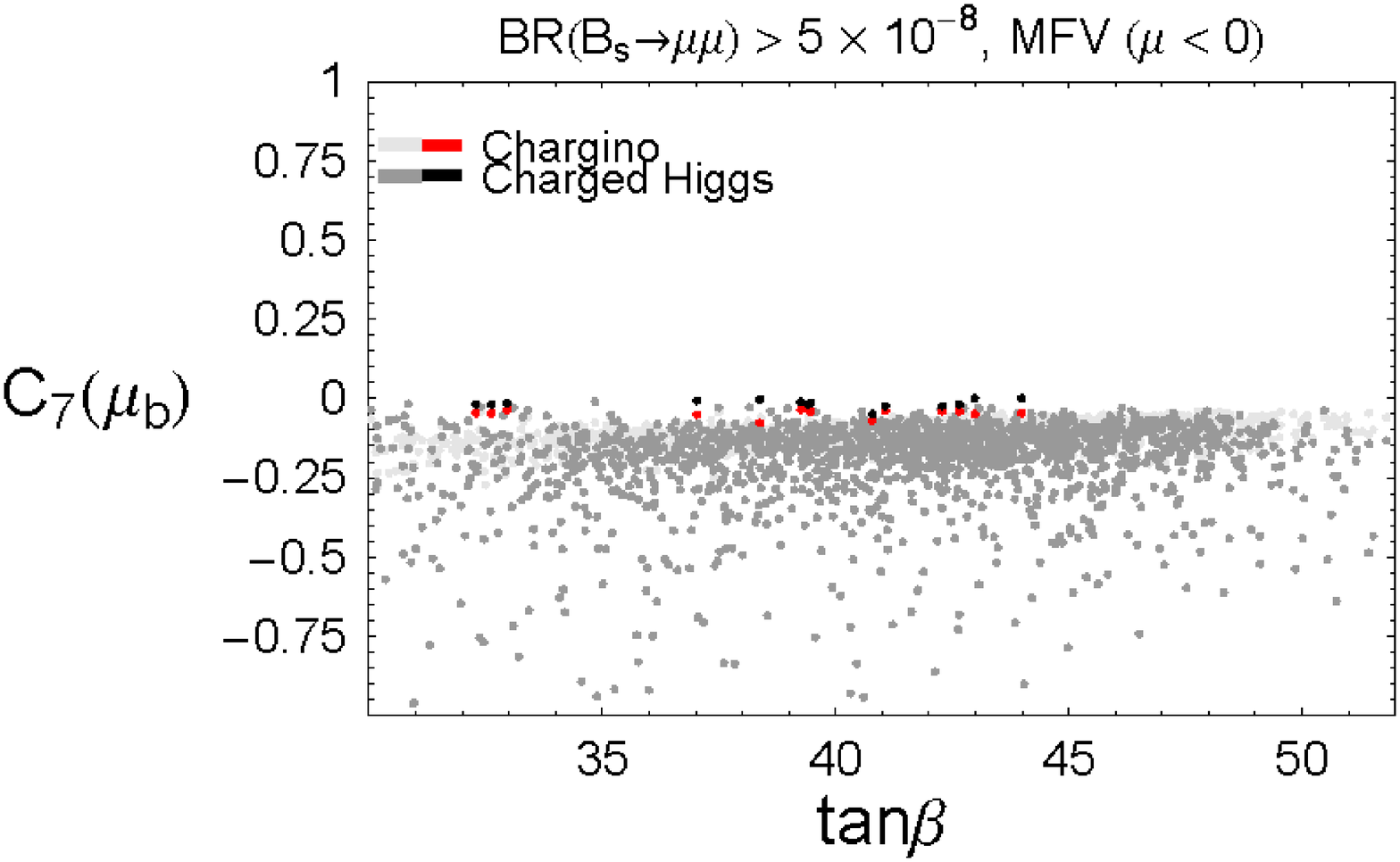}&\includegraphics[width=8cm]{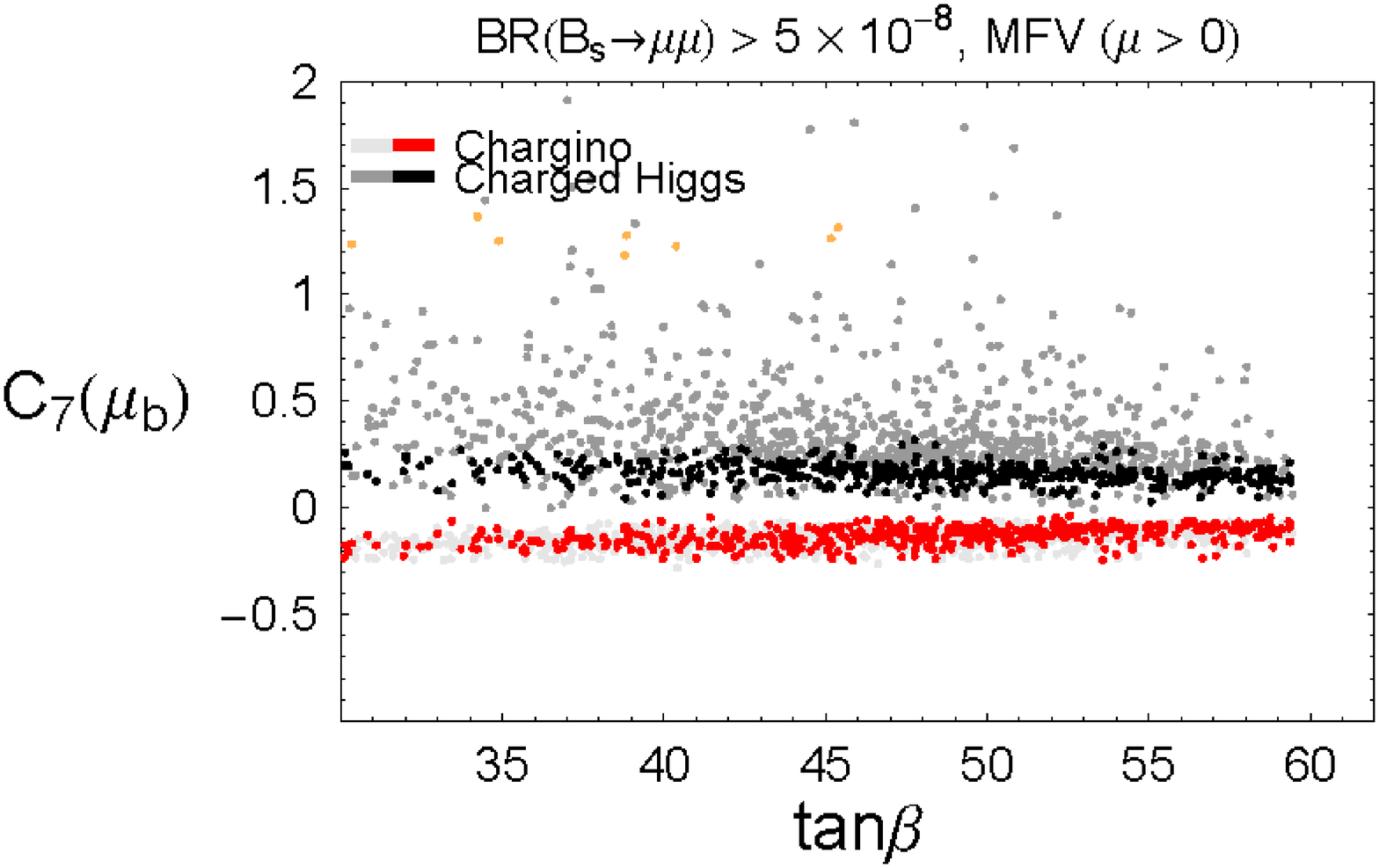}
 \cr
 \end{tabular} \caption{Charged Higgs and chargino contributions to the $C_7$
 Wilson coefficient in MFV as function of $\tan \beta$. The lower
 plots correspond to a large $B_s\to \mu\mu$ scenario. Red points
 refer to chargino and black points to charged Higgs contributions.
 Orange points are points where $C_7^{\rm tot}$ changes sign although they
 are not allowed by the  BR($b \to s l^+ l^-$) constraint.
 Grey points do not survive the FCNC constraints.} \label{fig:C7MFV}
 \end{figure}

 How the $b \to s \gamma$ and $a_\mu$ constraints are
 satisfied in this case?. The
 main difference between the CMSSM and the MFV model is the fact that now
 different scalar masses are independent and therefore it is possible to
 have light Higgs masses with heavy sfermions. Apart from this difference,
 the case of positive $\mu$ is qualitatively similar to the CMSSM.
 In Fig.~\ref{fig:C7MFV} we present the values of the Wilson coefficients
 $C_7^H$ and $C_7^{\tilde \chi}$ for $\mu>0$ and $\mu<0$. For $\mu>0$ the situation
 is very similar to the situation we found in the CMSSM. The only difference
 is that we can find points with  BR$(B_s \to \mu^+ \mu^-) > 5 \times 10^{-8}$
 for comparatively smaller values of $\tan \beta$ and that we do find
 several points with $C_7^{\tilde \chi} >1$ that change the total sign of
 $C_7$, although they are always forbidden by the $B \to X_s l^+ l^-$
 constraint as expected \cite{misiak}.

 Regarding the $B_s -\bar B_s$ mass difference, we see in
 Fig.~\ref{fig:DMBS} that for $\mu>0$, the MFV case is still very similar to
 the CMSSM one. However we can see that now it is possible to
 change the sign of Re$(M_{12}^s)$ if we disregard the $(B_s \to \mu^+\mu^-)$
 constraint. Once we impose the new CDF and D\O~constraint on
 BR$(B_s \to \mu^+\mu^-)$, no sizeable change in $\Delta M_{B_s}$
 is possible and taking into account the theoretical errors as done in
 Fig.~\ref{fig:DMBS2} the allowed points always agree with the recent
experimental result from the CDF collaboration.

 In all these plots, the main difference between the CMSSM and the MFV
 cases can be seen for $\mu<0$. Once more $\mu<0$ implies that both
 $\epsilon_0$ and $\epsilon_j$ are negative. From
 Figs.~\ref{fig:mabsmm} and~\ref{fig:BRtgb} we can see that, as opposed
 to the CMSSM case, we can reach BR$(B_s\to \mu^+\mu_-) \simeq 10^{-6}$
 with $\mu<0$ consistently with all the constraints and for values of
 $\tan \beta$ as low as 40.  This is due to the negative sign of the
 parameters $\epsilon_0$ and $\tilde \epsilon_3$. In fact for $\mu>0$
 the factor $ 1/\left({ (1+\epsilon_0 \tan \beta)(1+\tilde \epsilon_3
 \tan \beta)}\right)$ can amount to a suppression by a factor 10 for
 $\tan \beta =50$ and $\epsilon_0 = 0.012$ while it represents an
 enhancement factor of 20 for $\tan \beta =40$ and $\epsilon_0 = -
 0.012$. Therefore, for the same values of $M_A$ and $\tan \beta$ we
 can get an enhancement of two orders of magnitude when changing from
 $\mu>0$ to $\mu<0$.

  The main question is how are the different indirect constraints
 satisfied for these points.  From Fig.~\ref{fig:C7MFV}, we see that,
 in this case, both $C_7^H$ and $C_7^{\tilde \chi}$ are negative as expected and
 both of them very small as required by the $b \to s \gamma$ constraint
 summing up to, at most, $0.05$. This is
 possible only with a rather heavy spectrum although we want to keep the Higgses
 as light as possible in order to have a large BR$(B_s \to \mu^+ \mu^-)$.
 Similarly, even taking a conservative $a_{\mu}$ bound at
 $3 \sigma$, only a rather heavy SUSY spectrum is allowed. However, in
 this model the Higgs masses are independent from the rest of the SUSY
 spectrum and hence we can still find sizable Higgs contributions to
 the decay $(B_s \to \mu^+ \mu^-)$. In any case it is clear from the
 density of point with large branching ratio that all these points
 require a certain degree of fine-tuning to satisfy these requirements.

 Finally we discuss  the dark matter constraints. From Figs. \ref{fig:mamn} and
\ref{fig:mstaumn} we can see the regions of parameter space allowed by the
dark matter constraints. The  main difference between the CMSSM case
and the general MFV scenario is the appearance of a new set of allowed points
that correspond to neutralinos with a sizeable higgsino component. This has
also been noted previously in models where the Higgs-mass parameter has been
decoupled from the sfermion parameters
\cite{Dermisek:2003vn,Baek:2004et,Ellis2006}.  Now the
coupling of the Z-boson to an LSP pair is enhanced and the neutralinos
annihilate through a Z-boson in the s channel. This situation is similar to
the focus point region in the CMSSM. However, in contrast to the focus point
situation, the sfermions and Higgs masses can be still of the order of a few
hundred GeVs. This leads to the observed smear-out of the different regions
in the
MFV scenario.

\section{Large BR$(B_s \to \mu^+ \mu^-)$ collider phenomenology}
\label{sec:collider}

In the first part of this section we assume that a branching ratio
for the decay $B_s \to \mu^+ \mu^-$ has been measured
between $8 \times 10^{-8}$ and $10^{-7}$ and explore the consequences for the
phenomenology of high energy colliders~\cite{Dedes:2004yc}. In the
second part we will briefly comment on the occurrence of flavour
violating SUSY decays.

\begin{figure}
\begin{tabular}{lr}
\includegraphics[width=8cm]{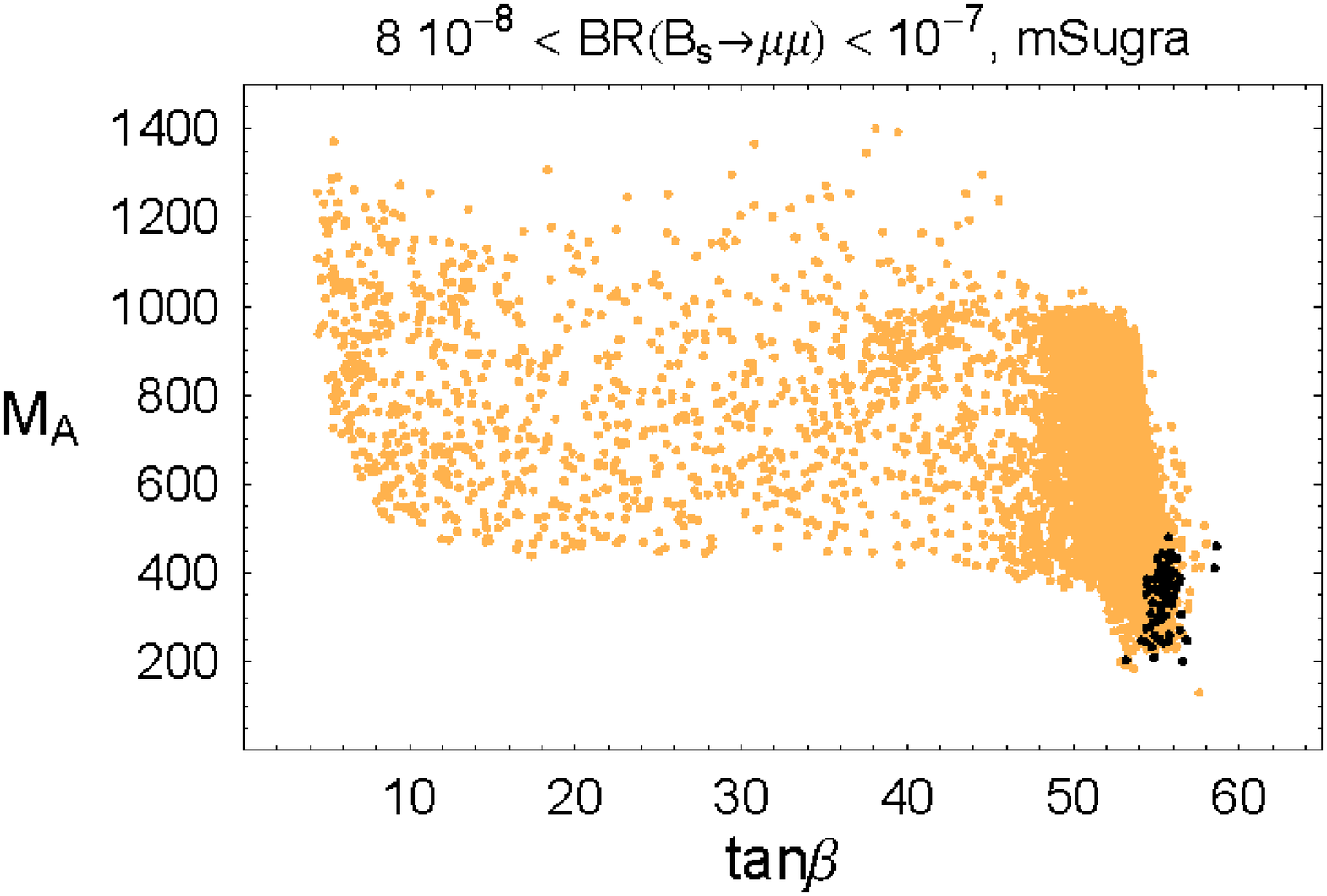}
&
\includegraphics[width=8cm]{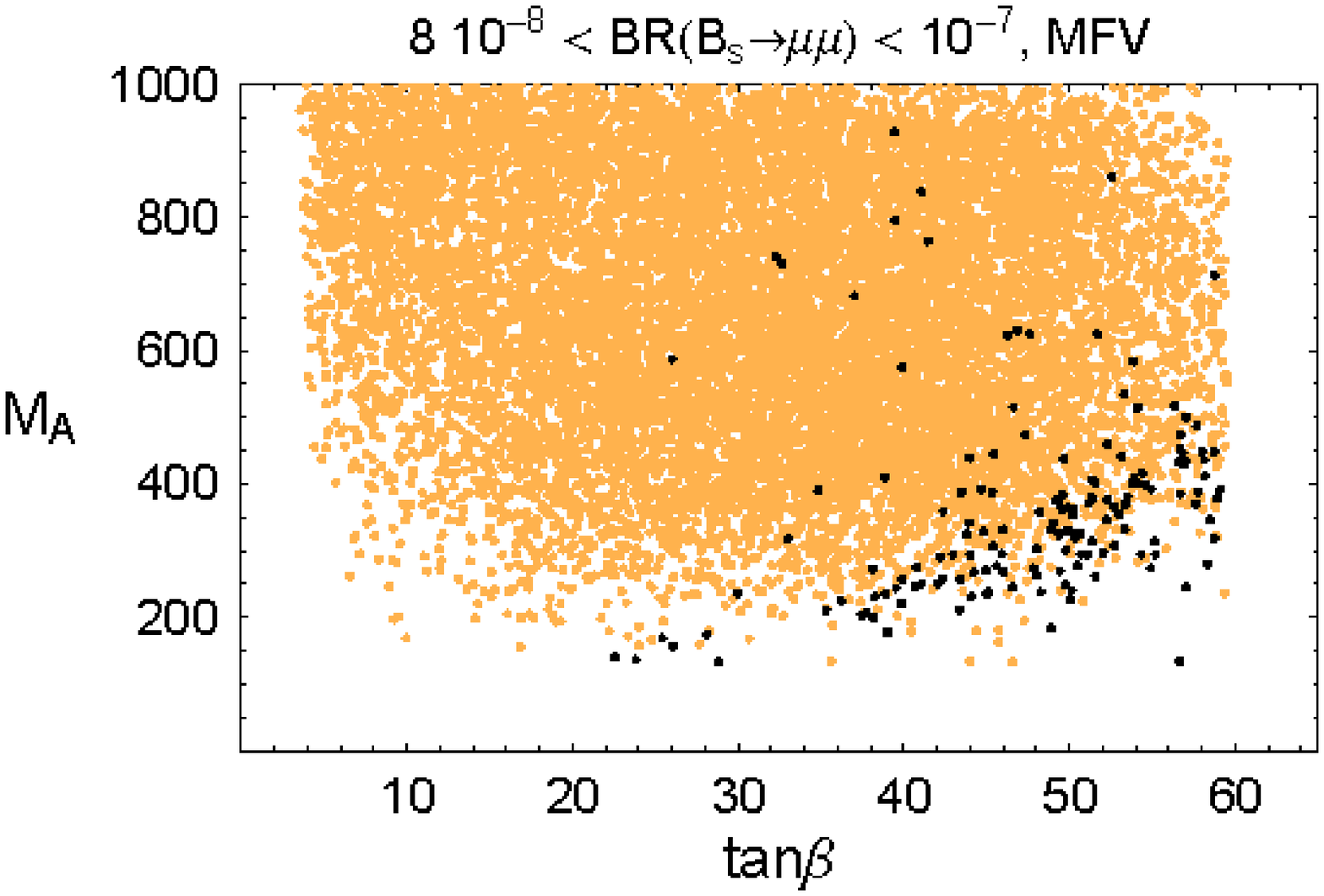}
\cr
\end{tabular} \caption{Correlation between $\tan \beta$ and $M_A$ in a large
$B_s\to \mu\mu$ scenario. Orange points have ${\rm BR}
(B_s\to \mu\mu) < 8 \times 10^{-8}$; blue points have $8 \times
 10^{-8} < {\rm BR} (B_s\to \mu\mu) < 10^{-7}$.} \label{fig:tbmalarge}
\end{figure}

\begin{figure}
\begin{tabular}{lr}
\includegraphics[width=8cm]{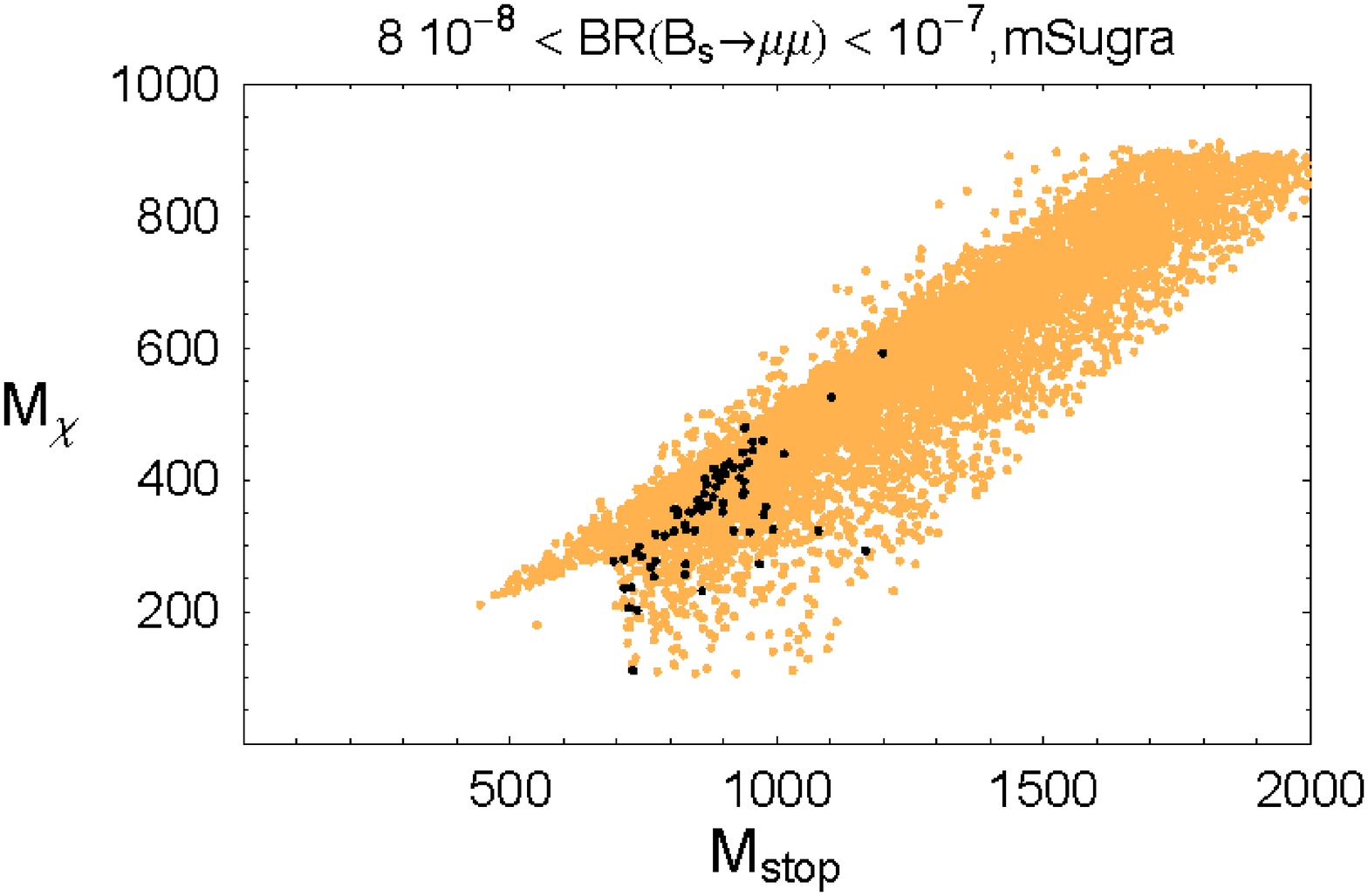}
&
\includegraphics[width=8cm]{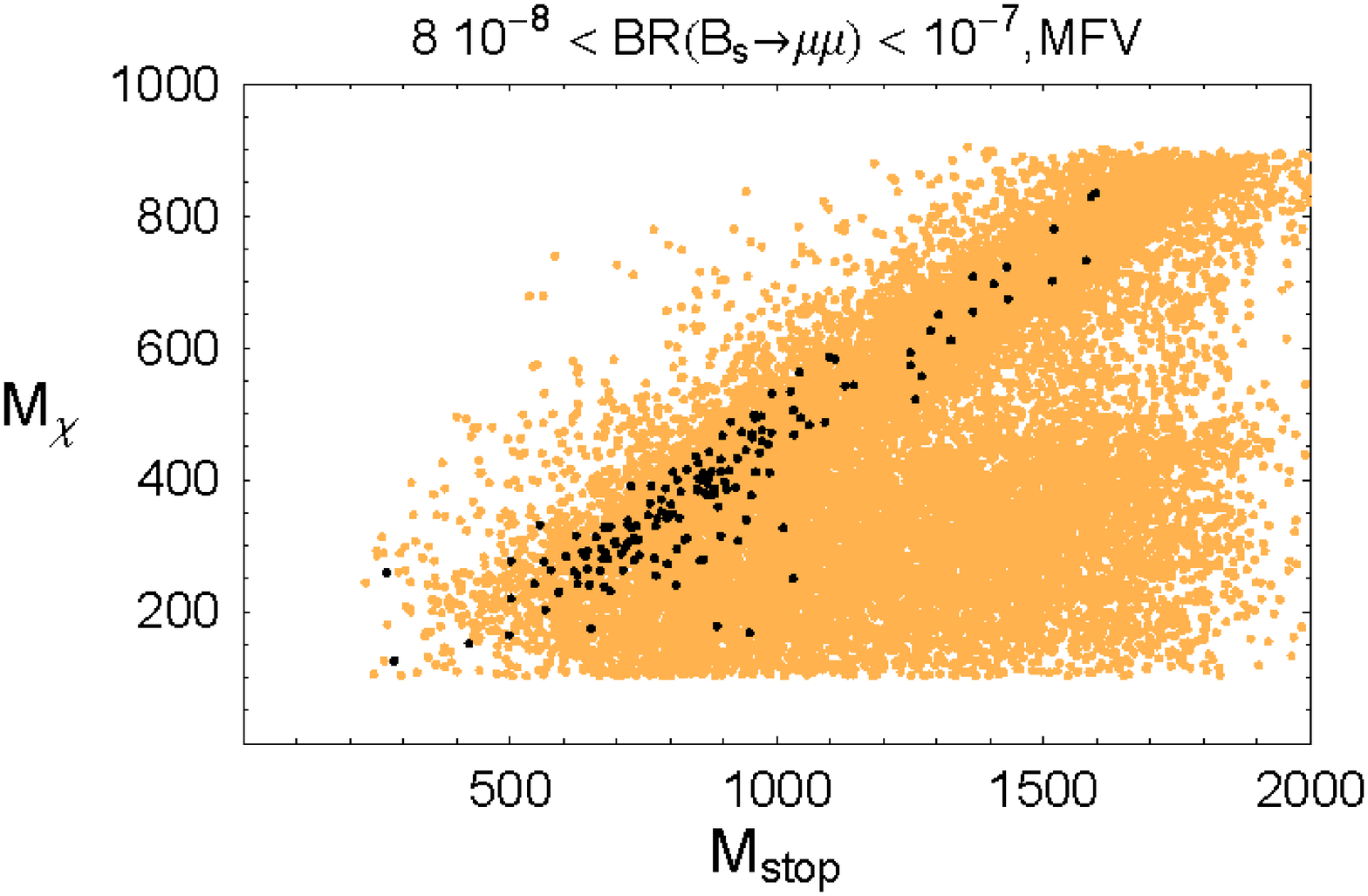}
\cr
\end{tabular} \caption{Correlation between $M_{\tilde t}$ and
$M_{\tilde \chi^\pm}$ in a large
$B_s\to \mu\mu$ scenario. Orange points have ${\rm BR}
(B_s\to \mu\mu) < 8 \times 10^{-8}$; blue points have $8 \times
 10^{-8} < {\rm BR} (B_s\to \mu\mu) < 10^{-7}$.} \label{fig:mstopmchlarge}
\end{figure}

We have seen in the previous section, that such a large branching ratio for
$B_s \to \mu^+ \mu^-$ requires a  light pseudoscalar Higgs
boson and  large $\tan\beta$ as it scales as $\tan^6\beta /
m^4_{A^0}$.  In the case of the CMSSM this indeed implies an upper
bound of about 570~GeV and
a low bound on $\tan\beta$ of about 50 as can be seen in the left plot of
Fig.~\ref{fig:tbmalarge} where the black points are those satisfying all
low energy bounds while yielding the large branching ratio. In the right plot of
Fig.~\ref{fig:tbmalarge} we see that these bounds get considerable
weaker in the MFV case. The reason is that the tight correlations
between the various SUSY masses is broken up and it is easier to satisfy the low
energy constraints, in particular $b\to s \gamma$.
Nevertheless we find a lower bound of about 20 for $\tan\beta$ and
an upper one of about 950 GeV for $m_{A^0}$. In the CMSSM upper bounds
on the masses of other particles are obtained, e.g.~the chargino has to
be lighter than 600 GeV as can be seen from Fig.~\ref{fig:mstopmchlarge} where
$m_{\tilde \chi^+_1}$ is shown versus $m_{\tilde t_1}$. In the case of the MFV
scenario the bound on $m_{\tilde \chi^+_1}$ is about 850~GeV.
Note that the upper
bound on $m_{\tilde \chi^+_1}$ implies in CMSSM (MFV) also an upper bound on
$m_{\tilde g}$ of about 1.7 TeV (2.4 TeV).
As a consequence the discovery of
SUSY \cite{Asai:2002xv} is guaranteed in both scenario due to gluino production.
An important question is if the heavier Higgs boson can be detected at LHC.
In the CMSSM it turns out that in the region with large BR($B_s \to \mu^+ \mu^-$)
the heavier neutral Higgs bosons can be detected via their the decays into
the $\tau^+ \tau^-$ final state with 30 fb$^{-1}$,
see \cite{Djouadi:2005gj} and references therein.
In the region with $M_{A0} \lsim 350$~GeV also the $\mu^+ \mu^-$ final state
can be used in case of an integrated luminosity of 60 fb$^{-1}$. Also the
charged Higgs bosons should be visible via its decay into $\tau \nu_\tau$
 \cite{Djouadi:2005gj}. In the case of the general MFV scenarios the regions
with $\tan\beta \lsim 25$ and/or $m_{A0} \gsim 800$~GeV will most likely require
larger statistics for detecting the heavy Higgs bosons. In the MFV scenario
and to a much lesser in extend in the CMSSM scenario
there are some points with $m_{H^+} < m_t - m_b$ implying that  the
branching ratio $t \to b \tau^+ \nu_\tau$ will be modified. Numerically
we find that there are 3 (10) data points (out of 40000 points)
in the region with
BR($B_s \to \mu^+ \mu^- > 8 \cdot 10^{-8}$)
(BR($B_s \to \mu^+ \mu^- < 8 \cdot 10^{-8}$)) with a maximal value of
BR($t \to b \tau^+ \nu_\tau$) of 12 (21) \%.

We have also checked if the running from $M_{\rm GUT}$ down to $M_{EWSB}$
can induce any sizable flavour violating branching ratio in the decays
of SUSY particles. As expected these decay branching
ratios are in general very small, of the order $10^{-5}$ and below. There
are a few points in the MFV scenario where BRs of the order $10^{-3}$ can
be obtained e.g.~for $\tilde b_1 \to \tilde \chi^-_1 c$
due to a kinematical suppression of $\tilde b_1 \to \tilde \chi^-_1 t$ and
$\tilde b_1 \to \tilde \chi^0_2 b$. In these points $\tilde b_1$ is mainly
a $\tilde b_L$
and thus the coupling $\tilde b_1$-$\tilde \chi^0_1$-$b$ is reduced due to
the small hypercharge of $\tilde Q_L$ implying a reduction in the
 decay mode
$\tilde b_1 \to \tilde \chi^0_1 b$.
However, it has
to be expected that the SUSY background of
$\tilde s_L \to \tilde \chi^-_1 c$ will be too large to observe this mode
at LHC.

\begin{figure}
\begin{center}
\includegraphics[width=8cm]{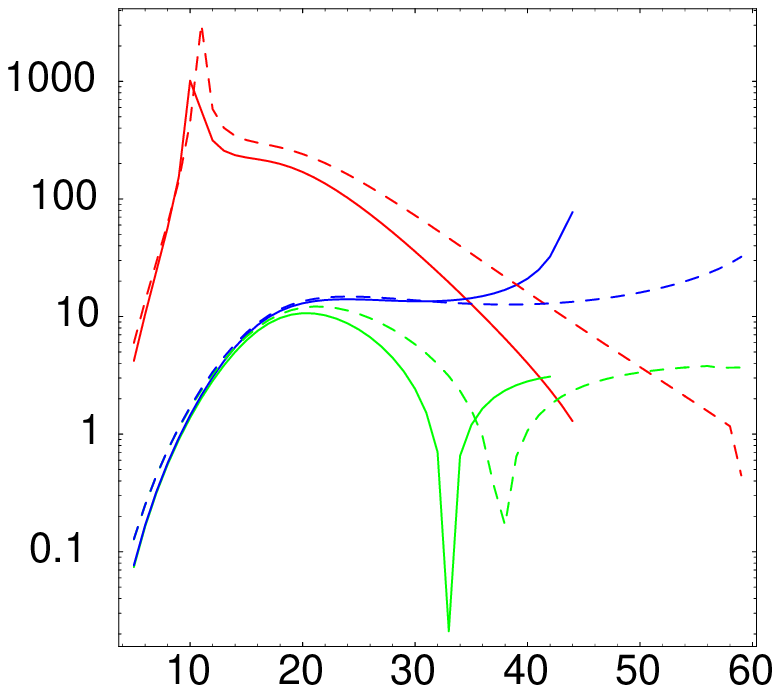}
\includegraphics[width=8cm]{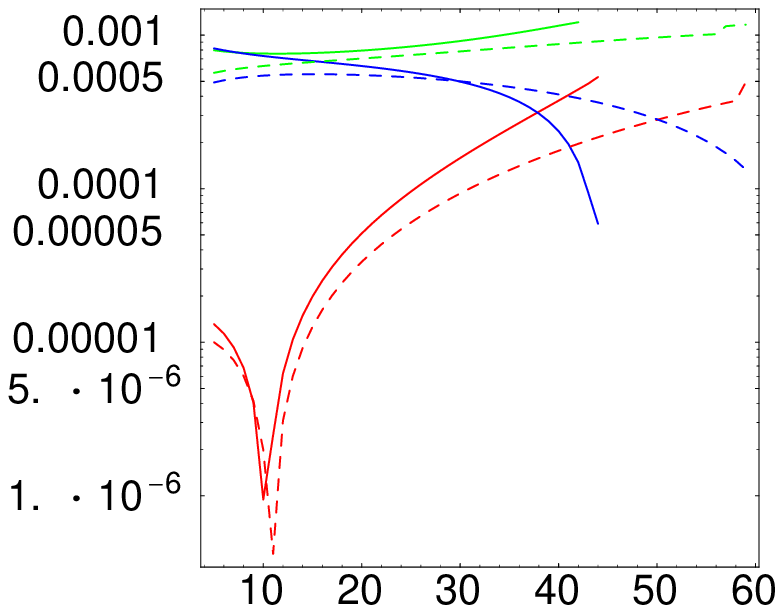}
\end{center}
\caption{Upper plot: approximate $\tilde t_1$-$c$-$\tilde \chi^0_1$ coupling over the
exact one as a function of $\tan\beta$ in the CMSSM. Lower plot: exact
$\tilde t_1$-$c$-$\tilde \chi^0_1$ couplings a function of
$\tan\beta$ in the CMSSM.  The SUSY parameters are fixed as
$m_{1/2}=300$ GeV, $m_0=500$ GeV and the blue lines correspond to
$A_0=500$ GeV blue, green lines to $A_0=0$ GeV and red lines to $A_0=-
500$ GeV. Full lines correspond to $\mu<0$ and dashed lines to
$\mu>0$.}
\label{fig:cplchitgb}
\end{figure}

In the literature (for an incomplete list see
\cite{Baer:1994xr,Kon:1994uc,Sender:1996qc,Porod:1996at,Hosch:1997vf%
,Porod:1998yp,Boehm:1999tr,Restrepo:2001me,Djouadi:2001dx,Das:2001kd%
,Porod:2003um,Das:2003pe,Muhlleitner:2003vg,Han:2003qe})
often an approximate solution of the
$\tilde t_1$-$c$-$\tilde \chi^0_1$ coupling is used which results from
a one-loop integration of the flavour violating RGEs for $M^2_{Q,23}$ and
$A_{u,23}$ \cite{Hikasa:1987db} resulting in a mixing between
$\tilde c_L$ and $\tilde t_1$. The projection $\epsilon$
of $\tilde c_L$ onto $\tilde t_1$
can be written as:
\begin{eqnarray}
\epsilon &=& \frac{\Delta_L \cos \theta_t + \Delta_R \sin \theta_t}
                {m^2_{{\tilde c}_L} - m^2_{{\tilde t}_1}} \\
\Delta_L & = & - \frac{g^2}{16 \pi^2} \ln \left(\frac{M_X^2}{m_W^2} \right)
V^{\ast}_{tb} K_{cb} Y_b^2
( M_{\tilde Q,3}^2 + M_{\tilde d_R,3}^2 + M_{H_1}^{2} + A_b^2 )
\label{eq:deltal} \\[0.2cm]
\Delta_R & = & \frac{g^2}{16 \pi^2} \ln \left(\frac{M_X^2}{m_W^2}
\right)
V^{\ast}_{tb} K_{cb} Y_b^2  m_t A_b
\label{eq:deltar}
\end{eqnarray}
resulting in the $\tilde t_1$-$c$-$\tilde \chi^0_1$ interaction
\begin{eqnarray}
{\cal L} = - \sqrt{2} \left( \frac{g' N_{11}}{6} + \frac{g N_{12}}{2} \right)
 \epsilon \bar{c} P_R \tilde \chi_1 \tilde t_1  + h.c.
\end{eqnarray}
Here $\theta_t$ is the mixing angle between left- and right stops.
All the quantities should be taken at the $M_{\rm GUT}$ as this is an
one-step integration of the RGES. Due to the largeness of $m_t$ it
can happen that all two-body decays of $\tilde t_1$ are kinematically suppressed
or even
forbidden at tree-level and that the main decay mode is
$\tilde t_1 \to \tilde \chi^0_1 c$. In the MFV scenario we find indeed
a couple of points where the tree-level decays are suppressed. These are
the points shown in the lower left corner of Fig.~\ref{fig:mstopmchlarge}.
Depending on the coupling strength it might be possible that $\tilde
t_1$ hadronizes before decaying and therefore it is important to check
the quality of this approximation. In Fig.~\ref{fig:cplchitgb} we show
the ratio of the approximate coupling over the exact calculation using
numerical solutions of the RGEs. One sees that for non-zero $A_0$ the
approximation gives roughly the order of magnitude but can be of by a
factor 100 in certain cases. Note, that this corresponds to a factor
$10^4$ in the partial width.  In the case $A_0 = 0$ the difference can
even be larger which is due to the smallness of the exact
coupling. The minimum of the exact coupling around $\tan\beta=10$ is
due to a cancelation between the $\tilde t_L$-$\tilde c_L$ mixing and
the $\tilde t_R$-$\tilde c_L$ which is not obtained within the
approximation because the $\tilde t_R$-$\tilde c_L$ is zero in this
case.
We have used the approximation formula for the calculation of the decay
$\tilde t_1 \to c \, \tilde \chi^0_1$ getting a maximal branching ratio
of $7 \cdot 10^{-3}$ which is about a factor of 200 larger than the exact
calculation. From this we conclude that results obtained using this
approximate formula have to be taken with some care. From this we conclude
that an observation of this decay mode $\tilde t_1 \to c \, \tilde \chi^0_1$
would be a clear indication of the existence of non-minimal flavour violation
structures.

\section{Conclusions}
\label{sec:conclu}

We have analysed flavour changing processes in a minimal flavour violating
supersymmetric model at the GUT scale. Two examples of this kind have
been studied and compared in more detail, the usual Constrained MSSM and a
completely generic MFV model defined at the GUT scale. In the analysis of
these models we have used two loop renormalization group equations taking into
account for the first time the complete flavour structure in the running of
the parameters.
We have applied all the updated constraints available at the moment including
the last measurement of the $B_s$ mass difference by CDF and the dark matter
relic abundance from the 3rd year results of the WMAP experiment.

The main objective of our analysis has been the study of the $B_s \to \mu^+
\mu^-$ decay. This decay rate is enhanced by additional
powers of $\tan \beta$ when mediated by neutral Higgses. In both the generic
MFV and even the CMSSM scenario
we indeed find that enhancements of the BR($B_s \to \mu^+ \mu^-$) of several
orders of magnitude are still possible consistently with all other
constraints. This implies that the present bound from the D\O~ and CDF
experiments can already rule out a large part of large $\tan \beta$ and
small $M_A$ region in the parameter space of these models.

In the CMSSM, large values (close to the current bound) for the $B_s \to
\mu^+ \mu^-$ decay would necessarily imply a value of $\tan \beta \geq 50$,
a light pseudoscalar, $M_A\leq 570$ GeV and positive sign of the $\mu$
parameter. These points of the parameter space can still satisfy the stringent
$b\to s \gamma$ constraints due to the opposite sign of the chargino and
charged-Higgs contributions. We have also checked explicitly that the new
constraint from BR($B\to X_s l^+ l^-$) eliminate all the points of the
parameter space where a large chargino contribution is able to reverse the
sign of the $C_7$ Wilson coefficient. Unfortunately the new experimental
results on the $B_s$  mass difference are not effective to constrain the
parameter space of the CMSSM once the  $B_s \to \mu^+ \mu^-$ bounds are
applied. This is due to the large theoretical uncertainty in the calculation
of the  $B_s$ mass
difference. With respect to the dark matter constraints, we have seen that the
allowed points with sizeable flavour changing effects correspond to the funnel
and coannihilation regions.

In the generic MFV model many of the well-defined regions observed in the
CMSSM scenario are smeared out due to the fact that the Higgs masses are now
independent from the other soft-breaking parameters. Now, having $8 \times
10^{-8} \leq {\rm BR}(B_s \to \mu^+ \mu^-) \leq 1 \times 10^{-7}$ would only
require $\tan \beta \geq 20$ and $M_A \leq 950$ GeV. Moreover, it is still
possible to obtain these large branching ratios with negative values of the
$\mu$ parameter at the expense of some fine-tuning to fulfil the $b\to s
\gamma$ constraint and satisfying the $\delta a_\mu$ bound at the 3 $\sigma$
level. As in the CMSSM case, the lack of more precise hadronic parameters in
$B_s$ decays prevents us to use the precise experimental results to further
constrain the allowed parameter space. Furthermore we have seen that in a
generic MFV scenario at the GUT scale a new set of points with a correct dark
matter  abundance appears, corresponding to neutralinos with a large higgsino
component. These points can still have relatively light Higgses and sfermions
compared to the focus point region of the CMSSM, and therefore their
contribution to FCNC processes can be sizeable.

Finally we have also explored the phenomenology to be expected at collider
experiments as LHC if a  $8 \times
10^{-8} \leq {\rm BR}(B_s \to \mu^+ \mu^-) \leq 1 \times 10^{-7}$ is
measured. The main feature of this scenario is that the SUSY spectrum is
relatively light and SUSY should be discovered at LHC. The neutral and
charged Higgses in this scenario should also be visible at the LHC. In some
case with light $m_{H^+}$ the branching ratio of the top quark into $b \tau^+
\nu_\tau$ will be modified from the SM expectations due to charged Higgs
mediation. We have studied the flavour violating decay $t \to  \chi^0_1
c$ and we have found that large branching ratios are not found in
contrast with the results in the literature obtained using approximate
formulas.

\section*{Acknowledgements}

E.L.~and W.P.~thank the ``Insitut f\"ur Theoretische Physik'' of the University
of Z\"urch where this work has been started for its hospitality.
W.P. is supported by a
MEC Ramon y Cajal contract,  by Spanish grant FPA2005-01269, by the
European Commission Human Potential Program RTN network
MRTN-CT-2004-503369 and by the European Science Foundation network
grant N.86
O.V. acknowledges support from the Spanish MEC and FEDER under contract
FPA2005/01678 and Generalitat Valenciana under contract GV05/267.
Research partly supported by the Department of Energy under Grant
DE-AC02-76CH030000. Fermilab is operated by Universities Research
Association Inc., under contract with the U.S. Department of Energy.

\appendix

\section{Sfermion Mass Matrix}

The $6\times 6$ sfermion mass matrices are given by
\begin{equation}
M^{2}_{\tilde{f}}
=
\left(\begin{array}{cc}
  M^{2}_{\tilde{f}_L}
+ ( T^{3}_{I} - Q_{f} \sin^{2}\theta_{W} ) \cos 2\beta \, m_{Z}^{2}
+ M_{f}^{2}
 & Y_{A_f}^*  \displaystyle {v \over \sqrt{2}}\ \Omega (\beta) -
M_{f}\mu \Theta(\beta)
\\
Y_{A_f} \displaystyle {v \over \sqrt{2}}\ \Omega (\beta) -
M_{f}\mu^{*} \Theta(\beta)
 & M^{2}_{\tilde{f}_R}
+ Q_{f} \sin^{2}\theta_{W} \cos 2\beta \, m_{Z}^{2}
+ M_{f}^{2}
\end{array}\right)
\enspace ,
\end{equation}
where
\begin{eqnarray}
\cases{\Theta(\beta)= \cot\beta,\ \Omega(\beta)=\sin\beta &
 for $T^{3}_{I} =  \frac{1}{2}$ \cr
                    \Theta(\beta)= \tan\beta,\ \Omega(\beta)=\cos\beta &
for $T^{3}_{I} = -\frac{1}{2}$ \cr}
\enspace ,
\end{eqnarray}
and $Y_{A_f}$ are the trilinear matrices equal at $M_{GUT}$ to
$Y_{A_f}= Y_f A_f$.
These matrices are diagonalized by the $6\times 6$ unitary matrices
$\Gamma_{f}$:
\begin{equation}
\mbox{diag} (M_{\tilde f_1},\dots,M_{\tilde f_6}) =
\Gamma_{\tilde f}^{} \cdot M_{\tilde f}^2 \cdot \Gamma_{\tilde f}^\dagger \enspace .
\end{equation}
The $6\times 3$ left and right block components of the mixing matrices are
defined as:
\begin{equation}
\Gamma_{\tilde f}^{6\times 6} = \left( \Gamma^{6\times 3}_{\tilde f L}  \ \
      \Gamma^{6\times 3}_{\tilde f R} \right) \enspace .
\end{equation}

In the flavour blind scenario, the most important off--diagonal entry
in the above squared mass matrices is the third generation LR
mixing. Below we present the analytic expressions of the $2\times 2$
stop system:
\begin{equation}
M^{2}_{\tilde{t}}
=
\left(\begin{array}{cc}
  M^{2}_{\tilde{t}_{LL}}
 & e^{-i \varphi_{\tilde{t}}} M^{2}_{\tilde{t}_{LR}}
\\
  e^{i \varphi_{\tilde{t}}} M^{2}_{\tilde{t}_{LR}}
 & M^{2}_{\tilde{t}_{RR}}
\end{array}\right)
\enspace ,
\label{s1}
\end{equation}
where
\begin{eqnarray}
  M^{2}_{\tilde{t}_{LL}}
&=&
  m^{2}_{Q_3}
+ ( {1\over 2} - {2\over 3} \sin^{2}\theta_{W} ) \cos 2\beta \, m_{Z}^{2}
+ m_{t}^{2}
\enspace ,
\\
  M^{2}_{\tilde{t}_{RR}}
&=&
  m^{2}_{U_3}
+ {2\over 3} \sin^{2}\theta_{W} \cos 2\beta \, m_{Z}^{2}
+ m_{t}^{2}
\enspace ,
\\
  M^{2}_{\tilde{t}_{LR}}
&=&
  m_{t} | A_{t} - \mu^{*} \Theta(\beta) |
\enspace ,
\\
\label{sfphase}
  \varphi_{\tilde{t}}
&=&
  \arg [ A_{t} - \mu^{*} \Theta(\beta) ]
\enspace ,
\end{eqnarray}
The eigenvalues are given by
\begin{equation}
2 m^{2}_{\tilde t_1, \tilde t_2}
= ( M^{2}_{\tilde{t}_{LL}} + M^{2}_{\tilde{t}_{RR}} )
\mp \sqrt{ ( M^{2}_{\tilde{t}_{LL}} - M^{2}_{\tilde{t}_{RR}} )^{2}
         + 4 ( M^{2}_{\tilde{t}_{LR}} )^{2}}
\enspace ,
\end{equation}
with $m^2_{\tilde t_1} \le m^{2}_{\tilde t_2}$.
We parametrize the mixing matrix ${\mathcal R}^{\tilde{t}}$ so that
\begin{equation}
\label{s2}
\left(\begin{array}{c}
  \tilde{t}_{1} \\ \tilde{t}_{2}
\end{array}\right)
=
{\mathcal R}^{\tilde{t}}
\left(\begin{array}{c}
  \tilde{t}_{L} \\ \tilde{t}_{R}
\end{array}\right)
=
\left(\begin{array}{cc}
  e^{\frac{i}{2} \varphi_{\tilde{t}}} \cos \theta_{\tilde{t}}
 & e^{-\frac{i}{2} \varphi_{\tilde{t}}} \sin \theta_{\tilde{t}}
\\
  - e^{\frac{i}{2} \varphi_{\tilde{t}}} \sin \theta_{\tilde{t}}
 & e^{-\frac{i}{2} \varphi_{\tilde{t}}} \cos \theta_{\tilde{t}}
\end{array}\right)
\left(\begin{array}{c}
  \tilde{t}_{L} \\ \tilde{t}_{R}
\end{array}\right)
\enspace ,
\end{equation}
where $\varphi_{\tilde{t}}$ is given in Eq.~(\ref{sfphase}) and
\begin{eqnarray} &&
\cos\theta_{\tilde{t}}
=
\frac{-M^{2}_{\tilde{t}_{LR}}}{\Delta}
\leq 0
\enspace , \quad
\sin\theta_{\tilde{t}}
=
\frac{M^{2}_{\tilde{t}_{LL}} - m^2_{\tilde t_1}}{\Delta}
\geq 0
\enspace ,
\nonumber \\ &&
\Delta^{2}
=
  ( M^{2}_{\tilde{t}_{LR}} )^{2}
+ ( m^2_{\tilde t_1} - M^{2}_{\tilde{t}_{LL}} )^{2}
\enspace .
\end{eqnarray}


\section{Chargino Mass Matrix}

The chargino mass matrix
\begin{equation}\label{charmass}
M^{\tilde{\chi}^{+}}_{\alpha\beta} =
\left(
\begin{array}{cc}
  M_2                        & m_{W} \sqrt{2} \sin\beta  \\
  m_{W} \sqrt{2} \cos\beta & \mu
\end{array}
\right)
\end{equation}
can be diagonalized by the biunitary transformation
\begin{equation}
U^{*}_{j\alpha} M^{\tilde{\chi}^{+}}_{\alpha\beta} V^{*}_{k\beta}
= m_{\tilde{\chi}_{j}^{+}} \delta_{jk}
\enspace ,
\end{equation}
where $U$ and $V$ are unitary matrices such that
$m_{\tilde{\chi}_{j}^{+}}$ are positive and
$m_{\tilde{\chi}_{1}^{+}} < m_{\tilde{\chi}_{2}^{+}}$.


\section{Neutralino Mass Matrix}

We define $N_{\alpha j}$ as the unitary matrix which makes the complex
symmetric neutralino mass matrix diagonal with positive diagonal
elements:
\begin{equation}
N_{\alpha j} M^{\tilde{\chi}^{0}}_{\alpha\beta} N_{\beta k}
= m_{\tilde{\chi}^{0}_{j}}\delta_{jk}
\enspace ,
\end{equation}
where $m_{\tilde{\chi}^{0}_{j}} < m_{\tilde{\chi}^{0}_{k}}$ for $j<k$.
In the basis:
\begin{equation}
\psi_{\alpha} =
\{ \tilde{B},\tilde{W}^0,\tilde{H}^{a},\tilde{H}^{b} \}
\enspace ,
\end{equation}
the complex symmetric neutralino mass matrix has the form
\begin{equation}\label{neutmass}
M^{\tilde{\chi}^{0}}_{\alpha\beta} =
\left(
\begin{array}{cccc}
M_1 & 0 &  - m_{Z}\cos \beta  \sin \theta_W  &  m_{Z} \sin \beta \sin \theta_W \\
0 & M_2 &  m_{Z}\cos \beta  \cos \theta_W & - m_{Z} \sin \beta  \cos \theta_W  \\
 - m_{Z}\cos \beta  \sin \theta_W & m_{Z}\cos \beta  \cos \theta_W & 0 & - \mu \\
m_{Z} \sin \beta \sin \theta_W & - m_{Z} \sin \beta  \cos \theta_W  & - \mu & 0
\end{array}
\right)
\enspace .
\end{equation}


\section{Loop Functions}

In this appendix, we collect the different loop functions in the text.

\begin{eqnarray}
H_2(x,y)&=& \Frac{x \log x}{(1-x)(x-y)} + \Frac{y \log y}{(1-y)(y-x)} \\
S_0(x)&=&{4 x - 11 x^2 +x^3 \over4 (1-x)^2} -{ 3x^3\over 2(1-x)^3}\log x
\\
f_1(x) &=& {-7 + 5 x + 8x^2  \over 6(1-x)^3} - {2 x - 3 x^2 \over (1-x)^4}
  \log x \\
f_2(x) &=& {3x-5x^2 \over 2(1-x)^2} + {2x-3x^2 \over (1-x)^3}
  \log x \\
  f_3(x)   &=&
        \frac{1}{2(x-1)^3}(x^2-4x+3+2\ln x),\\
Y(x)&=& {-3 x^3 +2 x^2 \over 2 (x-1)^4}\log x + {8 x^3+5 x^2-7x \over 12 (x-1)^3}\\
W(x)&=& {-32 x^4+38x^3+15 x^2-18 x\over 18 (x-1)^4}
\log x +\nn \\
&&{-18 x^4 +163 x^3-259x^2+108x\over 36(x-1)^3}\\
f_5(x) &=& {x\over 1-x} + {x\over (1-x)^2} \log x\\
f_6(x) &=& {38x-79x^2+47x^3\over 6(1-x)^3} +{4x-6x^2+3 x^4 \over (1-x)^4} \log x\\
f_7(x) &=& {52-101 x+43 x^2\over 6(1-x)^3} +
           {6-9x+2x^3\over (1-x)^4} \log x\\
c_0(m_1^2,m_2^2,m_3^2)&=&
-\Bigg[ {m_1^2 \log {m_1^2\over\mu^2} \over
       (m_1^2-m_2^2)(m_1^2-m_3^2)} + (m_1
       \leftrightarrow m_2) + (m_1 \leftrightarrow m_3) \Bigg]  \\
c_2(m_1^2,m_2^2,m_3^2)&=& {3\over 8} -{1\over 4} \Bigg[
{m_1^2 \log {m_1^4\over\mu^2} \over
       (m_1^2-m_2^2)(m_1^2-m_3^2)} + (m_1
       \leftrightarrow m_2) + (m_1 \leftrightarrow m_3) \Bigg]\\
d_2(m_1^2,m_2^2,m_3^2,m_4^2)&=&-{1\over 4} \Bigg[
{m_1^4 \log {m_1^4\over\mu^2} \over
       (m_1^2-m_2^2)(m_1^2-m_3^2)(m_1^2-m_4^2)} + (m_1
       \leftrightarrow m_2) + (m_1 \leftrightarrow m_3) \nonumber\\
        & & +
       (m_1 \leftrightarrow m_4) \Bigg]
\end{eqnarray}


\begin{thebibliography}{99}




\bibitem{fcncreview} For a recent review, see
\\
A.~Masiero and O.~Vives,
Ann.\ Rev.\ Nucl.\ Part.\ Sci.\  {\bf 51} (2001) 161
[arXiv:hep-ph/0104027];
\\
A.~Masiero and O.~Vives,
New Jour.\ Phys.\  {\bf 4} (2002) 4.
%

\bibitem{MFV}
  G.~D'Ambrosio, G.~F.~Giudice, G.~Isidori and A.~Strumia,
  Nucl.\ Phys.\ B {\bf 645}, 155 (2002)
  [arXiv:hep-ph/0207036].

\bibitem{flavour}
M.~Dine, R.~G.~Leigh and A.~Kagan,
Phys.\ Rev.\ D {\bf 48} (1993) 4269
[arXiv:hep-ph/9304299];
\\
  G.~G.~Ross, L.~Velasco-Sevilla and O.~Vives,
  Nucl.\ Phys.\ B {\bf 692} (2004) 50
  [arXiv:hep-ph/0401064].
\\
For a review and further references see:\\
G.~G.~Ross, ``Models of Fermion masses'', Published
in \textit{TASI 2000} ed. J.~L.~Rosner (World Scientific,
New Jersey, 2001)

\bibitem{Babu:1999hn}
  K.~S.~Babu and C.~F.~Kolda,
  Phys.\ Rev.\ Lett.\  {\bf 84} (2000) 228
  [arXiv:hep-ph/9909476].

\bibitem{Choudhury:1998ze}
  S.~R.~Choudhury and N.~Gaur,
  Phys.\ Lett.\ B {\bf 451} (1999) 86
  [arXiv:hep-ph/9810307].

\bibitem{Isidori:2001fv}
  G.~Isidori and A.~Retico,
  JHEP {\bf 0111} (2001) 001
  [arXiv:hep-ph/0110121].

\bibitem{Bobeth:2002ch}
  C.~Bobeth, T.~Ewerth, F.~Kruger and J.~Urban,
  Phys.\ Rev.\ D {\bf 66} (2002) 074021
  [arXiv:hep-ph/0204225].

\bibitem{Chankowski:2000ng}
  P.~H.~Chankowski and L.~Slawianowska,
  Phys.\ Rev.\ D {\bf 63} (2001) 054012
  [arXiv:hep-ph/0008046].

\bibitem{Buras:2002wq}
  A.~J.~Buras, P.~H.~Chankowski, J.~Rosiek and L.~Slawianowska,
  Phys.\ Lett.\ B {\bf 546} (2002) 96
  [arXiv:hep-ph/0207241].

\bibitem{Buras:2002vd}
  A.~J.~Buras, P.~H.~Chankowski, J.~Rosiek and L.~Slawianowska,
  Nucl.\ Phys.\ B {\bf 659} (2003) 3
  [arXiv:hep-ph/0210145].

 \bibitem{Isidori:2006pk}
  G.~Isidori and P.~Paradisi,
  arXiv:hep-ph/0605012.

\bibitem{Carena:2006ai}
  M.~Carena, A.~Menon, R.~Noriega-Papaqui, A.~Szynkman and C.~E.~M.~Wagner,
  arXiv:hep-ph/0603106.

\bibitem{Baek:2002wm}
  S.~Baek, P.~Ko and W.~Y.~Song,
  JHEP {\bf 0303} (2003) 054
  [arXiv:hep-ph/0208112].

\bibitem{Dedes:2001fv}
  A.~Dedes, H.~K.~Dreiner and U.~Nierste,
  Phys.\ Rev.\ Lett.\  {\bf 87} (2001) 251804
  [arXiv:hep-ph/0108037].

\bibitem{Arnowitt:2002cq}
  R.~Arnowitt, B.~Dutta, T.~Kamon and M.~Tanaka,
  Phys.\ Lett.\ B {\bf 538} (2002) 121
  [arXiv:hep-ph/0203069].

\bibitem{Ibrahim:2002fx}
  T.~Ibrahim and P.~Nath,
  Phys.\ Rev.\ D {\bf 67} (2003) 016005
  [arXiv:hep-ph/0208142].

\bibitem{Mizukoshi:2002gs}
  J.~K.~Mizukoshi, X.~Tata and Y.~Wang,
  Phys.\ Rev.\ D {\bf 66} (2002) 115003
  [arXiv:hep-ph/0208078].

\bibitem{Dedes:2002zx}
  A.~Dedes, H.~K.~Dreiner, U.~Nierste and P.~Richardson,
  arXiv:hep-ph/0207026.

\bibitem{Baer:2002gm}
  H.~Baer, C.~Balazs, A.~Belyaev, J.~K.~Mizukoshi, X.~Tata and Y.~Wang,
  JHEP {\bf 0207} (2002) 050
  [arXiv:hep-ph/0205325].

\bibitem{Blazek:2003hv}
  T.~Blazek, S.~F.~King and J.~K.~Parry,
  Phys.\ Lett.\ B {\bf 589} (2004) 39
  [arXiv:hep-ph/0308068].

\bibitem{Kane:2003wg}
  G.~L.~Kane, C.~Kolda and J.~E.~Lennon,
  arXiv:hep-ph/0310042.

\bibitem{Ellis:2005sc}
  J.~R.~Ellis, K.~A.~Olive and V.~C.~Spanos,
  Phys.\ Lett.\ B {\bf 624} (2005) 47
  [arXiv:hep-ph/0504196].

\bibitem{Baek:2004et}
  S.~Baek, Y.~G.~Kim and P.~Ko,
  JHEP {\bf 0502} (2005) 067
  [arXiv:hep-ph/0406033].

\bibitem{Dermisek:2003vn}
  R.~Dermisek, S.~Raby, L.~Roszkowski and R.~Ruiz De Austri,
  JHEP {\bf 0304} (2003) 037
  [arXiv:hep-ph/0304101].

\bibitem{Ellis2006}
  J.~Ellis, K.~A.~Olive, Y.~Santoso and V.~C.~Spanos,
  arXiv:hep-ph/0603136.

\bibitem{Bartl:2001}
  A.~Bartl, T.~Gajdosik, E.~Lunghi, A.~Masiero, W.~Porod, H.~Stremnitzer and O.~Vives,
  Phys.\ Rev.\ D {\bf 64}, 076009 (2001)
  [arXiv:hep-ph/0103324].

\bibitem{Polonsky}
 N.~Polonsky and A.~Pomarol,
 Phys.\ Rev.\ D {\bf 51}, 6532 (1995)
 [arXiv:hep-ph/9410231].


\bibitem{Ibanez}
 L.~E.~Iba\~nez, D.~Lust and G.~G.~Ross,
  Phys.\ Lett.\ B {\bf 272} (1991) 251
  [arXiv:hep-th/9109053];
\\
 G.~G.~Ross,
  arXiv:hep-ph/0411057.


\bibitem{Martin:1993zk}
 S.~Martin and M.~Vaughn, Phys.~Rev.~{\bf D50}, 2282 (1994);
Y.~Yamada,  Phys.~Rev.~{\bf D~50}, 3537 (1994);
I.~Jack, D.R.T.~Jones, Phys.~Lett.~{\bf B333} (1994) 372.

\bibitem{wmap3rdyear}
  D.~N.~Spergel {\it et al.},
  arXiv:astro-ph/0603449.



\bibitem{Abazov:2006dm}
  V.~Abazov  [D0 Collaboration],
  arXiv:hep-ex/0603029.

\bibitem{CDFBs:2006}
`` Measurement of the  Bs - Bs Oscillation Frequency'', available at CDF's
B-physics  webpage:
http://www-cdf.fnal.gov/physics/new/bottom/bottom.html

\bibitem{Bennett:2004pv}
  G.~W.~Bennett {\it et al.}  [Muon g-2 Collaboration],
  Phys.\ Rev.\ Lett.\  {\bf 92}, 161802 (2004)
  [arXiv:hep-ex/0401008].

\bibitem{Abe:2005.qqqq}
  M.~Iwasaki {\it et al.}  [Belle Collaboration],
  arXiv:hep-ex/0503044.

\bibitem{Aubert:2004it}
B.~Aubert {\it et al.}  [BABAR Collaboration],
Phys.\ Rev.\ Lett.\  {\bf 93} (2004) 081802
[arXiv:hep-ex/0404006].

\bibitem{Eidelman:2004wy}
  S.~Eidelman {\it et al.}  [Particle Data Group],
  Phys.\ Lett.\ B {\bf 592}, 1 (2004).


\bibitem{CDF-d0}
    [CDF Collaboration],
  hep-ex/0507091.

\bibitem{last}
CDF Collaboration, Public note 8176.\\
D\O~Collaboration, note 5009-CONF.


\bibitem{Dedes:2002er}
  A.~Dedes and A.~Pilaftsis,
  Phys.\ Rev.\ D {\bf 67} (2003) 015012
  [arXiv:hep-ph/0209306].

\bibitem{Brignole:2003iv}
  A.~Brignole and A.~Rossi,
  Phys.\ Lett.\ B {\bf 566} (2003) 217
  [arXiv:hep-ph/0304081].

\bibitem{Paradisi:2005tk}
  P.~Paradisi,
  JHEP {\bf 0602} (2006) 050
  [arXiv:hep-ph/0508054].

\bibitem{Masiero:2005wr}
  A.~Masiero, P.~Paradisi and R.~Petronzio,
  arXiv:hep-ph/0511289.

\bibitem{Curiel:2003uk}
  A.~M.~Curiel, M.~J.~Herrero, W.~Hollik, F.~Merz and S.~Penaranda,
  Phys.\ Rev.\ D {\bf 69} (2004) 075009
  [arXiv:hep-ph/0312135].



\bibitem{Kanemura:2004cn}
  S.~Kanemura, K.~Matsuda, T.~Ota, T.~Shindou, E.~Takasugi and K.~Tsumura,
  Phys.\ Lett.\ B {\bf 599} (2004) 83
  [arXiv:hep-ph/0406316].

\bibitem{Ibrahim:2004cf}
  T.~Ibrahim, P.~Nath and A.~Psinas,
  Phys.\ Rev.\ D {\bf 70} (2004) 035006
  [arXiv:hep-ph/0404275].

\bibitem{Ibrahim:2004gb}
  T.~Ibrahim and P.~Nath,
  Phys.\ Rev.\ D {\bf 71} (2005) 055007
  [arXiv:hep-ph/0411272].

\bibitem{Hollik:2005as}
  W.~Hollik, S.~Penaranda and M.~Vogt,
  arXiv:hep-ph/0511021.

\bibitem{prev}
  A.~Abulencia {\it et al.}  [CDF Collaboration],
  Phys.\ Rev.\ Lett.\  {\bf 95} (2005) 221805
  [Erratum-ibid.\  {\bf 95} (2005) 249905]
  [arXiv:hep-ex/0508036];
\\
  V.~M.~Abazov {\it et al.}  [D0 Collaboration],
  Phys.\ Rev.\ Lett.\  {\bf 94} (2005) 071802
  [arXiv:hep-ex/0410039].


\bibitem{Isidori:2006jh}
  G.~Isidori and P.~Paradisi,
  arXiv:hep-ph/0601094.

\bibitem{others}
  A.~J.~Buras, P.~H.~Chankowski, J.~Rosiek and L.~Slawianowska,
  Nucl.\ Phys.\ B {\bf 619} (2001) 434
  [arXiv:hep-ph/0107048].


\bibitem{misiak}
  P.~Gambino, U.~Haisch and M.~Misiak,
  Phys.\ Rev.\ Lett.\  {\bf 94}, 061803 (2005)
  [arXiv:hep-ph/0410155].

\bibitem{HLMW}
  T.~Huber, E.~Lunghi, M.~Misiak and D.~Wyler,
  arXiv:hep-ph/0512066.


\bibitem{Ciuchini:1997xe}
   M.~Ciuchini, G.~Degrassi, P.~Gambino and G.~F.~Giudice,
   Nucl.\ Phys.\ B {\bf 527}, 21 (1998)
   [arXiv:hep-ph/9710335].

\bibitem{Buras:2002tp}
 A.~J.~Buras, A.~Czarnecki, M.~Misiak and J.~Urban,
 Nucl.\ Phys.\ B {\bf 631}, 219 (2002)
 [arXiv:hep-ph/0203135];
 A.~J.~Buras and M.~Misiak,
 Acta Phys.\ Polon.\ B {\bf 33}, 2597 (2002)
 [arXiv:hep-ph/0207131].

\bibitem{Borzumati:1998tg}
  F.~M.~Borzumati and C.~Greub,
  Phys.\ Rev.\ D {\bf 58} (1998) 074004
  [arXiv:hep-ph/9802391].

\bibitem{Borzumati:1998nx}
  F.~M.~Borzumati and C.~Greub,
  Phys.\ Rev.\ D {\bf 59} (1999) 057501
  [arXiv:hep-ph/9809438].

\bibitem{Bobeth:1999ww}
  C.~Bobeth, M.~Misiak and J.~Urban,
  Nucl.\ Phys.\ B {\bf 567} (2000) 153
  [arXiv:hep-ph/9904413].

\bibitem{Degrassi:2000qf}
   G.~Degrassi, P.~Gambino and G.~F.~Giudice,
   JHEP {\bf 0012}, 009 (2000)
   [arXiv:hep-ph/0009337].

\bibitem{Carena:2000uj}
  M.~Carena, D.~Garcia, U.~Nierste and C.~E.~M.~Wagner,
  Phys.\ Lett.\ B {\bf 499}, 141 (2001)
  [arXiv:hep-ph/0010003].

\bibitem{Degrassi:2006eh}
  G.~Degrassi, P.~Gambino and P.~Slavich,
  arXiv:hep-ph/0601135.

\bibitem{cleobsg} CLEO Collaboration (S. Chen {\it et al.}), \prl{87}, 251807 (2001).

\bibitem{bellebsg1} Belle Collaboration (K. Abe {\it et al.}), Phys. Lett/ B {\bf 511}, 151 (2001).

\bibitem{bellebsg2} Belle Collaboration (P. Koppunberg {\it et al.}), \prl{93}, 061803 (2004).

\bibitem{babarbsg1} BaBar Collaboration (B. Aubert {\it et al.}), \prd{72}, 052004 (2005).

\bibitem{babarbsg2} BaBar Collaboration (B. Aubert {\it et al.}), hep-ex/0507001.

\bibitem{hfag}
  Heavy Flavor Averaging Group,
  arXiv:hep-ex/0603003.

\bibitem{Hurth:2003dk}
  T.~Hurth, E.~Lunghi and W.~Porod,
  Nucl.\ Phys.\ B {\bf 704}, 56 (2005)
  [arXiv:hep-ph/0312260].

\bibitem{chomisiak}
  P.~L.~Cho, M.~Misiak and D.~Wyler,
  Phys.\ Rev.\ D {\bf 54}, 3329 (1996)
  [arXiv:hep-ph/9601360].

\bibitem{Bertolini:1990if}
  S.~Bertolini, F.~Borzumati, A.~Masiero and G.~Ridolfi,
  Nucl.\ Phys.\ B {\bf 353}, 591 (1991).

\bibitem{haber}
  H.~E.~Haber and G.~L.~Kane,
  Phys.\ Rept.\  {\bf 117}, 75 (1985).

\bibitem{moroi}
  T.~Moroi,
  Phys.\ Rev.\ D {\bf 53}, 6565 (1996)
  [Erratum-ibid.\ D {\bf 56}, 4424 (1997)]
  [arXiv:hep-ph/9512396].

\bibitem{BNLexp}
  G.~W.~Bennett  [Muon Collaboration],
  arXiv:hep-ex/0602035.

\bibitem{Jegerlehner:2003qp}
  F.~Jegerlehner,
  Nucl.\ Phys.\ Proc.\ Suppl.\  {\bf 126} (2004) 325
  [arXiv:hep-ph/0310234].


\bibitem{unknown:2006cr}
    [ALEPH Collaboration],
  arXiv:hep-ex/0602042.


\bibitem{refs}
  J.~R.~Ellis, K.~A.~Olive, Y.~Santoso and V.~C.~Spanos,
  Phys.\ Lett.\ B {\bf 565}, 176 (2003)
  [arXiv:hep-ph/0303043].

\bibitem{kraml}
  G.~Belanger, S.~Kraml and A.~Pukhov,
  Phys.\ Rev.\ D {\bf 72}, 015003 (2005)
  [arXiv:hep-ph/0502079].

\bibitem{Mackenzie:2006un}
  P.~B.~Mackenzie,
  arXiv:hep-ph/0606034.

\bibitem{Pierce:1996zz}
D.~M.~Pierce et al.,
Nucl.\ Phys.\ B {\bf 491} (1997) 3.

\bibitem{PorodNew}
W.~Porod, in preparation

\bibitem{Degrassi:2001yf}
G.~Degrassi, P.~Slavich and F.~Zwirner,
Nucl.\ Phys.\ B {\bf 611} (2001) 403.

\bibitem{brignole}
A.~Brignole, G.~Degrassi, P.~Slavich and F.~Zwirner,
Nucl.\ Phys.\ B {\bf 631} (2002) 195.

\bibitem{Brignole:2002bz}
A.~Brignole, G.~Degrassi, P.~Slavich and F.~Zwirner,
Nucl.\ Phys.\ B {\bf 643} (2002) 79.

\bibitem{Dedes:2003km}
A.~Dedes, G.~Degrassi and P.~Slavich,
Nucl.\ Phys.\ B {\bf 672} (2003) 144.
%

\bibitem{DedesSlavich}
A.~Dedes and P.~Slavich,
Nucl.\ Phys.\ B {\bf 657} (2003) 333.

\bibitem{ADKPS}
  B.~C.~Allanach, A.~Djouadi, J.~L.~Kneur, W.~Porod and P.~Slavich,
  JHEP {\bf 0409} (2004) 044.

\bibitem{Bobeth:2001jm}
  C.~Bobeth, A.~J.~Buras, F.~Kruger and J.~Urban,
  Nucl.\ Phys.\ B {\bf 630} (2002) 87
  [arXiv:hep-ph/0112305].

\bibitem{Goto:1996dh}
  T.~Goto, Y.~Okada, Y.~Shimizu and M.~Tanaka,
  Phys.\ Rev.\ D {\bf 55} (1997) 4273
  [Erratum-ibid.\ D {\bf 66} (2002) 019901]
  [arXiv:hep-ph/9609512].

\bibitem{Baek:2001kh}
  S.~Baek, T.~Goto, Y.~Okada and K.~i.~Okumura,
  Phys.\ Rev.\ D {\bf 64} (2001) 095001
  [arXiv:hep-ph/0104146].

\bibitem{Ibrahim:1999aj}
  T.~Ibrahim and P.~Nath,
  Phys.\ Rev.\ D {\bf 62} (2000) 015004
  [arXiv:hep-ph/9908443].


\bibitem{Ball:2006xx}
  P.~Ball and R.~Fleischer,
  arXiv:hep-ph/0604249.

\bibitem{Dedes:2004yc}
  A.~Dedes and B.~T.~Huffman,
  Phys.\ Lett.\ B {\bf 600} (2004) 261
  [arXiv:hep-ph/0407285].

\bibitem{Asai:2002xv}
S.~Asai  [ATLAS and CMS Collaborations],
Eur.\ Phys.\ J.\ directC {\bf 4S1} (2002) 17.

\bibitem{Djouadi:2005gj}
  A.~Djouadi,
  arXiv:hep-ph/0503173.

\bibitem{Baer:1994xr}
  H.~Baer, J.~Sender and X.~Tata,
  Phys.\ Rev.\ D {\bf 50} (1994) 4517
  [arXiv:hep-ph/9404342].

\bibitem{Kon:1994uc}
  T.~Kon and T.~Nonaka,
  Phys.\ Rev.\ D {\bf 50} (1994) 6005
  [arXiv:hep-ph/9405327].

\bibitem{Sender:1996qc}
  J.~Sender,
  Phys.\ Rev.\ D {\bf 54} (1996) 3271.

\bibitem{Porod:1996at}
  W.~Porod and T.~W\"ohrmann,
  Phys.\ Rev.\ D {\bf 55} (1997) 2907
  [Erratum-ibid.\ D {\bf 67} (2003) 059902]
  [arXiv:hep-ph/9608472].

\bibitem{Hosch:1997vf}
  M.~Hosch, R.~J.~Oakes, K.~Whisnant, J.~M.~Yang, B.~l.~Young and X.~Zhang,
  Phys.\ Rev.\ D {\bf 58} (1998) 034002
  [arXiv:hep-ph/9711234].

\bibitem{Porod:1998yp}
  W.~Porod,
  Phys.\ Rev.\ D {\bf 59} (1999) 095009
  [arXiv:hep-ph/9812230].

\bibitem{Boehm:1999tr}
  C.~Boehm, A.~Djouadi and Y.~Mambrini,
  Phys.\ Rev.\ D {\bf 61} (2000) 095006
  [arXiv:hep-ph/9907428].

\bibitem{Restrepo:2001me}
  D.~Restrepo, W.~Porod and J.~W.~F.~Valle,
  Phys.\ Rev.\ D {\bf 64} (2001) 055011
  [arXiv:hep-ph/0104040].

\bibitem{Djouadi:2001dx}
  A.~Djouadi, M.~Guchait and Y.~Mambrini,
  Phys.\ Rev.\ D {\bf 64} (2001) 095014
  [arXiv:hep-ph/0105108].

\bibitem{Das:2001kd}
  S.~P.~Das, A.~Datta and M.~Guchait,
  Phys.\ Rev.\ D {\bf 65} (2002) 095006
  [arXiv:hep-ph/0112182].

\bibitem{Porod:2003um}
  W.~Porod,
  Comput.\ Phys.\ Commun.\  {\bf 153} (2003) 275
  [arXiv:hep-ph/0301101].

\bibitem{Das:2003pe}
  S.~P.~Das, A.~Datta and M.~Guchait,
  Phys.\ Rev.\ D {\bf 70} (2004) 015009
  [arXiv:hep-ph/0309168].

\bibitem{Muhlleitner:2003vg}
  M.~Muhlleitner, A.~Djouadi and Y.~Mambrini,
  Comput.\ Phys.\ Commun.\  {\bf 168} (2005) 46
  [arXiv:hep-ph/0311167].

\bibitem{Han:2003qe}
  T.~Han, K.~I.~Hikasa, J.~M.~Yang and X.~m.~Zhang,
  Phys.\ Rev.\ D {\bf 70} (2004) 055001
  [arXiv:hep-ph/0312129].

\bibitem{Hikasa:1987db}
K.~Hikasa and M.~Kobayashi,
Phys.\ Rev.\ D {\bf 36} (1987) 724.


\end{thebibliography}
\end{document}